\RequirePackage[backend=biber]{biblatex}
\documentclass[UKenglish,a4paper,thm-restate]{lipics-v2021}
\nolinenumbers

\newif\ifdebug\debugfalse
\newif\ifshowsketches\showsketchestrue

\newif\ifparlineno\parlinenofalse

\RequirePackage{nameref}

\usepackage{suffix}

\usepackage[utf8]{inputenc}
\usepackage[T1]{fontenc}

\usepackage{phoenician}

\usepackage{stmaryrd}
\newcommand\contradiction{\enspace$\lightning$}

\usepackage{hyperref,csquotes,hyphenat}
\usepackage{comment}
\usepackage{modular}
\usepackage{amsmath,amsthm,amssymb,mathtools,thmtools}
\usepackage{dsfont,interval,braket}
\usepackage[alignedforall]{lpform}
\usepackage{acronym}
\usepackage{multicol}
\usepackage{xparse,enumerate,needspace}
\usepackage [disable]
{todonotes}
\newcommand\progress[1]{}
\usepackage{float}
\usepackage{algorithm,algorithmicx}
\usepackage[noend]{algpseudocode}
\usepackage{pgfplots}
\usepackage{tikz,subcaption,placeins}
\usetikzlibrary{decorations.pathreplacing,calligraphy,patterns,arrows.meta,external}
\usetikzlibrary{math} \usetikzlibrary{arrows,decorations.pathreplacing,patterns}
\makeatletter

\tikzset{drawscheduleapplystyle/.code={\tikzset{#1}}}

\def\schedule@draw<#1>[#2] at(#3) #4;{
  \edef\globalopts{#2}
\if\relax\detokenize{#1}\relax
\let\schedule@list\empty
    \foreach \x in {#4} {
      \ifx\schedule@list\empty
        \xdef\schedule@list{\x}
      \else
        \xdef\schedule@list{\x,\schedule@list}
      \fi
    }
  \else
\xdef\schedule@list{#4}
  \fi
\xdef\drawscheduleWidth{0.0}
  \begin{scope}[shift={(#3)}]
  \foreach [count=\i from 0] \machine/\w/\machineopts in \schedule@list {
    \edef\newWidth{\drawscheduleWidth+\w}
    \xdef\drawscheduleTime{0.0}
    \ifdim \w pt > 0 pt
      \foreach \t/\labl/\jobopts/\nodeopts in \machine {
        \edef\newTime{\drawscheduleTime+\t}
        \if\relax\detokenize{#1}\relax
          \draw [drawscheduleapplystyle/.expand once=\globalopts,drawscheduleapplystyle/.expand once=\machineopts,drawscheduleapplystyle/.expand once=\jobopts] (\drawscheduleTime,\drawscheduleWidth) rectangle (\newTime,\newWidth) node[pos=.5,drawscheduleapplystyle/.expand once=\nodeopts] {\labl};
        \else
          \draw [drawscheduleapplystyle/.expand once=\globalopts,drawscheduleapplystyle/.expand once=\machineopts,drawscheduleapplystyle/.expand once=\jobopts] (\drawscheduleWidth,\drawscheduleTime) rectangle (\newWidth,\newTime) node[pos=.5,drawscheduleapplystyle/.expand once=\nodeopts] {\labl};
        \fi
        \xdef\drawscheduleTime{\newTime}
      }
      \xdef\drawscheduleWidth{\newWidth}
    \else
      \xdef\drawscheduleWidth{\drawscheduleWidth-\w}
    \fi
  }
  \end{scope}
}

\newcommand\schedule{}
\def\schedule{\schedule@checkorient}
\def\schedule@checkorient{\pgfutil@ifnextchar u{\schedule@up}{\schedule@checkops<>}}
\def\schedule@up up#1;{\schedule@checkops<u>#1;}
\def\schedule@checkops<#1>{\pgfutil@ifnextchar[{\schedule@ops<#1>}{\schedule@checkat<#1>[]}}
\def\schedule@checkat<#1>[#2]{\pgfutil@ifnextchar a{\schedule@at<#1>[#2]}{\schedule@draw<#1>[#2] at(0,0) }}
\def\schedule@ops<#1>[#2]#3;{\schedule@checkat<#1>[#2]#3;}
\def\schedule@at<#1>[#2]#3at#4(#5) #6;{\schedule@draw<#1>[#2] at(#5) #6;}

\newcommand\schedule@drawlines[4][]{
  \foreach \t/\labl/\drawopts/\nodeopts in {#4} {
    \if\relax\detokenize{#1}\relax
      \draw [dash pattern={on 1.5pt off 0.8pt},drawscheduleapplystyle/.expand once=\drawopts] (\t,#3+#2) -- (\t,-#2) node [below,drawscheduleapplystyle/.expand once=\nodeopts] {\labl};
    \else
      \draw [dash pattern={on 1.5pt off 0.8pt},drawscheduleapplystyle/.expand once=\drawopts] (-#2,\t) -- (#3+#2,\t) node [right,drawscheduleapplystyle/.expand once=\nodeopts] {\labl};
    \fi
  }
}

\newcommand\vlines[3][.2]{\schedule@drawlines{#1}{#2}{#3}}
\newcommand\hlines[3][.2]{\schedule@drawlines[h]{#1}{#2}{#3}}

\makeatother \usepackage{tabulary}

\usepackage[backend=biber]{biblatex}
\AtEveryBibitem{ \clearname {editor} \clearfield {series} \clearfield {url} \clearfield {issn} \clearfield {isbn} \clearfield {eprint} }
\usepackage[
]{hauke_lncs}

\docsvlist{theorem, observation, lemma, proposition, corollary, definition}

\declaretheorem[style=stmt-style, numberwithin=section, name=Invariant]{invariant}

\renewcommand{\lpindent}{\hspace{10pt}}

\usepackage{thm-restate}

\AtBeginDocument{

}
\usepackage{appendix}

\newcommand\aref[1]{\hyperref[#1]{\autoref*{#1} (\nameref*{#1})}}
\usepackage{lineno}
\let\lineref=\relax

\newcommand\lineref[1]{\hyperref[#1]{Line~\ref*{#1}}}

\usepackage{algorithmicx,algpseudocode}
\usepackage{colonequals}

\newcommand{\Oh}{\mathcal{O}}\newcommand{\NP}{\textsc{np}}\newcommand{\veps}{\varepsilon}

\newcommand{\paramrt}{2^{\O((d + \epsinv \log(\epsinv \pmax)) \cdot \log(\epsinv \pmax))}}
\newcommand{\paramsens}{2^{\O(d \log(\epsinv \pmax))}}
\newcommand{\roundedrt}{2^{\O(\epsinv \log(\epsinv) \log\log(\epsinv))}}
\newcommand{\roundedsens}{2^{\O(\epsinv \log(\epsinv) \log\log(\epsinv))}}

\newcommand{\kpconstr}{knapsack constraint \eqref{ip:dyn:constr:blue_area}}

\acrodef{lexmin}[lex. min.]{lexicographically minimal}
\acrodef{confip}[\sc{ConfIP}]{Configuration IP}
\acrodef{EPTAS}[\textsc{eptas}]{Efficient Polynomial-Time Approximation Scheme}
\acrodefplural{EPTAS}[\textsc{eptas}s]{Efficient Polynomial-Time Approximation Schemes}
\acrodef{FPTAS}[\textsc{fptas}]{Fully Polynomial-Time Approximation Scheme}
\acrodefplural{FPTAS}[\textsc{fptas}s]{Fully Polynomial-Time Approximation Schemes}
\acrodef{PTAS}[\textsc{ptas}]{Polynomial-Time Approximation Scheme}
\acrodefplural{PTAS}[\textsc{ptas}s]{Polynomial-Time Approximation Schemes}
\acrodef{FPT}[\textsc{fpt}]{fixed-parameter tractable}
\acrodefplural{FPT}[\textsc{fpt}]{fixed-parameter tractable}
\acrodef{EPAS}[\textsc{epas}]{Efficient Parameterized Approximation Scheme}
\acrodefplural{EPAS}[\textsc{epas}s]{Efficient Parameterized Approximation Schemes}

\newcommand\tsmall{\text{\normalfont small}}

\newcommand\tsuper[2]{\super {\text{\normalfont #1}} #2}
\providecommand\tred{\mathrm{red}}
\providecommand\tblue{\mathrm{blue}}
\providecommand\textcite[1]{\citet{#1}}
\newcommand\at[1]{{\nabla^{(#1)}}}
\NewDocumentCommand{\lexobj}{s}{{\tsuper {\IfBooleanTF#1{mod-lex}{lex}}	\varphi }}
\newcommand\realOPT{\OPT*}
\newcommand\fracOPT{\OPT^{\text{frac}}}

\providecommand\mc{\mathcal}
\providecommand\var{\operatorname{\texttt{var}}~}
\providecommand\param{\operatorname{\texttt{param}}~}

\NewDocumentCommand\ipimprov{o}{\text{\normalfont (ILP)}\xspace }

\newcommand\ipdyn{\text{\normalfont (ILP\textsuperscript{dynamic})}\xspace}
\newcommand\ipdynred{\text{\normalfont (ILP\textsuperscript{dynamic}\textsubscript{decr})}\xspace}
\newcommand\ipdyncmin{\text{\normalfont (ILP\textsuperscript{dynamic}($\Cmin$))}\xspace}

\newcommand\ipconf{\text{Configuration-ILP}\xspace}

\providecommand\rtb{$\mathrm{red} \to \mathrm{blue}$\xspace}
\providecommand\btr{$\mathrm{blue} \to \mathrm{red}$\xspace}
\newcommand\lb\ell\newcommand\mb\Lambda

\providecommand\nub[1][\relax]{\super \tblue {\expandafter #1 \nu}}
\providecommand\mub[1][\relax]{\super \tblue {\expandafter #1 \mu}}
\providecommand\mur[1][\relax]{\super \tred {\expandafter #1 \mu}}

\newcommand\thecrit{\operatorname{critical-type}}
\NewDocumentCommand\crit{s O{} d()}{
	\IfNoValueTF{#3}{
		\dot t
	}{
		\thecrit
		\IfNoValueTF{#1}{
			\enpar[#2]{#3}
		}{
			\enpar*[#2]{#3}
		}
	}
}

\providecommand\crit{\operatorname{critical-type}}

\let\OPT=\relax
\NewDocumentCommand\OPT{t* t+ t-}{
	\let\suffix=\relax
	\IfBooleanT#1{
		\newcommand\suffix{^\star}
	}
	\IfBooleanT#2{
		\newcommand\suffix{^{\uparrow}}
	}
	\IfBooleanT#3{
		\newcommand\suffix{^{\downarrow}}
	}
	\operatorname{OPT\suffix}
}
\NewDocumentCommand\newlname{m m}{
	\IfNoValueTF{#1}{
		,name=\texttt{#2}
	}{
		,name={#1\texttt{[#2]}}
	}
}
\NewDocumentCommand\lname{s m}{\IfBooleanTF#1{{~\texttt{[#2]}}}{\texttt{#2}}}

\newcommand\inv{\bgroup-1\egroup}

\def\eg{e.g.\xspace} \def\ie{i.e.\xspace}  
\newcommand\obda{wlog.\xspace} 

\usepackage{hyperref}
\newcommand{\appendixref}[1]{\hyperref[#1]{Appendix~\ref*{#1}}}

\newcommand\nfoldN{N}
\newcommand\nfoldR{R}
\newcommand\nfoldS{S}
\newcommand\nfoldT{H}
\newcommand\nfoldn\nfoldN\newcommand\nfoldr\nfoldR\newcommand\nfolds\nfoldS\newcommand\nfoldt\nfoldT
\newcommand\nfold{$\nfoldn$-fold\xspace}
\title{Robust Scheduling on Uniform Machines\thanks{Supported by the German Research Foundation (DFG) project \enquote{Fine-grained complexity and algorithms for scheduling and packing} (JA 612/25-1).}}
\subtitle{New Results Using a Relaxed Approximation Guarantee}
\titlerunning{Robust Scheduling on Uniform Machines}

\author{Hauke Brinkop}{Kiel University, Kiel, Germany}{hab@informatik.uni-kiel.de}{https://orcid.org/0000-0002-7791-2353}{}
\ifx\authoranonymous\relax\else
\author{David Fischer}{Helmut Schmidt University, Hamburg, Germany}{fischerd@hsu-hh.de}{https://orcid.org/0000-0001-8402-1818}{}
\author{Klaus Jansen}{Kiel University, Kiel, Germany}{kj@informatik.uni-kiel.de}{https://orcid.org/0000-0001-8358-6796}{}
\authorrunning{H.\@ Brinkop \and D.\@ Fischer \and K.\@ Jansen}
\fi


\begin{document}

	\maketitle
\begin{abstract}
		We consider the problem of scheduling $n$ jobs on $m$ uniform machines while minimizing the makespan ($Q||C_{\max}$) and maximizing the minimum completion time ($Q||C_{\min}$) in an online setting with migration of jobs.
		In this online setting, the jobs are inserted or deleted over time, and at each step, the goal is to compute a near-optimal solution while reassigning some jobs, such that the overall processing time of reassigned jobs, called migration, is bounded by some factor $\beta$ times the processing time of the job added or removed.
		
		We propose Efficient Polynomial Time Approximation Schemes (EPTASs) with an additional load error of $\mathcal{O}(\varepsilon p_{\max})$ for both problems, with constant amortized migration factor $\beta$, where $p_{\max}$ is the maximum processing time in the instance over all steps.
		As an intermediate step, we obtain Efficient Parameterized Approximation Schemes (EPASs) for both problems, $(1+\varepsilon)$-competitive algorithms parameterized by $p_{\max}$ and the number of different processing times $d$ in an instance, with $\beta$ bounded in a function of $p_{\max}$, $d$ and $\varepsilon$.
		
		This is the first result in the direction of a polynomial time approximation  scheme in the field of online scheduling with bounded reassignment on uniform machines; before, such results were known only for the considered problems on identical machines.
		Crucial to our result is a division of the machines into large and small machines depending on the current approximate objective value, allowing for different approaches on either machine set, as well as a new way of rounding the instance that does not depend on the current objective value.
	\end{abstract}
	
	\thispagestyle{empty}\clearpage
	\setcounter{page}{1}

\section{Introduction}
In this work, we consider the \emph{online} versions of the classical scheduling problems on uniform machines $\QCmax$ and $\QCmin$, where $n$ jobs of various sizes have to be assigned to $m$ machines of differing speeds, such that the total processing time of jobs assigned to each machine is balanced in some way.
In online scheduling, these jobs are inserted or deleted over time, and the assignment of jobs to machines must incorporate these changes in the job set immediately.
The run time of algorithms for online problems is usually measured as the time it takes to process one step, \eg add or remove a single job.
In the classical online setting, as first considered by \textcite{G69}, as soon as a job is added to or removed from the instance, one must assign it to a machine, without changing the assignment of any other jobs.
As \textcite{SSS09} noted, it is impossible to maintain an optimal solution already for $\PCmax$, where all machines have the same speed, even if jobs are only inserted, unless $\textsc{p} = \textsc{np}$.

For online problems, the quality of solutions is thus instead measured by the competitive ratio $\alpha$ of an algorithm, where an algorithm is $\alpha$-competitive if it always maintains solutions within a factor $\alpha$ of the optimal offline solution.
By this definition, $\alpha$-competitive algorithms return $\alpha$-approximate solutions at each step. 
For $\QCmax$, it is known that no better competitive ratio than $2.564$ can be achieved without reassignment of jobs \cite{ES15}.
If one does allow a (small) number of jobs to be reassigned at each step however, the goal now becomes to design an algorithm with small competitive ratio $\alpha$, while simultaneously bounding this reassignment in some measure.
The bound on the reassignment is often done in terms of the processing time of the added and removed jobs \cite{W00, AGZ99, SSS09, EJ13, SV16}, as first introduced by \textcite{W00}, and called \emph{migration}.

Online scheduling problems with bounded migration on parallel identical or uniform machines appear in various applications.
For example, when minimizing the makespan, the machines may represent communication channels with identical or uniformly related bandwidth, and the jobs requests for bandwidth on these channels, where the goal is to minimize the maximum load, \ie makespan, over all channels~\cite{AGZ99, W00}.
The same problem also arises in distributed database platforms~\cite{AGZ99, W00}, where machines represent distributed databases storing the same data, and jobs represent database accesses.
The completion time of a machine then models the time required to complete all database accesses, and the aim is to minimize the maximum access response time.
In both scenarios, jobs arrive or depart over time, and may be reassigned to any other machine for some cost that can be modeled in the migration framework~\cite{W00}.
An application for minimizing the maximum completion time can be found in storage area networks (SAN) design~\cite{SSS09}, where storage disks, modeled as jobs, are assigned to subservers, modeled as identical machines, and the same data is stored on each subserver to maximize fault tolerance.
Adding a new disk to, or removing a faulty disk from, the system should not lead to a complete reconfiguration of the system, but instead should be reasonably expected to depend on the size of the disk, which corresponds to bounded migration in the scheduling model \cite{SSS09}.
The generalization to uniform machines seems natural for this application, as subservers may have different capacity with respect to the total load of diskspace they are able to handle efficiently.

Formally, an (offline) instance of both problems $\QCmax$ and $\QCmin$ consists of a set $\mc J$ of $n$ jobs and $m$ uniform machines, where each job $J$ has a processing time $p_J \in \N_{>0}$, and each machine $i$ has a speed $s_i \in \Q_{>0}$.
A job $J$ needs time $\frac{p_J}{s_i}$ to be completed on a machine $i$.
A \emph{schedule} $\sigma \colon \mc J \to \iv m$ is an assignment of jobs to machines.
For $\QCmax$, the goal is then to find a schedule $\sigma$ such that the makespan, the maximum completion time of any machine in $\sigma$, is minimized. This problem is a generalization of the problem of scheduling jobs on identical machines $\PCmax$, which has the same parameters, except that $s_i = 1$ for all machines $i$ in the instance.
$\PCmax$ is already strongly \NP{}-hard in the offline setting, but both machine settings admit an offline \ac{EPTAS} \cite{JR22,berndt2023new}.
For $\QCmin$, the goal is to find a schedule $\sigma$ such that the minimum completion time of any machine in $\sigma$ is maximized.
This problem is a generalization of $\PCmin$, the problem on identical machines, which is also already strongly \NP{}-hard in the offline setting.
While there exists an \ac{EPTAS} for $\PCmin$~\cite{SV16}, for $\QCmin$, currently only a \ac{PTAS} is known~\cite{AE98}.
In the online versions of these problems, the set of jobs to be scheduled is not fixed, but instead extended or reduced in discrete steps $q$.
At the start ($q = 0$), we are given an initial job set $\mc J_0$, and at any step thereafter ($q \geq 1$), a single job is added and its processing time revealed, or a single job is removed from the job set.
We call this setting for online scheduling problems (fully) dynamic.
One may also restrict the setting so that jobs may only be added, which we call permanent.

Let $\Delta$ denote the difference in value of a variable, set or function compared to itself from one specified step to another, and let $\Delta \mc J$ be the difference of the job set $\mc J$ between steps $q$ and $q+1$.
An online algorithm has \emph{bounded amortized migration} with amortized migration factor $\beta$, if for any step, the overall processing time $\xi$ of jobs moved (from one machine to another) satisfies $\xi \leq \beta\enpar*{\sum_{j \in \Delta\mc J} p_j + \Delta \Phi}$ for some non-negative potential function $\Phi$ that is $0$ at the beginning.
If $\abs {\Delta \Phi}$ is upper bounded by some value $\beta'$, we say that the algorithm has bounded reassignment potential $\beta'$.
This amortized notion of the migration factor is \eg used by \textcite{SV16}. If the reassignment potential is not bounded, this can lead to large changes at a single step, as \eg is the case for the \ac{PTAS} for bin packing in the permanent setting by \textcite{IL09}, where sometimes the whole instance has to be repacked.
If $\beta' = 0$, then the algorithm has \emph{bounded migration} with \emph{migration factor $\beta$}, as used \eg by \textcite{SSS09}. This restricted definition implies that $\xi \leq \beta \sum_{j \in \Delta\mc J} p_j$ for any step.

\noindent\emph{Related work.} The first to consider approximation schemes in the migration framework were \textcite{SSS09}, who gave an $(1+\veps)$-competitive algorithm for permanent $\PCmax$, for all $\veps > 0$, with migration factor $\beta \in 2^{\Oh(\epsinv (\log\epsinv)^2)}$ and run time $2^{2^{\O(\epsinv (\log \epsinv)^2)}} + \O(n)$, with the run time later being improved to $2^{\O(\epsinv[2] (\log\epsinv)^3)} + \O(n)$ by \textcite{EJ13}.
Due to the $(1+\veps)$-competitiveness, this algorithm constitutes an online \ac{EPTAS} with migration factor $\beta$.
Aiming to bound the migration factor of an online \ac{PTAS} in $f(\eps)$ for some computable function $f$ seems natural, as then, for fixed $\veps$, not only is the run time polynomial, but the migration factor is constant.
To this end, \textcite{SSS09} introduce the term of \emph{robustness}. An online scheduling algorithm is called robust if it has a constant (amortized) migration factor for any fixed $\eps > 0$.
\textcite{SV16} managed to expand this result to both dynamic $\PCmax$ and dynamic $\PCmin$ by allowing some amortization with bounded reassignment potential.
After showing that there exist no \acp{PTAS} for general instances of dynamic $\PCmin$ with constant (non-amortized) migration factor, the authors give a robust \ac{EPTAS} for dynamic $\PCmax$ and dynamic $\PCmin$ with implicitly given amortized migration factor $\beta \in 2^{\O(\epsinv[3] (\log \epsinv)^3)}$, reassignment potential $\beta'$ bounded by an $\eps$ fraction of the optimal objective value, and run time $2^{\O(\epsinv[3] (\log \epsinv)^3)} + \O(n)$.
Note that our definition of amortized migration factor slightly differs from theirs, but is compatible.\footnote{\textcite{SV16} define the amortized migration factor $\beta$ for each step $q$ as $\sum_{t=0}^q \xi_t \leq \beta \sum_{t=0}^q \sum_{j \in J_t} p_j$, where $\xi_t$ is the total processing time of jobs migrated at step $t$. They then define the accumulated reassignment potential as $K_t := \max(\xi_t/\beta - \sum_{j \in J_t} p_j, 0)$. It is easy to see that by setting $K_t = \Phi_t - \Phi_{t-1}$, where $\Phi_t$ is the potential just after step $t$, our definition of amortized migration with bounded reassignment implies theirs.}

Similar results also exist for related problems, such as Bin Packing in the permanent \cite{IL09,EL09} as well as the dynamic setting \cite{FFGGKRW18, BDGJK19, BJK20}, and dynamic Knapsack \cite{BGJJK21}, with newest results for both dynamic Bin Packing \cite{BJK20} and dynamic Knapsack \cite{BGJJK21} achieving both fully polynomial run times and migration factors.
In the world of scheduling, \textcite{WSV20} recently gave a robust algorithm for permanent $\PCmin$ with polynomial run time and migration factor, but relaxed competitive ratio of $(4/3 + \eps)$.
The latest migration-related result for dynamic $\QCmax$ is by \textcite{M23}, using a strategy employed in previous results without reassignment \cite{AAFPW93, BFN00} together with the migration framework to achieve a $(8/3+\eps)$-competitive robust algorithm with amortized migration factor in $\O(\epsinv)$, and a $(4+\eps)$-competitive robust algorithm with non-amortized migration factor in $\O(\epsinv)$, the latter for $\eps \leq 0.8$.
To the best of our knowledge however, there are no known results on robust \acp{PTAS} for either $\QCmax$ nor $\QCmin$ on uniform machines, neither in the permanent nor in the dynamic setting.
For $\QCmin$, even in the offline setting it is not know if there exists an \ac{EPTAS} at all.
A way to tackle such gaps that has become popular in recent years are so-called \acp{EPAS}~\cite{BHI02,M07,JM19,LSS22,ABBCGKMSS23}, $(1+\O(\eps))$-approximation algorithms whose run time is in $f(k,\eps) \poly(\abs{I})$, for some parameter $k$ and the encoding length of the instance $\abs{I}$.
Such algorithms can thus be seen as \ac{FPT} \acp{EPTAS}, parameterized in $k$.
As \textcite{ABBCGKMSS23} remark, these algorithms are not directly comparable to \acp{PTAS}.

Other notable ways to measure reassignment in an online setting are in the number of jobs reassigned per step, as done \eg by \textcite{FGK24} for Fully-Dynamic Load Balancing in a restricted machine environment and by \textcite{GKKP17} for Dynamic Set Cover, fixed movement costs per item, as \eg used by \textcite{FFGGKRW18} for Fully-Dynamic Bin Packing, or job-individual assignment costs, as used by \textcite{BEM22}.

\subsection{Our contributions}
Let $\pmax$ be the maximum processing time of any job occurring in a scheduling instance.
We assume to know $\pmax$ a priori, a common assumption in online scheduling~\cite{CKZ01}, which can also be viewed as a particular kind of advice~\cite{BKKKM09,D15,RRS15}.
This is also natural for the applications regarding communication channels, distributed database platforms and storage area networks, where $\pmax$ relates to the maximum allowed bandwidth per request, the maximum time required per access of a database and the maximum size of a single disk, respectively, all of which can be reasonably assumed to be upper bounded even in an online setting. 
As our main contribution, we achieve a robust \ac{EPTAS} with amortized migration factor and bounded reassignment potential, but an additional load error of $\O(\veps \pmax)$ on the solution, for dynamic $\QCmax$ and dynamic $\QCmin$.

An additional load error on a solution means that the loads, \ie the total processing time of jobs on a machine, of the solution deviate by at most this error from $(1+\O(\eps))$-approximate solutions.
The algorithm can thus also be considered an \emph{additive approximation scheme} as defined by \textcite{BRVW21} with parameter $h = \max(\realOPT,\pmax)$ and optimal makespan $\realOPT$.
\begin{restatable}[label=cor:noparam_general]{THEOREM}{noparamgeneral}
	There exist robust \acp{EPTAS} for both dynamic $\QCmax$ and dynamic $\QCmin$ with additional load error $\O(\eps \pmax)$, run time $\roundedrt + \O(\abs{I})$, amortized migration factor $\roundedsens$, and reassignment potential bounded by $3 \eps \pmax$.
\end{restatable}
For specific instances (instances with no jobs $< \veps \pmax$), the result is a true robust \ac{EPTAS} with constant migration factor.
\begin{restatable}[label=cor:noparam_nosmall]{THEOREM}{noparamnosmall}
	For instances of both dynamic $\QCmax$ and dynamic $\QCmin$ without jobs of processing time less than  $\eps \pmax$, there exist robust \acp{EPTAS} with run time \\ $\roundedrt + \O(\abs{I})$, and migration factor $\roundedsens$. 
\end{restatable}
For a detailed description of these results, we refer to \appendixref{sec:noparam}.
While no true robust \acp{EPTAS} for general instances, to the best of our knowledge, these results are the first in the direction of robust approximation schemes for both problems on uniform machines, as well as the first result in the direction of an efficient approximation scheme for offline $\QCmin$.
This addresses an open question posed by \textcite{V12} on the existence of algorithms with constant migration or reassignment factor for more general machine settings than identical machines, citing uniformly related machines as a natural generalization, as well as addresses an open question posed by \textcite{M23} on the existence of algorithms with competitive ratio better than $2$ for scheduling on uniformly related machines in the migration framework.

It is also important to note that the need for amortization and the additional load error of \autoref{cor:noparam_general} on the load are purely due to the small jobs with processing time $< \veps \pmax$.
The use of amortization is in fact necessary here, as \textcite{SV16} have shown that already $\PCmin$ does not admit a robust \ac{PTAS} with constant, non-amortized migration factor.
To handle the small jobs in the dynamic setting of general instances, we develop a general framework on how to place and migrate these jobs in an online setting, given a robust algorithm for large jobs ($p_j \geq \eps \pmax$). This framework for small jobs is similar to the widely used offline approach \cite{H97}, and thus might be of independent interest for online scheduling.
\begin{restatable}[
	, name  = Dynamic Grouping
	, label = stmt:dynamic_grouping
]{THEOREM}{stmtDynamicGrouping}
Given a robust algorithm for \QCmax (resp. \QCmin) with run time $g$ for jobs of size $\geq \eps \pmax$, where migration is bounded by a function $\beta(p_J)$ for each job $J$, there exists a robust algorithm for \QCmax (resp. \QCmin) for all jobs with amortized migration bounded by a function $\bar{\beta}(p_J) = 2 \cdot \beta(3 p_J)$, bounded reassignment potential $3 \veps \pmax$, run time at most $\O(g \cdot \poly(\epsinv))$, and with additional load error at most $\veps \pmax(4 + \ceil*{\frac 3 m}) \in \O(\veps \pmax)$.
\end{restatable}
As an intermediate step, we also give robust \acp{EPAS} for both $\QCmax$ and $\QCmin$, in both the permanent and the dynamic setting, parameterized in the maximum processing time $\pmax$ and the number of different processing times $d$ occurring in the instance.
We hereby extend the notion of robustness to \acp{EPAS} by allowing the (amortized) migration factor to be parameterized.
It suffices to a priori know upper bounds for $\pmax$ and $d$ to achieve the same result (with dependencies on the bounds instead of the actual values of $\pmax$ and $d$).
Note that $d \leq \pmax$ is a (trivial) upper bound that can always be used.
\begin{restatable}[
	, name  = \acp{EPAS} for $\QCmax$ and $\QCmin$ in the Dynamic Case
	, label = stmt:main_theorem
	]{LEMMA}{stmt:main_theorem}
	There exist robust \acp{EPAS} for dynamic $Q||\gamma$ with parameters $\pmax$ and $d$, run time \\ $\paramrt + \O(\abs{I})$, and parameterized migration factor $\paramsens$,
	\begin{enumerate}[(i)]
		\item to minimize the maximum completion time (makespan), $\gamma = \Cmax$ \label{stmt:param_dynamic_algorithm}
		\item to maximize the minimum completion time (machine covering), $\gamma = \Cmin$ \label{stmt:param_qcmin_algorithm} \stopphier
	\end{enumerate}
	\leavevmode
\end{restatable}

The general idea behind obtaining the algorithms stated in \autoref{cor:noparam_nosmall} and \autoref{cor:noparam_general}, is to first round the input values between $\eps \pmax$ and $\pmax$ in such a way that $\pmax,d \in \O(\epsinv[c])$ for some constant $c \in \N$, and afterwards, to apply the \ac{EPAS} as stated in \autoref{stmt:main_theorem} with the appropriate objective, directly implying these results.
This rounding is done via a more sophisticated version of the classical rounding scheme~\cite{HS87} introduced by \textcite{BDJR22}, in order to get the claimed run times and migration factors.

The basic concept behind the \acp{EPAS} for \QCmax and \QCmin is to first guess the objective value $T$  at each step $q$.
Then, all machines in an instance whose capacity (speed times $\realOPT$)  is large enough to allow an additive error of $\Omega(\pmax)$ can be covered in an ILP formulation by a single knapsack-like constraint.
Their actual assignment is then constructed by a simple greedy algorithm, essentially a robust version of the well-known list-scheduling by \textcite{G69} (see \appendixref{sec:additive}).
Machines whose capacities are not large enough then have a capacity bounded by a polynomial in $\pmax$ and $\epsinv$, which allows the efficient usage of a configuration ILP based approach.
The main difficulties to overcome here are the combination of both greedy and configurations based approaches for different machines in the same step, and the transitioning of machines whose capacity either becomes large enough, or too small, to be handled by the knapsack-like constraint through the addition/removal of jobs from the instance.
This is the main technical contribution.

\noindent \emph{Organization of this paper.}
We start by giving some preliminary notions, firstly stating the basic notation used throughout this paper, and secondly introducing the most important concepts used in robust scheduling.
We proceed by introducing our \ac{EPAS} for dynamic $\QCmax$, also called the \enquote{parameterized robust algorithm}, first describing it's two main components separately, and then combining these two components into the desired \ac{EPAS}, obtaining \autoref{stmt:main_theorem},~(\ref{stmt:param_dynamic_algorithm}).
Afterwards, we describe how we use and adjust these ideas and techniques to get the \ac{EPAS} for dynamic $\QCmin$ as well.
Due to space constraints, all technical details of these results, and the full description of
\autoref{stmt:main_theorem},~(\ref{stmt:param_qcmin_algorithm}), the \ac{EPAS} for $\QCmin$, can be found in \appendixref{sec:combine_online_full} and \appendixref{sec:combine_online_cmin}, respectively.
We continue by describing how to use these algorithms, together with the modified rounding scheme, to obtain our main results, namely \autoref{cor:noparam_nosmall} and \autoref{cor:noparam_general}, of which a full formal description can be found in \appendixref{sec:noparam}.
Finally, we give a short conclusion and pose some interesting open questions raised by the findings of this work.
Also to be found in the appendix are minor results and further technical details which we use for our main theorems, but which are either easily verifiable or just slightly modified results from the literature, as well as the result for a dynamic grouping technique for small jobs (\autoref{stmt:dynamic_grouping}) in \appendixref{sec:frames}.

\section{Preliminaries}
\subsection{Notation}
We use migration, (amortized) migration factor and reassignment factor as defined in the introduction, where the (amortized) migration factor is called \enquote{parameterized} if $\beta$ is dependent on some parameter $k$.
For minimization problems, a robust \ac{EPTAS} is an online algorithm, such that for any $\veps > 0$, the algorithm computes an $(1+\O(\veps))$-approximate solution with (amortized) migration factor dependent only on $\eps$, and whose run time is bounded by $f(\eps) \cdot \poly \abs I$, for some computable function $f$.
For maximization problems, a robust \ac{EPTAS} is defined to be $(1-\O(\veps))$-competitive.
Similarly, for minimization problems, a robust \ac{EPAS} is an online algorithm, such that for any $\veps > 0$, the algorithm computes an $(1+\O(\veps))$-approximate solution with parameterized (amortized) migration factor dependent only on $\eps$ and $k$, and whose run time is bounded by $f(k,\eps) \cdot \poly \abs I$, for some parameter $k$ and some computable function $f$, and defined to be $(1-\O(\veps))$-competitive for maximization problems as well.
We assume \obda that for any occurring $\veps > 0$, $\epsinv$ is integral.
Let $\uvec j$ denote the $j$-th unit vector, i.e. $\uvec j _ j = 1$ and $\uvec j _ {j'} = 0$ for all $j' \neq j$.
We use the Bourbaki interval notation, \eg write $\rinterval 0 1$ for $\set { x | 0 \leq x < 1}$.
We extend this notation to integer intervals, \eg write $\lInterval 0 4$ for $\sing{ 1, 2, 3, 4}$.
As shorthand, we use $\iv x \defeq \Interval 1 x$.
We use the operator $\at q$ to denote a variable, set or function at a specific step $q$.
Usually when $\Delta$ is used for two different steps $q$ and $q'$, these steps are clear from context.

Let $p \in \Interval \pmin \pmax^d$ be the vector of (different) job processing times and $\nu \in \N_0^d$ the corresponding multiplicity vector, \ie for $j \in \iv d$ there are $\nu_j$ jobs of processing time $p_j$ in the instance. 
We redefine a schedule $\sigma$ to be a map $\sigma \colon [m] \times [d] \rightarrow \N$, mapping a machine $i \in [m]$ and a job type $j \in [d]$ to the number of jobs of type $j$ that are scheduled on a machine $i$, such that $\sum_{i \in \iv m}\sigma(i,j) = \nu_j$ for all $j \in \iv d$.
Such an encoding of the jobs and the schedule is referred to as a \emph{high-multiplicity} encoding.
Note that for this definition, jobs of the same type have become indistinguishable.
We justify the use of this encoding in \appendixref{sec:conversion}.
If not stated otherwise, from now on we exclusively deal with high-multiplicity instances and schedules as redefined above.
Adding a job to or removing it from the job set $\mc J$ is akin to adding/removing a job of this type to/from the multiplicity vector $\nu$.
The makespan/minimum completion time of a schedule is the maximum completion time/minimum completion time of any machine $i$ in $\sigma$. 
Depending on the context, the minimal makespan/maximal minimum completion time over all schedules for some multiplicity vector $\at q \nu$ at step $q$ is called $\realOPT(\at q \nu)$, $\at q {\OPT*}$ if the job set is evident, or simply $\realOPT$ if both job set and step are evident.
The goal of dynamic $\QCmax$/$\QCmin$ is then to find a schedule with minimal makespan/maximal minimum completion time for every step $q$ over the duration, where either $\at {q+1} \nu = \at q \nu + \uvec j$, for some $j \in \iv d$, or $\at {q+1} \nu = \at q \nu - \uvec j$, for some $j \in \iv d$ and $\at q \nu_j > 0$.

We make use of known results for \nfold ILPs, which are a specific type of block structured ILPs.
More precisely, $\nfoldN$-fold ILPs are ILPs with an arbitrary objective vector $c$ and  constraint matrix $A$ of size $\nfoldr + \nfolds\nfoldt \times \nfoldn\nfoldt$ for $\nfoldr,\nfolds,\nfoldt,\nfoldN \in \N$ consisting of $\nfoldr$ arbitrary rows, with a block diagonal matrix below where each block has size $\nfolds \times \nfoldt$.
Such ILPs can be solved in time $2^{\O(\nfoldr\nfolds^2)}\enpar{\nfoldr\nfolds\norm A_\infty}^{\O\enpar*{\nfoldr^2\nfolds + \nfolds^2}} \enpar{\nfoldn\nfoldt}^{1 + \mathrm o(1)}$ using the algorithm of \textcite{DBLP:conf/soda/CslovjecsekEHRW21}.

We also formally define \enquote{configurations}, \ie all ways to place jobs with total processing time at most some upper bound $u$, which are a helpful tool in formulating ILPs for the considered problems.
\begin{definition}[Configurations]
	\[
		\mc C \colon \Q_{\geq 0} \to \mathds P(\N_0^d)
		\enspace , \qquad
		u \mapsto \Set { c \in \N_0^d | p^\intercal c \leq u }
		\enspace . \qedhere
	\]
\end{definition}

\subsection{Robust Scheduling on Uniform Machines: General Concepts}
\label{sec:difficulty}

In this section, we introduce the most important concepts and tools commonly used in the robust scheduling literature, which we also employ in part to obtain our results for robust scheduling on uniform machines.
Robust algorithms often make use of \emph{sensitivity} of a subproblem.
Formally, sensitivity measures how much a solution has to change in relation to a change of the input. 
A very strong tool that is often applied in the design of such algorithms (as in the case of the aforementioned robust algorithms for \PCmax) is the following classical result by \textcite{DBLP:journals/mp/CookGST86} for general ILPs.
\begin{lemma}[
	, name={\textcite{DBLP:journals/mp/CookGST86}}
	, label=stmt:Cook
]
Let $A \in \Z^{M \times N}$ and $c \in \Z^N$.
Given an optimal integer solution $x^\star$ to the ILP $\min c^\intercal x$ subject to $Ax = b$, $x \geq \zero$, if the modified ILP $\min c^\intercal x$ subject to $Ax = b'$, $x \geq \zero$ is feasible, then there is an optimal solution $\tilde x$ to the latter such that
\[
	\norm { \tilde x - x^\star }_1 \leq \norm { b - b'}_1  N \delta
	\enspace ,
\]
where $\delta$ the largest absolute value of any subdeterminant of $A$.
\end{lemma}

It is well-known that via slack variables one can also allow inequalities at the cost of increasing only the number of variables and constraints by a constant factor.
Such bounds are generally referred to as \emph{sensitivity bounds} and are characterized by the factor on the right hand side, which in case of the above result is $N \delta$.
However, for uniform machines, as far as we are concerned, there is no ILP formulation known where the largest absolute value of any subdeterminant and the change of the right hand side to accommodate insertion/deletion of a job is known to reasonably bounded.
The general simple estimation for the largest absolute value of any subdeterminant would be $\norm A_\infty ^{\min\sing{M,N}}$ (which can be strengthened a little using the Hadamard inequality, but this leads to the very same problems).

The two most common ILP formulations to decide feasibility of a makespan $T$ for (offline) $\QCmax$, both of which can easily be adapted to decide feasibility of a minimum completion time for $\QCmin$, are as follows; for the sake of shortness, we write $s_i = s_t$ for a machine $i$ of type $t$.
\begin{multicols}{2}
	\renewcommand{\lpindent}{\hspace{0pt}}
	\underline{Assignment-ILP}
	\begin{lpformulation}
	\lplabel{ip:assign}
		\lpeq*{\sum_{i \in \iv m} x_{i,j} = \nu_j}{j \in \iv d}
		\lpeq*{\sum_{j \in \iv d} p_j x_{i,j} \leq \floor{s_i T}}{i \in \iv m}
		\lpeq*{\var x_{i,j} \in \N_0}{(i,j) \in \iv m \times \iv d} 
	\end{lpformulation}
	\underline{Configuration-ILP}
	\begin{lpformulation}
	\lplabel{ip:conf}
		\lpeq*{\sum_{t \in \iv \tau} \sum_{c \in \mc C(s_t T)} x_{t,c} \cdot c = \nu}{}
		\lpeq*{\sum_{c \in \mc C(s_t T)} x_{t,c} = \mu_t}{t \in \iv \tau}
		\lpeq*{\var x_{t,c} \in \N_0}{t \in \iv \tau, c \in \mc C(s_t T)}
	\end{lpformulation}
\end{multicols}
If we were to apply the Cook bound to known formulations after some standard rounding procedure depending on $T$, the result is always a sensitivity bound that is not only dependent on $\epsinv$ -- which is what we would need if we were to use the resulting bound to design a robust algorithm.
A different way to deal with such problems is to utilize special structural properties of ILPs that admit smaller bounds with respect to the run time and sensitivity in the ILP's parameters.
One such kind of structural property is the aforementioned $\nfoldN$-fold block structure.
Such ILPs are particularly interesting in our case, as they admit a small sensitivity bound in the block parameters.
\begin{lemma}[
	, name={\textcite[Theorem 16]{JLR20} }
	, label=stmt:JLR
]
	An \nfold ILP with a linear objective and a constraint matrix $A \in \Z^{(\nfoldr + \nfoldn \nfolds) \times (\nfoldn \nfoldt)}$ that consists of $N$ blocks of size $\nfoldr \times \nfoldt$ and additionally $\nfoldr$ arbitrary constraints admits a sensitivity bound of $\O(\nfoldr \nfolds \norm{A}_\infty)^{\nfoldr \nfolds}$.
\end{lemma}
It is well known that the Assignment-ILP formulation of (offline) $\QCmax$ and $\QCmin$ have $\nfoldN$-fold structure \cite{KK18}.
This formulation can be adapted to represent both dynamic $\QCmax$ and dynamic $\QCmin$ via simple and straightforward modifications incorporating \emph{lexicographically minimal solutions} as used in \cite{V12} that particularly ensure that the changes of the right hand side are bounded.\footnote{We thank an anonymous reviewer for pointing out this approach and the necessary modifications to us.}
Using the algorithm by \textcite{DBLP:conf/soda/CslovjecsekEHRW21}, one can then solve both problems optimally in time $(\pmax)^{\O(d^2)} \O(n)$, and with migration factor $(\pmax)^{\O(d)}$ as given by the special sensitivity bound for $\nfoldN$-fold ILPs (\autoref{stmt:JLR}).
For details, we refer to \autoref{sec:simple}.
However, directly applying a standard rounding scheme (\eg \cite{AAWY97, H97}) on this approach in order to obtain an \ac{EPTAS}, as was successfully done before for the offline problems, suffers from the same problems as before, even with the special sensitivity: The sensitivity bound is not only dependent on $\epsinv$, and the resulting algorithm therefore not robust.

Our results now stem from a new $\nfoldn$-fold-ILP formulation for the considered problem with additionally overall small dimension and coefficient, and a non-standard rounding scheme depending on $\pmax$ instead of $T$.
This allows us to use the sensitivity result by \textcite{JLR20} to accommodate changes of the right hand side by a bounded change of the solution.
We combine this with a carefully designed technique to bound the changes of the right hand side in terms of the size of the inserted/removed job to obtain bounded migration.
Moreover, the small overall dimensions allow us to use a state-of-the-art algorithm for general ILPs, which yields better run time in the parameters in the approximate parameterized setting (\autoref{stmt:main_theorem}) compared to the exact approach, as well as an overall smaller run time in the non-parameterized, purely approximate setting.
  \section{Construction of the parameterized robust algorithm}
	In this section, we aim to give a high-level idea on the proof of \autoref{stmt:main_theorem},~(\ref{stmt:param_dynamic_algorithm}), \ie the \ac{EPAS} for dynamic $\QCmax$.
	Due to space constraints, all technical details of the proof can be found in \appendixref{sec:combine_online_full}.
We show the Lemma by iteratively constructing an algorithm and proving all the claimed properties for it; we refer to this algorithm simply as \enquote{the robust parameterized algorithm}.
	Afterwards, we give an overview on how to modify these ideas to achieve the \ac{EPAS} for dynamic $\QCmin$, \ie how to proof \autoref{stmt:main_theorem},~(\ref{stmt:param_qcmin_algorithm}), as well.
	
By definition of $\realOPT$, any solution where the load of any machine $i$ is smaller or equal to its capacity $s_i \realOPT$ is an optimal solution.
	The construction of the parameterized robust algorithm is subdivided as follows:
	First, we show how to obtain such an algorithm if we are ensured that each machine exceeds a given minimum capacity.
	Then, we show how to do this if all machines subceed this very threshold.
	Eventually we show how we can combine these approaches to get rid of assumptions about machine capacities, which is by far the most difficult part.
The upcoming (sub-)sections make use of the threshold
\[  \ell \defeq (1 + \eps)^{1 + \ceil {\log_{1+\eps} (2 \epsinv \pmax)}} \enspace . \]
Note that
\[
	\eps (1 + \eps)^\inv \ell \geq \eps \ell \geq 2 \pmax \enspace , \qquad \tand \ell \leq \poly(\epsinv, \pmax) \enspace .
\]
In the following, assume that machine speeds are all powers of $(1 + \eps)$.
We can achieve this property by rounding each machine speed up to the next power of $(1 + \eps)$.
This process introduces an additional multiplicative error of at most $(1 + \eps)$ on the solution.
Let $\tau$ denote the number of different machine speeds after rounding.
Let $s \in \Q^\tau$ be the vector of rounded machine speeds and $\mu \in \N_0^\tau$ the corresponding multiplicity vector.
We write $\smax$ for the largest speed after rounding and $\smin$ for the smallest.
\subsection{Case 1: all machines (always) have large capacity}
In this subsection, we assume that at all times, every machine has a capacity of \emph{at least} $\lb$ w.r.t $\realOPT$.
To compute approximate solutions, here, we use
a simple adaptation of the list-scheduling algorithm~\cite{G69} that allows for insertion and deletion of jobs with migration bounded in $p_j + \pmax$ for insertion/deletion of a job $j$, returning a solution at every step $q$ where each machine has completion time at most $\frac{p^\intercal \nu}{s^\intercal \mu} + \pmax \leq \realOPT{} + \pmax$.
Since $\veps \lb \geq \pmax$, and all machines of each type $t$ have capacity $s_t \realOPT \geq \lb$ by assumption, this is a $(1 + \eps)$--approximate solution at every step $q$ due to
\[
	(1 + \eps) s_t \realOPT{}
	\geq s_t \realOPT{} + \eps \lb
	\geq s_t \realOPT{} + \pmax \enspace .
\]
Details can be found in \appendixref{sec:additive}.

\subsection{Case 2: all machines (always) have small capacity}
For this section, we assume that at all times, every machine has a capacity of \emph{at most} $\lb$ w.r.t $\realOPT$.
Then, all we need is that $\ell \leq \poly(\epsinv, \pmax)$, i.e. $\ell$ is bounded by some fixed polynomial in $\epsinv$ and $\pmax$.
To obtain a solution at step $q$, we could set up a \ipconf with $u = T$ in the configuration set for a current guess on the makespan $T$, discard machine types where $s_t T < 1$, resulting in at most $1 + \log_{1+ \eps} \ell \in \O(\epsinv \log(\epsinv \pmax))$ machine types that can be assigned a non-zero configuration.
We could then solve this using a standard ILP algorithm parameterized by the number of constraints ($\leq d + 1 + \log_{1+ \eps} \ell$) and the largest coefficient ($\leq \ell + 1$), as all capacities are bounded by $\lb$.
However, doing this makes going from step $q$ to $q+1$ -- in contrast to the case where all machines have large capacities -- not that straightforward.
This is only due to removal: If we were only allowed to add jobs, we could simply increase $T$ when necessary, possibly include some newly available configurations into the ILP, convert our old solution to a solution of the new ILP by adding \enquote{$0$}'s for the new entries (those that correspond to newly introduced configurations) and then use the result by \textcite{JLR20} to find a solution nearby.
However, if we can remove jobs, we cannot convert our old solution to one with smaller $T$ in general, as we might have to change the assignment of $\Theta(m)$ machines, if they are assigned a configuration that has to be removed from the ILP due to the decrease of $T$.

This problem also occurs for identical machines, and \textcite{SV16} solved it in the following way: They list a superset of necessary configurations and use a linear objective $\lexobj$ that encodes \enquote{Minimizing the makespan}.
We can proceed similarly here as -- due to the assumption that capacities never exceed $\ell$ -- we can simply list all configurations whose load does not exceed $\ell$.
We additionally need to generalize the construction of the linear objective such that it can be applied to uniform machines with a few machine types, which is rather straightforward (see \appendixref{sec:lexmin}).
Recall that at most $1 + \log_{1+ \eps} \ell \in \O(\epsinv \log(\epsinv \pmax))$ machines can be assigned a non-zero configuration at each step, if we list only machine types that -- before or after the operation -- can actually be assigned a job, so this number is still $\O(\epsinv \log(\epsinv \pmax))$.
We then achieve the desired result using the sensitivity bound as stated in \autoref{stmt:JLR}.
Note that, since the objective constructed leads to unique optimal solutions, to compute a new solution it suffices to simply \emph{solve} the ILP, while only implicitly bounding the changes of variables.
Details on this approach, both for the sensitivity bound and the actual solving of the ILP, can be found in \appendixref{sec:ilpsolve}.
This is also where the run time and migration factor of the algorithms come from, as solving the ILP and bounding the change in parameters is the bulk of the computational work.

\subsection{Combining both cases}
Let us call the algorithm for only large capacities \enquote{blue algorithm}, and the one for only small capacities \enquote{red algorithm}.
Our goal is now to partition the jobs and machines into red and blue jobs and machines at each step $q$, respectively, and use the blue algorithm for blue jobs and blue machines, and the red algorithm for red jobs and machines to compute feasible approximate solutions for both sets in the partition, and by that, for the whole instance.
Let the vector $\mub \in \N^{\tau}$ be the vector of blue machines, $\nub \in \N^d$ be the vector of blue jobs at some step $q$.

Consider the following ILP that is parameterized by $\nu$, $\mub$ and some value $\alpha \in \Q_{\geq 0}.$\footnote{A technical detail: for ILPs, the RHS is usually constrained to be integral.
We round it down before applying \autoref{stmt:JLR} or running an ILP solver, covering the details where this occurs.
For the rest, we work with fractional values which simplifies our statements.}
This ILP amounts to a modified version of the \ipconf (with upper bound $\lb$ on the configurations of each machine type), extended by a knapsack-like constraint with capacity $\alpha$ dependent on the total capacity of the blue machines.
The total capacity on the blue machines is itself dependent on the current number of blue machines $\mub$.
The parameter $\nu$ gives us the current multiplicity of each job type in the instance.
The objective function $\lexobj$ again encodes \enquote{Minimizing the makespan} on the given set of configurations -- details on this can be found in \appendixref{sec:lexmin}.

\vspace{-2em}
\begin{multicols}{2}
\begin{lpformulation}[{\mathrlap{\ipdyn}}]\lpeq*{}{}
	\lplabel{ip:dyn}
\lpobj*{min}{\lexobj(x)}
	\lpeq[ip:dyn:constr:blue_area]{
		p^\intercal \super \tblue \nu
		\leq \alpha
	}{}
	\lpeq[ip:dyn:constr:red_count]{\sum_{c \in \mc C(\ell)}  x_{t,c} = \mu_t - \super \tblue \mu_t}{t \in \iv \tau}
	\lpeq[ip:dyn:constr:blue_red]{\sum_{\substack{t \in \iv \tau \\ c \in \mc C(\ell)}}  x_{t,c} \cdot c  + \super \tblue \nu = \nu}{}
	\end{lpformulation}
	\begin{lpformulation}
	\lpeq*{}{}
	\lpeq*{\text{\normalfont \underline{with variables}}}{}
	\lpeq*{\var x_{t,c} \in \N_0}{t \in \iv \tau, c \in \mc C(\ell)}
	\lpeq*{\var \super \tblue \nu \in \N_0^d}{}
	\lpeq*{\text{\normalfont \underline{with parameters}}}{}
	\lpeq*{\param \nu \in \N_0^d}{}
	\lpeq*{\param \super \tblue \mu \in \N_0^\tau}{}
	\lpeq*{\param \alpha \in \Q_{\geq 0}}{}
	\stopphier
\end{lpformulation}\end{multicols}
Variables $x_{t,c}$ represent the configurations of red machines, and thus encode all jobs placed on these machines.
Variables $\nub$ represent all jobs placed on blue machines instead.
Constraint~\ref{ip:dyn:constr:blue_area} ensures that the total load of jobs on blue machines represented by $\nub$ does not exceed the currently available capacity on blue machines $\alpha$.
Constraint~\ref{ip:dyn:constr:red_count} guarantees that the number of configurations of each machine type $t$ in any solution matches the current number of red machines $\mu_t - \mub_t$ of this type.
Finally, Constraint~\ref{ip:dyn:constr:blue_red} makes sure that each job is assigned via a configuration to a red machine, or assigned to the blue machines via $\nub$.

Let $T$ be a guess on the optimal makespan at some step $q$.
We are now able to compute an approximate solution for dynamic $\QCmax$ at step $q$ by setting $\mub_t : = \mu_t$ if $s_t T \geq \lb$ and $\mub_t := 0$ otherwise, for each $t \in \iv \tau$, and setting $\alpha := T {s^\intercal \mub}$, \ie the total capacity of the blue machines.
Using the red algorithm on such a modified ILP then computes a solution with makespan $T$ on the red machines for some subset of the jobs $\nu - \nub$, such that the load of jobs not placed on the red machines $p^\intercal \nub$ is smaller or equal to the total capacity on the blue machines $T {s^\intercal \mub}$, if it exists.
If a solution exists, using the blue algorithm to compute an assignment of $\nub$ on $\mub$ then returns a solution with makespan at most $(1+\eps) T$ by its definition.
Combining these solutions gives us a schedule with a makespan of at most $(1+\eps) T$ over all machines.
If no solutions exists to the ILP with makespan $T$, we can correctly conclude there exists no solution with this makespan.

However, to go from step $q$ to $q+1$, we need to repartition the set of machines into blue and red machines depending on the possibly different optimal makespan, a process we call \emph{recolouring}.
This recolouring affects $\mub$, changing the right hand side of \ipdyn, and thus impacting the bound on the sensitivity and the migration factor.
In fact, setting up \ipdyn in such a way that, at each step $q$, using the parameterized red and blue algorithms to compute \emph{robust} solutions turns out to be quite challenging.
The two main problems are (i) that we might have to recolour $\Omega(m)$ machines, if their capacities cross the threshold $\ell$ due to changes in the optimal makespan; and (ii) the right hand side of the knapsack constraint is highly dependent on the optimal makespan: even a tiny change in the optimum can trigger a $\omega_{\eps,\pmax,d}(1)$ change.
Both scenarios would directly lead to unbounded migration.

\begin{figure}[t]
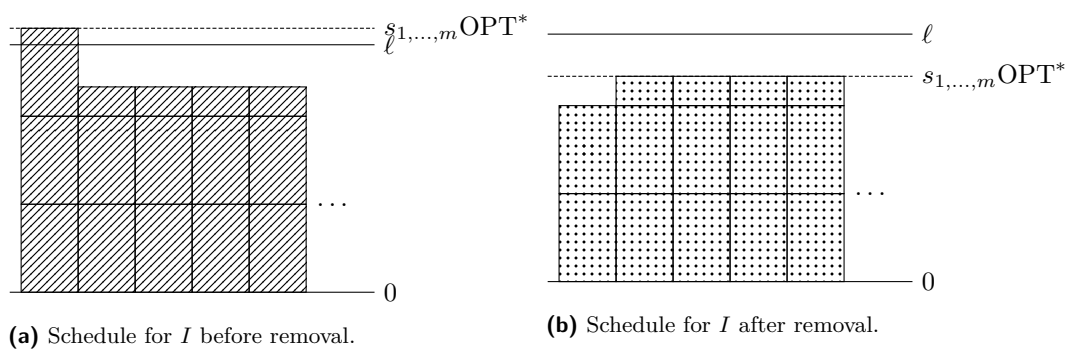

	\begin{subfigure}[c]{0.5\linewidth}
    \tikz[xscale=0.75,yscale=3.5]{

\tikzmath{\m = 6;}
\hlines \m {
    {1}/$s_{1,\dots,m} \text{OPT}^*$//,
    {1-1/16}/$\lb$/solid/,
    0/$0$/solid/
};
\schedule up {
    {
        {1/3}//pattern=north east lines/,
        {1/3}//pattern=north east lines/,
        {1/3}//pattern=north east lines/
    }/1/,
        {
        {1/3}//pattern=north east lines/,
        {1/3}//pattern=north east lines/,
        {1/9}//pattern=north east lines/
    }/1/,
        {
        {1/3}//pattern=north east lines/,
        {1/3}//pattern=north east lines/,
        {1/9}//pattern=north east lines/
    }/1/,
        {
        {1/3}//pattern=north east lines/,
        {1/3}//pattern=north east lines/,
        {1/9}//pattern=north east lines/
    }/1/,
        {
        {1/3}//pattern=north east lines/,
        {1/3}//pattern=north east lines/,
        {1/9}//pattern=north east lines/
    }/1/
};
\node at (\m-0.5,1/3) {$\dots$}; }
    \caption{Schedule for $I$ before removal.}
    \label{fig:1}
    \end{subfigure}
\begin{subfigure}[c]{0.5\linewidth}
    \tikz[xscale=0.75,yscale=3.5]{\tikzmath{\m = 6;}
\hlines \m {
    {1-2/9}/$s_{1,\dots,m} \text{OPT}^*$//,
    {1-1/16}/$\lb$/solid/,
    0/$0$/solid/
};
\schedule up {
    {
        {1/3}//pattern=dots/,
        {1/3}//pattern=dots
    }/1/,
        {
        {1/3}//pattern=dots/,
        {1/3}//pattern=dots/,
        {1/9}//pattern=dots/
    }/1/,
        {
        {1/3}//pattern=dots/,
        {1/3}//pattern=dots/,
        {1/9}//pattern=dots/
    }/1/,
        {
        {1/3}//pattern=dots/,
        {1/3}//pattern=dots/,
        {1/9}//pattern=dots/
    }/1/,
    {
        {1/3}//pattern=dots/,
        {1/3}//pattern=dots/,
        {1/9}//pattern=dots/
    }/1/
};
\node at (\m-0.5,1/3) {$\dots$}; }
    \caption{Schedule for $I$ after removal.}
    \label{fig:2}
    \end{subfigure}
    \label{fig:both}
    \caption{Example of potentially unbounded parameter change for an instance $I$.}
\end{figure}

To gain some more insight into this issue, consider the following example: Let $I$ be an instance with speeds $s_1 = \dots = s_m = 1$, $2m+1$ jobs of size $\pmax$, and $m-1$ jobs of size $\frac{\pmax}{3}$.
Let $\eps$ be chosen such that $3 \pmax > \lb$ but $2 \pmax + \frac{\pmax}{3} < \lb$.
A schedule might look as pictured in \autoref{fig:1}: By pigeon-hole, there must be at least one machine $i$ with at least $3$ jobs of size $\pmax$.
Thus, the capacity of all machines for makespan $\realOPT$ must be greater than $\lb$.
All machines are coloured blue (pictured as north-east-line patterned).
The placement of the remaining jobs may be any solution that results from a greedy assignment, \eg each machine has scheduled $2$ jobs of size $\pmax$ and one job of size $\frac{\pmax}{3}$.
Now assume $1$ job of size $\pmax$ is removed from $i$, \ie the machine with $3$ jobs of size $\pmax$: Then, $\realOPT$ immediately drops to $\pmax + \frac{\pmax}{3}$.
In such a case, we get a situation as pictured in \autoref{fig:2}: All $m$ machines have to be recoloured red (pictured as dotted) and $\alpha$ drops from $> m \lb$ to $0$.
This leads to and unbounded change in parameters and thus unbounded migration as measured by sensitivity as well, even though for a lexicographically minimal solution on the newly red machines, no jobs have to be migrated.

Our approach to tackle this is twofold: (i) we discretize the makespan guarantees in order to not be forced to react to small changes all the time; and (ii) we distribute the work that needs to be done when the makespan guarantee changes over the preceding steps, such that in the step where this change occurs, the remaining necessary changes are properly bounded.
The latter means that  we recolour machines over time such that when our guarantee changes, there are at most a few machines left to recolour; and analogously we change the right hand side of the knapsack constraint over time.
First, let us capture the restriction on the machine colouring by an invariant:
\begin{invariant}[
	, name = Machine Colouring Invariant
	, label = invar:machine_colouring
]
	Given a bound $T \in \Q_{\geq 0}$ on the objective value, for any blue machine of type $t$ we have $s_t T \geq \ell$; and for any red machine of type $t$ we have $s_t T \leq \ell$.
\end{invariant}

Since our ILP is parameterized by the value $\alpha$ that represents the part of the overall capacity on blue machines that may be used by the ILP, our next step is to formalize this interaction, \ie ensure that we restrict ourselves to values of $\alpha$ that do not exceed the overall total capacity of the blue machines, for which we now allow another additional multiplicative error of $(1+\veps)$.
\begin{invariant}[
	, name = Compatible Colouring Invariant
	, label = invar:comp_colouring
]\[ 
	0 \leq \alpha \leq (1 + \eps) T s^\intercal \super \tblue \mu \enspace . \stopphier \]
\end{invariant}

For fixed $T$ and $\mub$, we will also call $(1 + \eps) T s^\intercal \super \tblue \mu$ the \enquote{available blue area (for $T$ and $\mub$)}.
As mentioned before, the direct application of \autoref{stmt:JLR} does not lead to a sufficiently small sensitivity bound, due to possible large differences in the right hand side $b$ and $b'$ -- this is the main issue to overcome.
Given a (approximate) makespan $T$, we require the red schedule to not exceed $T$, while for the blue schedule we only require a makespan of $(1 + \eps) T$.
This will later allow us to bound the number of machine recolourings by some kind of witness argument.

One property of the objective $\lexobj$ is that, if applied to all variables, optimal solutions are unique as long as feasible solutions can be bounded by some fixed value.
This property also applies to \ipdyn, which we show in \appendixref{sec:lexmin}.
Our main focus thus does not rely on finding solutions that meet the requirements on the migration, but to adjust the parameters in such a way that the unique optimal solution to these parameters does so.
Thus, instead of classifying solutions, we classify triples of parameter values, called \enquote{states}, calling those \enquote{$T$-valid} where the optimal solutions meets our requirements (for makespan $T$):

\begin{definition}[
	, name = {States, associated solutions, ILP-makespan, and $T$-validity}
	, label = defi:state
]
	A triple of the ILP parameters $(\alpha, \mub, \nu)$ is called \emph{state}.
	The (unique) optimal solution $(x, \nub)$ of \ipdyn for these parameters is called \emph{the associated solution}.
	The \emph{(ILP) makespan of a state} is the (ILP) makespan of the associated solution, \ie $\Cmax(x)$.
	If the ILP is infeasible, we define the makespan to be $\infty$.
	
	A state $S$ is \emph{$T$-valid} iff \autoref{invar:machine_colouring} and \autoref{invar:comp_colouring} both hold for $T$; and the associated solution of $S$ has (ILP) makespan at most $T$.
\end{definition}

Note that if there is a $T$-valid state, there is also $T'$-valid state for any $T' \geq T$.
Solving \ipdyn for a $T$-valid state directly yields a solution with makespan at most $(1+\eps)^2 T$, with the error being introduced by the additional $\eps T$ error on the blue area, and the actual placement of blue jobs on blue machines (see \appendixref{sec:greedy}).
As we want to solve the ILP for (approximately) the optimal makespan $\realOPT$, we introduce the following definition.

\begin{definition}[
	, name = Optimal ILP Makespan
	, label = defi:opt
]
	Given a job vector $\nu \in \N_0^d$, $\OPT \in \Q_{\geq 0}$ is the smallest number such that there is a $\OPT$-valid state $S$.
	Moreover, we define $\OPT-$ to be the largest power of $(1 + \eps)$ strictly smaller than $\OPT$; and $\OPT+ \defeq (1 + \eps) \OPT-$.
\end{definition}
\begin{definition}[
	, name = Valid States
	, label = defi:valid
]
	A state $S$ is valid if it is $\OPT+$-valid.
\end{definition}

Note that there is not necessarily just one, but often times many valid states.
We will exploit this heavily, by only converting a valid state to another valid state with \enquote{nice} properties, \ie bounded migration.
By the definition of $\OPT$ and \autoref{invar:comp_colouring}, $\OPT* \geq \OPT$. Thus, computing a solution to the ILP for any valid state always returns an at most $(1+\eps)^3 \in (1+\O(\eps))$-approximate solution.
Now, we can discuss the general idea of achieving bounded migration using this framework of \ipdyn, states and associated solutions.
Let $S = (\alpha, \mub, \nu)$ be a valid state and $\bar\nu$ a changed job set.
We assume \obda that $\bar\nu$ differs from $\nu$ by exactly one job.
Our goal is to always find a valid state $\bar S = (\bar\alpha, \mub[\bar], \bar\nu)$ such that $\abs{\Delta \alpha} + \norm{\Delta\mub}_1 \leq \poly(\pmax,\epsinv)$.
As we show in \appendixref{sec:ilpsolve} (\autoref{stmt:sensitivity}~(\ref{stmt:sensitivity:param})), this directly ensures that the unique associated solutions have bounded $\ell_1$-distance and thus we obtain bounded migration.
Since we require $\bar S$ to be valid, the associated solution to this state is then again a $(1+\O(\eps))$-approximate solution to the new job set, and the algorithm therefore is a robust \ac{EPAS}.
We construct such an $\bar S$ by carefully recolouring machines and changing $\alpha$ from $S$ to $\bar S$.
To do this, we only consider machines of capacity exactly $\lb$ for recolouring in each step: Due to the discretization of the makespan to $\OPT+$, these are the only types of machines that are \enquote{critical}, \ie they may be forced to be either red (if $\OPT+$ decreases) or blue (if $\OPT+$ increases) at the next makespan change, but may not be appropriately coloured yet before this change.
At the same time, these critical machines are the only ones that are allowed to be coloured either red or blue for $\OPT+$ by \autoref{invar:machine_colouring}.

First, let us consider the case where $\Delta\OPT+ = 0$, \ie the makespan guarantee does not change.
When some job $j$ is added, we only increase the value of $\alpha$ if necessary -- thus, we act some kind of \enquote{lazy}. 
If an increase is necessary but we cannot increase $\alpha$ due to \autoref{invar:comp_colouring}, we can conclude that there is some red machine that can be recoloured blue -- otherwise $\OPT+$ would have increased.
By doing so, due to the makespan guarantee on blue machines being only $(1+ \eps) \OPT+$, we gain an additional space of at least $\eps \ell \geq \pmax$ that can be used for the necessary increase of $\alpha$.
Thus, recolouring a single machine suffices in this case.
When some job $j$ is removed, as long as there is a blue machine that must not be blue after $\OPT+$ decreases, we recolour a single machine if possible without ending up in an invalid state, to prepare for an eventual decrease of $\OPT+$.
If there is no such machine, we decrease $\alpha$ by $\O(\pmax)$ if possible (and the maximal possible value otherwise), while ensuring that $\alpha$ does not subceed a specific threshold.
The latter is necessary to ensure validity once has $\OPT+$ dropped, i.e. just after a drop we want $\alpha$ to be the maximal possible value that it may take under the new $\OPT+$-value and its implied blue area.

An increase of $\OPT+$ only occurs when a job is added and no more critical machines can be recoloured blue. Since an increase of $\OPT+$ increases the available blue area on every blue machine by at least $\eps \OPT+ \geq \eps \lb$, we can simply increase $\alpha$ by as much.
A decrease of $\OPT+$ implies the removal of a job, and we recolour the machines that we need to recolour (some of those that are currently blue but may not be allowed to be blue under the new value of $\OPT+$), and reduce $\alpha$ to ensure that it does not exceed its maximal value in respect to the new, reduced $\OPT+$-value.
Due to our continuous \btr recolouring at each job removal, we can show that this affects only a small number of machines and a small change in $\alpha$ at the point of $\OPT+$ decrease, as otherwise, they could have (and would have) been recoloured before, or were never coloured blue for this $\OPT+$-value in the first place.
Crucial for proving this is the fact that, by the continuous recolouring, $\alpha$ always uses almost all available blue area, implying that if the recolouring and $\alpha$ changes were too big, already the job set previous to the removal would have been valid for a decreased $\OPT+$-value -- a contradiction to the validity of this state.
We thus show that we go from valid state to valid state when adding or removing jobs, while changing the state parameters only in terms of $\poly(\epsinv,\pmax)$.
As the parameters are the only changes in the right hand side of \ipdyn, besides the job set being changed by $1$ for the inserted or removed job, this then directly implies a bound  $\poly(\epsinv,\pmax)$ on the change of the right hand side.

With this result in hand, and the fact that \ipdyn has \nfold structure, we can use the sensitivity result stated in \autoref{stmt:JLR} to get our bound on the sensitivity, and thus, migration factor.
As \ipdyn has only a relatively small number of constraints in total, the most efficient way to compute the (unique) solutions at each step $q$ is via an algorithm for solving ILPs with small number of constraints by \textcite{JR22}.
Details on the sensitivity and ILP solver can be found in \appendixref{sec:ilpsolve}.

\subsection{Extension to Robust $\QCmin$}

The construction of the parameterized robust algorithm necessary to proof \autoref{stmt:main_theorem},~(\ref{stmt:param_qcmin_algorithm}), \ie handling dynamic $\QCmin$, is almost symmetrical to the parameterized robust algorithm for $\QCmax$.
The high-level idea is in fact the same:
We split the machine set into blue and red machines depending on their capacity at some step $q$ w.r.t. a slightly larger fixed threshold $\lb$, using this split to cover large machines ($\geq \lb$) with a single dual knapsack-like constraint in an ILP formulation.
The blue machines keep their $\eps \lb$-increased capacity.
This may at first sound counter-intuitive, as for maximizing the minimum completion time, this means we now have to cover more capacity than strictly necessary when colouring a machine blue.
We can offset this by the allowed error however, and this will help us later to keep the changes bounded while colouring \btr.
We formalize this in a modified version of \ipdyn.
The main difference here is that, as we want to approximate the minimum completion time objective of a solution, we require the total processing time of jobs on these blue machines to be \emph{above} some parameter $\alpha$ that depends on the total capacity of the blue machines.
Thus, the knapsack constraint in \ipdyn is changed to the dual knapsack constraint $p^\intercal \super \tblue \nu \geq \alpha$.
We also use a modified $\lexobj$ function that maximizes the minimum completion time on the red machines by computing a lexicographically minimal solution to the dual order, i.e. to the reverse order of configurations that we used for \Cmax.
To make sure that the minimum completion time is large enough on the blue machines as well, we need to redefine the \enquote{compatible colouring} to $\alpha \geq (1 + \eps) T s^\intercal \super \tblue \mu$.
A solution to this modified $\ipdyn$ with makespan $\realOPT$ in the colouring is now an $(1-2\eps)$-approximate solution to $\QCmin$ for the job set at step $q$.

The main challenge, again, is now bounding the change in the RHS $\norm{b-b'}_1$ to keep sensitivity, and thus migration, small.
Like before, this is done via a discretization of the objective value, slow change of $\alpha$, and careful recolouring of critical machines in each step.
Here however, we restrict our ILP to $\OPT-$- instead of $\OPT+$-values, in order to get a proper approximation for the new objective.
The careful recolouring and changing of $\alpha$ is done via similar \enquote{Insert}- and \enquote{Remove}-functions as before, taking a state of the three parameters $\alpha$, $\mub$ and $\nu$, and modifying them in such a way that they do not change to much, but are valid for the new step $q+1$, \ie the corresponding ILP does admit a solution with the proper approximate minimum completion time. 
The rest of the proof then follows, but \enquote{dual}, as for dynamic $\QCmin$, the difficulty of these changes is switched.
As mentioned before, we now actually have to cover more capacity when colouring a machine blue.
This is necessary to make the removal of jobs rather easy, as recolouring a blue machine red always frees enough load to \enquote{fill the gap} left by the removal of job on any machine.
If we couldn't guarantee this property, removing a single job could trigger a decrease of $\OPT-$ while a number of critical machines are still coloured blue, and thus an unbounded \btr recolouring of machines.
With this property however, at a decrease of $\OPT-$, we can guarantee that all critical machines are already correctly coloured blue and $\alpha$ does not decrease too much.
Instead, now inserting a job requires continuous recolouring of machines \rtb, as to make sure all critical machines are recoloured before an increase of $\OPT-$ is triggered.  	

\section{Getting Rid of the Parameterization}
With the \acp{EPAS} in hand, we can now describe the necessary modifications made to any instance in order to achieve the \ac{EPTAS} results as stated in \autoref{cor:noparam_nosmall} and \autoref{cor:noparam_general}.
For this, we generally use the standard approach of reducing the complexity of an instance by preprocessing, \ie rounding and simplifying, the input at the cost of an additional error on the solutions.
In contrast to the usual approaches however, which do the rounding dependent on a guess of the optimal objective value $\realOPT$, our rounding depends on $\pmax$ instead, as $\realOPT$ itself is not fixed but dependent on the current job set at a specific step.

Our preprocessing follows a recent improvement over the well known rounding and scaling scheme for scheduling problems.
This improvement, as first described by \textcite{BDJR22}, aims to reduce the largest absolute entry in the configurations necessary to describe a scheduling problem from $(\epsinv)$ to $\log(\epsinv)$ after rounding.
They successfully apply this technique to reduce the $\ell_1$-norm of the columns, and thus the hereditary discrepancy of the matrix, from $\epsinv$ to $\log(\epsinv)$ in a \ipconf formulation for offline $\PCmax$.
Solving this ILP with the algorithm by \textcite{JR22}, whose run time depends on the hereditary discrepancy, then yields an \ac{EPTAS} with run time $2^{\O(\epsinv \log(\epsinv) \log\log(\epsinv))}$.
As \ipdyn is an extension of \ipconf, and we use the same algorithm by  \textcite{JR22} to solve the ILP, we also make use of this improved rounding scheme to not only reduce $\pmax$ and $d$ to values only dependent on $\epsinv$, but also to reduce the $\ell_1$-norm of the configuration part of the ILP.
Additionally however, we have to reduce the largest coefficient in the other constraints of \ipdyn to $\log(\epsinv)$ as well, in order to get the proper bound on the $\ell_1$-norm of the columns.
Luckily, after applying the rounding above, the only other time where a coefficient larger than $\log(\epsinv)$ appears in \ipdyn is in the (single) knapsack constraint.
To deal with this, we use a technique introduced by \textcite{KPW20} called \enquote{Coefficient Reduction} to reduce the largest coefficient of this constraint to $1$, while introducing an additional $\log{\epsinv}$ constraints and variables as a trade-off.
This such rounded and reduced ILP we call \ipdynred.
Details on this preprocessing can be found in \appendixref{sec:noparam}.
As our rounding scheme is dependent on $\pmax$, we first consider only instances with processing time values in $[\veps \pmax, \pmax]$.
Instances that include processing times $< \veps \pmax$ are dealt with using our dynamic grouping technique as stated in \autoref{stmt:dynamic_grouping}. 

Any solution to \ipdynred can be transformed to a $(1+\O(\eps))$-approximation of the original instance, and all relevant parameters for solving the ILP are now bound in terms of $\epsinv$ only.
Again, using the same functions for inserting and removing jobs as in the parameterized case, we transition from step $q$ to $q+1$.
Here, this keeps the change of the RHS bounded in terms of $\poly(\epsinv)$.
Thus, by again using \autoref{stmt:JLR} to bound sensitivity and the algorithm by \textcite{JR22} to solve the ILP, we achieve a run time of $\roundedrt + \O(\abs{I})$, matching the currently fastest \ac{EPTAS} for offline scheduling of jobs on identical machines \cite{BDJR22}, and a migration factor of $\roundedsens$.
Details on this can be found in \appendixref{sec:ilpsolve}, \autoref{stmt:ipdyn_solve},~(\ref{stmt:ipdyn_solve:rounded}) and \autoref{stmt:sensitivity}~(\ref{stmt:sensitivity:rounded}), respectively.
Note that by modifying a standard rounding scheme (\eg \cite{AAWY97, H97}) to be dependent on $\pmax$ instead of $T$, results with the same approximation guarantees can also be achieved by applying this rounding to the exact approach described in \autoref{sec:simple}.
However, this results in an algorithm with a worse run time of $2^{\O(\epsinv[2] \log^3(\epsinv))} + \O(\abs{I})$, and a worse migration factor $2^{\O(\epsinv \log^2(\epsinv))}$. 	\section{Conclusion}
In this paper, we considered the problem of scheduling jobs on uniform machines for two objectives, minimizing the makespan and maximizing the minimum completion time, in a dynamic setting, where jobs may be inserted or deleted from the instance.
For instances of these problems with $\pmin \geq \veps \pmax$, we gave robust \acp{EPTAS}, and, for general instances, robust \acp{EPTAS} with an additional load error of $\O(\eps \pmax)$, as well as an amortized migration factor with bounded reassignment potential.
The need for amortization is justified, as amortization is already required for dynamic $\PCmin$, the special case of the problem of maximizing the minimum completion time on identical machines, as proven by \textcite{SV16}.
These results address open questions posed by \textcite{V12} and \textcite{M23}.
As an intermediate step, we provided \acp{EPAS} for both problems, parameterized by the largest processing time $\pmax$ and the number of job types $d$.
Interestingly, while introducing an additional load error, the run time of our \ac{EPTAS} result is better than the run time of the currently fastest (true) \ac{EPTAS} for offline $\QCmax$ \cite{berndt2023new}, by a factor of $\O(\log^2)$ in the exponent, and matches the run time of the currently fastest \ac{EPTAS} for $\PCmax$ \cite{BDJR22}.
We achieved these results by introducing a partition of the machines into machines of small and large capacity (red and blue machines), capturing this idea in a novel ILP formulation with both \nfold structure and small dimension.
We then showed how to adjust this partition for the changing makespans occurring in the dynamic setting in such a way that the sensitivity of the ILP, and thus the migration factor, stay bounded.
We believe this idea can be used for other dynamic scheduling problems as well.

Our work suggests several open problems.
One interesting possible implication of our results are potential improvements in both run time and migration factor for problems $\PCmax$ and $\PCmin$ in the dynamic setting.
Scheduling on uniform machines is a generalization of scheduling on identical machines, and, for the case of identical machines, $\veps \pmax' \leq \veps \OPT*$ always holds, where $\pmax'$ is the current largest processing in the instance (in contrast to the overall largest processing time $\pmax$).
If one could adapt our approach to have an additional load error of $\O(\eps \pmax')$ instead of $\O(\eps \pmax)$, this error would be negated for scheduling on identical machines, and applying such a result would be a direct improvement on the run time and migration factor of the robust \ac{EPTAS} for $\PCmax$ and $\PCmin$ by \textcite{SV16}.
This requires a new partition of the jobs into large and small jobs any time $\pmax'$ changes by the insertion or deletion of a job however, which our current framework cannot handle directly.
We thus pose this issue as an open question.
Of even greater interest would be to get rid of the additional load error of $\O(\eps \pmax)$ altogether, to get a true robust \ac{EPTAS} for general instances of $\QCmax$ and $\QCmin$, which we pose as a second open question.

 	\clearpage
	\section*{References}\printbibliography[heading=none]
	\clearpage
	\begin{appendix}\label{sec:appendix}
\section{Technical Details of the Parameterized Robust Algorithm}
\label{sec:combine_online_full}

In this section, we give the full technical details necessary to proof \autoref{stmt:main_theorem},~(\ref{stmt:param_dynamic_algorithm}).

\subsection{Some Preliminaries}

Before we can formulate our procedure precisely, we need some more definitions.
For a schedule $\sigma$, we write $\sigma_i$ as the vector in $\N_0^d$ with $j$-th entry being $\sigma(i,j)$ for each $j \in \iv d$.
The completion time of a machine $i$ in a schedule $\sigma$ is $\sum_{j : \sigma_i(j)} p_j / s_i$.
The completion time of machine $i$ of type $t$ in a schedule $\sigma$ can then be described as $s_t^\inv p^\intercal \sigma_i$.
Let $\iv x_0 \defeq \iv x \ucup \sing 0$.

Now, we define the \enquote{freeness} of states, \eg how much space as measured in the parameter $\alpha$ is additionally allotted in some solution, even though a smaller value would suffice for a solution with the same makespan.

\begin{definition}[
	, name  = $\kappa$-free States
	, label = defi:kappa_free
]
A valid state $(\alpha, \mub, \nu)$ is \emph{$\kappa$-free} if $(\alpha - \kappa, \mub, \nu)$ is valid.
\end{definition}
One can interpret a $\kappa$-free state as a state where there is a solution valid to \ipdyn with the knapsack constraint \eqref{ip:dyn:constr:blue_area} having slack at least $\kappa$.
The general idea behind this measure is that we only want to compute states that act in a controlled manner regarding their \enquote{freeness}. By this, we can ensure that, when another job is inserted or removed in the next step, the parameters $\mub$ and $\alpha$ don't change too much.

We now intend to define a partial map that maps a makespan to the machine type $t \in \iv \tau$ whose real capacity (under the given makespan) is exactly $\lb$.
The map is only partial as there might simply be no type with appropriate capacities, since there does not have to be a machine of speed $(1+\eps)^k$ for every $k \in \Z$.
\begin{definition}[
	, name  = Critical Type
	, label = defi:critical_type
]
\[
		\thecrit \colon (1 + \eps)^\Z \pto \iv \tau
		\enspace , \qquad
		T \mapsto \begin{dcases}
			t & \tif \exists t \in \iv \tau\qsep s_t T = \lb 
			\\
			\bot & \totherwise
		\end{dcases}
		\enspace .
		\stopphier
	\]
\end{definition}
As a shorthand we write $\crit \defeq \crit(\OPT+)$ when $\OPT+$ is evident from context, and $\at q \crit \defeq \crit*(\at q {\OPT+})$.
Moreover, as a convention, if we consider multiple makespans an their critical types, we will always name those types based on the notation $\dot t$, e.g. $\bar {\dot t}$, $\tilde {\dot t}$ et cetera.

Machines of this type are the ones that might be necessary to be recoloured during a sequence of steps where $\OPT+$ does not change.
To see this, we need to show that $\OPT$ changes by a factor of at most $(1+\eps)$, which implies similar bounds for $\OPT+$ and $\OPT-$.
This then implies that recolouring all machines of this type is enough to maintain \autoref{invar:machine_colouring} and feasibility of the solution.
In general, there is no such bound -- even for identical machines, as \textcite{SV16} already noted.
Thus, instead of introducing a general measure on all instances, we bound the change for the relevant cases only, where at least one machine has capacity at least $\lb$ either before or after a change of the job set.
This suffices, since -- as described above -- we intend to use this bound to show we only need to recolour machines \rtb or \btr that are of critical type.
Clearly, if no machine can be coloured blue (without violating \autoref{invar:machine_colouring}), there is no need (nor possibility) to do a recolouring of any kind.
To show the bound on the change, we need the following weak estimation of $\OPT$:
\begin{proposition}[
	, name = A Weak Estimation Of $\OPT$
	, label = stmt:estimate_opt
]
	\hfill\hfill$\displaystyle
		\OPT \leq \smax^\inv p^\intercal \nu \enspace .
	$\hfill\,\!
\end{proposition}
\begin{proof}
Let $\sigma$ a schedule where all jobs are placed onto a single machine $i$ with $s_i = \smax$.
	Then $\sigma$ has makespan $\smax^\inv p^\intercal \nu$, so it is a feasible solution for this problem with $T \gets \smax^\inv p^\intercal \nu$.
	Thus, we know that there is a $\smax^\inv p^\intercal \nu$-valid state, and thus $\OPT \leq \smax^\inv p^\intercal \nu$ by \fullref{defi:opt}.
\end{proof}
With this simple estimate, we can show the following lemma.
\begin{lemma}[
	, name = Bounding Makespan Changes
	, label = stmt:makespan_change
]
Consider $\nu \in \N_0^d$ and $j \in \iv d$.
	Let $f(\cdot) \in \Sing{ \OPT(\cdot) \Ssep \OPT-(\cdot) \Ssep \OPT+(\cdot) }$.
	If $\smax\OPT+(\nu + \uvec j) \geq \lb$, then 
	\hfill$
		\frac{f(\nu + \uvec j)}{f(\nu)} \leq (1 + \eps)
	\enspace ,$ \hfill \, \\*
	and if $j \in \supp(\nu)$ and $\smax\OPT+(\nu) \geq \lb$, 
	\hfill$
		\frac{f(\nu - \uvec j)}{f(\nu)} \geq (1 + \eps)^\inv \geq (1 - \eps)
		\enspace .
		\stopphier
	$\hfill
\end{lemma}
\begin{proof}
	We start with the first inequality and $f(\cdot) = \OPT(\cdot)$.
	Let $\bar\nu \defeq \nu + \uvec j$.
	By assumption, we have $\smax \OPT+(\bar \nu) \geq \lb$.
	Hence
	\[
		(1 + \eps)^\inv \smax^\inv \lb
		\leq \OPT(\bar \nu)
		\leq \OPT(\nu) + \Delta\OPT
		\iff
		(1 + \eps)^\inv \smax^\inv \lb - \Delta\OPT \leq \OPT(\nu) \enspace .
	\]
	We can thus deduce that $\Delta\OPT \leq \smax^\inv p_j$ by transforming an optimal solution, placing $j$ onto the fastest machine, and transforming the schedule back into a $\OPT(\nu) + \smax^\inv p_j$-valid solution. 
	Moreover, obviously $\Delta\OPT \geq 0$.
	Now,
	\begin{align*}
		\frac{\OPT(\bar\nu)}{\OPT(\nu)}
		& =
		\frac{\OPT(\nu) + \Delta\OPT}{\OPT(\nu)}
		=
		1 + \frac{\Delta\OPT}{\OPT(\nu)}
		\leq 
		1 + \frac{\Delta\OPT}{(1 + \eps)^\inv \smax^\inv \lb - \Delta\OPT}
		\\&\leq
		1 + \frac{\smax^\inv p_j}{(1 + \eps)^\inv \smax^\inv \lb - \smax^\inv p_j}
		=
		1 + \frac{p_j}{(1 + \eps)^\inv \lb - p_j}
		\leq
		1 + \frac \pmax {(1 + \eps)^\inv \lb - \pmax}
		\\&\leq
		1 + \frac \pmax { 2 \pmax \eps^\inv - \pmax}
		= 1 + \frac \pmax {\pmax \eps^\inv + \underbrace{(\eps^\inv - 1)}_{\geq 0} \pmax}
		\\&\leq 1 + \eps 
	\end{align*}

	The cases $f(\cdot) = \OPT-(\cdot)$ and $f(\cdot) = \OPT+(\cdot)$ follow directly via \autoref{defi:opt} and the inequality on $\OPT$.
	The second inequality can be shown by using the first with $\nu \gets \nu - \uvec j$ and inverting both sides.
\end{proof}
While there is nothing inherently \enquote{forcing} us to only recolour machines of critical type, doing so seems quite natural.
As it simplifies the analysis and the design of our algorithms (there are simply fewer cases to consider), we capture this idea in an invariant:
\begin{invariant}[Recolouring Invariant]
	\label{invar:recolouring}
	At a step $q \to q+1$, only machines $i \in \iv m$ of critical type are recoloured. 
\end{invariant}

We are now ready to describe how we want to modify a state each time we add or remove a job, such that the new state has distance bounded in $\poly(\epsinv,\pmax)$.

\subsection{Permanent Case: Handling Insertion}
\label{sec:insertion}

We begin by looking at the permanent setting, where jobs can only be inserted. The change of our parameters $\mub$ and $\alpha$ depends on how much we have to increase the $\kappa$-freeness of the current state to \enquote{fit in} the new job. First, we make a general observation about solutions with enough $\kappa$-freeness.

\begin{lemma}[
	, name = Adding a job to a $\kappa$-free state
	, label = {stmt:kappa_free_add}
]
	Let $S = (\alpha, \mub, \nu)$ be valid and $\kappa$-free.
	Then for any $\nu^\Delta \in \N_0^d$ with $p^\intercal \nu^\Delta \leq \kappa$, $(\alpha, \mub, \nu + \nu^\Delta)$ is $\OPT+(\nu)$-valid.
\end{lemma}
\begin{proof}
	Let $S' \defeq (\alpha - \kappa, \mub, \nu)$.
	Then $S'$ is valid by \fullref{defi:kappa_free}.
	Let $(x, \super \tblue \nu)$ be the solution associated with $S'$.
	If we now increase the usable blue area back to $\alpha$, this solution remains feasible and has unchanged makespan.
	The knapsack constraint then has slack at least $\kappa$, so adding the job set $\nu^\Delta$ can be done by simply placing the jobs onto the blue machines.
	The new solution does not violate \autoref{invar:comp_colouring} nor \autoref{invar:machine_colouring}, as its makespan is no larger than the makespan of $(x , \super \tblue \nu)$.
\end{proof}

So if a state is at least $p_j$-free, we can directly add job $j$ to the ILP without changing any of the state parameters, and still stay valid.
If this is not possible, we may have increase the freeness somehow.
Luckily, when colouring a machine of $\crit(\OPT+(\nu))$ \rtb, we get at least $\veps s_{\crit} T = \veps \lb \geq 2 \pmax$ bonus capacity due to the definition of $\lb$ and the additional error allowed on blue machines by \autoref{invar:comp_colouring}, leading to the following lemma.

\begin{lemma}[
	,name=Recolouring a Red Machine
	,label=stmt:rtb_pmax_free
]
	If $S \defeq (\alpha, \mub, \nu)$ is valid and there is a red machine of type $t \in \iv \tau$ with $s_t \OPT+ \geq \lb$, then $S' \defeq (\alpha + s_t \OPT+ + \pmax, \mub + \uvec t, \nu)$ is valid and $\pmax$-free.
	Moreover, $\Cmax(S') \leq \Cmax(S)$.
\end{lemma}
\begin{proof}
	Let $i$ be machine to be recoloured and $t$ its type, let $\sigma_i$ be the assignment of jobs to $i$.
	Then $(\alpha + s_i \OPT+(\nu), \mub + \uvec t, \nu)$ has a makespan no larger than $(\alpha, \mub, \nu)$, as we can do the following modifications to the solution $(x, \nub)$ associated to $(\alpha, \mub, \nu)$:
	\begin{enumerate}
		\item Set $x_{t,\sigma_i} \gets x_{t,\sigma_i} - 1$, $\nu \gets \nu - \sigma_i$ and $\mu_t \gets \mu_t - 1$; then the modified solution is feasible and its makespan did not increase.
		\item Set $\alpha \gets \alpha + s_t \OPT+(\nu)$, $\nu \gets \nu + \sigma_i$, and $\nub \gets \nub + \sigma_i$; then the job set is the same as in the beginning, the makespan did not increase and the knapsack constraint is not violated as $p^\intercal \nub \leq \alpha$, so $p^\intercal (\nub + \sigma_i) \leq \alpha + p^\intercal \sigma_i \leq \alpha + s_i \OPT+$.
		\item $\mub_t \gets \mub_t + 1$ and $\mu_t \gets \mu_t + 1$, which is a change of parameters, but does not change the right hand side of the ILP where only $\mu - \mub$ appears.
		\item $\alpha \gets \alpha + \pmax$; then the state becomes $\pmax$-free if it was valid before.
	\end{enumerate}
	To prove the validity of the state before the last step, let us consider the maximally available blue area.
	Now, we know exactly about the change of it, which is
	\[
		(1 + \eps) \OPT+(\nu)  s^\intercal (\mub + \uvec t) - (1 + \eps) \OPT+(\nu)  s^\intercal \mub)
		= (1+\veps) s_t \OPT+ \enspace .
	\]
	Thus, by subtracting $(1+\veps) s_t \OPT+$ from each side, we have
	\begin{multline*}
		\alpha \leq (1 + \eps) \OPT+(\nu)  s^\intercal \mub = (1 + \eps) \OPT+(\nu)  s^\intercal (\mub + \uvec t) - (1+\veps) s_t \OPT+
		\\ \iff
		\alpha + (1+\veps) s_t \OPT+ \leq (1 + \eps) \OPT+(\nu)  s^\intercal (\mub + \uvec t)
		\enspace .
	\end{multline*}
	As $s_t \OPT+ + \pmax \leq (1+\veps) s_t \OPT+$ by the definition of $\lb$, $\alpha$ is increased by exactly $s_t \OPT+ + \pmax$ after the last step of our procedure, this implies that afterwards, \autoref{invar:comp_colouring} is not violated for $T \gets \OPT+$.
	Since before the last step, the value of $\alpha$ was even smaller, we get that in both situations, the current state is indeed valid.
	Hence, $(\alpha + s_t \OPT+ + \pmax, \mub + \uvec t, \nu)$ is $\pmax$-free by \fullref{defi:kappa_free}, and has makespan at most $\Cmax(S)$, which eventually shows the claim.
\end{proof}

By \autoref{invar:recolouring} we may only recolour machines of critical type \rtb to gain $\pmax$-freeness. What if all machines of this type have already been recoloured? We then can make the following observation about the validity and makespan of such solutions.

\newcommand\primed[1]{#1'}
\begin{lemma}[
	, label = stmt:opt_max
]
	Let $\nu$ be a job set and for all $t \in \iv \tau$, $\mub_t = \mu_t$ if $s_t \OPT+ \geq \lb$ and $0$ otherwise.
	Then $\enpar[\Big]{\floor*{(1 + \eps) \OPT+(\nu)  s^\intercal \mub}, \allowbreak\mub,\allowbreak \nu}$ is valid and its makespan is \emph{at most} $\OPT$.
\end{lemma}
\begin{proof}
	At first glance it might seems like a contradiction that the makespan could be less than $\OPT$.
	But this actually can happen, as \fullref{defi:opt} differs from \fullref{defi:valid} in terms of the parametrization of the invariants.
	In \fullref{defi:opt}, they have to hold for $\OPT$, while for a valid solution they are only required to hold for $\OPT+$.
	And since our definition of makespan of a state only considers $\Cmax(x)$ of the associated solution $(x, \nub)$, we can indeed have a makespan below $\OPT$.
	Let $\bar S \defeq (\bar\alpha, \mub[\bar], \nu)$ be an $\OPT$-valid state, i.e. such that the invariants \ref{invar:machine_colouring} and \ref{invar:comp_colouring} both hold for $T \gets \OPT$, and the associated solution has makespan at most $\OPT$.
	Such a state always exists by definition of $\OPT$.
	Then
	\begin{multline*}
		\bar\alpha
		\leq (1 + \eps) \OPT(\nu)  s^\intercal \mub[\bar] \leq (1 + \eps) \OPT+(\nu)  s^\intercal \mub[\bar]
		\\ = (1 + \eps) \OPT+(\nu)  s^\intercal \mub - (1 + \eps) \OPT+(\nu)  s^\intercal (\mub-\mub[\bar])
		\enspace .
	\end{multline*}
	Note that $\mub \geq \mub[\bar]$, as otherwise, by assumption on the form of $\mub$, under $\mub[\bar]$ either more than $\mu_t$ machines are coloured blue for some $t \in \iv \tau$, or a machine of some type $t \in \iv \tau$ with $s_t \OPT+ < \ell$ is coloured blue, which would violate \autoref{invar:comp_colouring}.
	So in particular, by adding $(1 + \eps) \OPT+(\nu)  s^\intercal (\mub-\mub[\bar])$ on each side of the inequality and rounding down, we have 
	\[
		\alpha' \defeq \floor*{\bar\alpha + (1+\veps) \OPT+ \cdot s^\intercal \enpar*{\mub - \mub[\bar]} } \leq \floor*{(1 + \eps) \OPT+(\nu)  s^\intercal \mub} \enspace .
	\]
	By inductively applying \fullref{stmt:rtb_pmax_free} it follows that $\Cmax(\alpha', \mub, \nu) \allowbreak \leq \OPT$.
	Thus, as $\alpha' \leq \floor*{(1 + \eps) \OPT+(\nu)  s^\intercal \mub}$, also $\Cmax(\floor*{(1 + \eps) \OPT+(\nu)  s^\intercal \mub} \skomma \mub \skomma \nu) \leq \OPT$.
\end{proof}
We can use this statement now to show the validity of states where a new job is added, and $\OPT+$ doesn't change.
\begin{lemma}[
	, name  = Existence of a Red Machine
	, label = stmt:exists_red
]
	Consider $j \in \iv d$.
	If $(\alpha, \mub, \nu)$ is valid and $\OPT+(\nu) = \OPT+(\nu + \uvec j)$, then either $(\alpha + \kappa, \mub, \nu + \uvec j)$ is valid for some $\kappa \in \iv {p_j}_0$, or there is a red machine of capacity at least $\lb$. 
\end{lemma}
\begin{proof}
	Assume that $(\alpha + \kappa, \mub, \nu)$ is invalid for all $\kappa \in \iv {p_j}_0$.
	Then for some $\kappa \in \iv {p_j-1}_0$ we have $\alpha + \kappa = \floor*{(1 + \eps) \OPT+(\nu)  s^\intercal \mub}$, as otherwise \fullref{defi:kappa_free} together with \fullref{stmt:kappa_free_add} lead to a contradiction for $\alpha \gets \alpha + p_j$.
	If there was no red machine of capacity at least $\lb$, we would have coloured all machines blue that we are allowed to under \autoref{invar:machine_colouring}, and with $\alpha \gets \alpha + \kappa$ we would have used every bit of area that is available under \autoref{invar:comp_colouring}.
	This is a contradiction to \autoref{stmt:opt_max}. 
\end{proof}

To now prove correctness and bounded distance, we need the following proposition.
\begin{proposition}[
	, name =  Increasing $\alpha$ for valid states
	, label = stmt:alpha_increase_valid
]
	If $S = (\alpha, \mub, \nu)$ is valid, then $S' = (\alpha', \mub, \nu)$ is valid for any $\alpha' \in \interval[scaled] \alpha {(1 + \eps) \OPT+(\nu)  s^\intercal \mub}$.
\end{proposition}
\begin{proof}
	Let $(x, \nub)$ be the solution associated to $S$.
	Clearly, $(x, \nub)$ is a feasible solution to $S'$.
	Moreover, $\Cmax(x) \leq \OPT+$ as $S$ is valid, and since $\OPT+$ is the same for $S'$ (as it only depends on $\nu$), $S'$ is valid.
\end{proof}

We are now ready to describe our algorithm for inserting a new job for a state $S$, such that the new job set is valid to the returned state, and the distance from $S$ to the returned state is bounded.

\begin{algorithm}[h]
\caption{Inserting a new job $j$, given the current state $S$}
\label{algo:insert}
\begin{algorithmic}[1]
	\Function{Insert}{$j$, $S$}
	\State $S$ is of the form $(\alpha, \mub, \nu)$
	\If{$\OPT+(\nu + \uvec j) = \OPT+(\nu)$}
		\If{$S$ is not $p_j$-free}
			\State$\alpha \gets \min \Sing{(1 + \eps) \OPT+(\nu + \uvec j)  s^\intercal \mub \ssep \alpha + p_j}$ \label{algo:insert:line:alpha_gets_min}
			\If{$S$ is (still) not $p_j$-free}
					\State $t \gets \crit(\OPT+(\nu))$ \Comment{Note that $t \neq \bot$}
					\State $\mub \gets \mub + \uvec t$ \Comment{Note that $\mub_t \neq \mu_t$}
					\State $\alpha \gets \alpha + s_t \OPT+(\nu) + \pmax$
			\EndIf
		\EndIf
	\Else
		\If{$(\alpha, \mub, \nu + \uvec j)$ is valid}
			\State$\alpha \gets (1 + \eps) \OPT-(\nu + \uvec j)  s^\intercal \mub$ \label{alpha_new_delta} \Comment{Only needed for \Call{Remove}{}}
		\Else
			\State$\alpha \gets \alpha + p_j$ \label{alpha_second_pj}
		\EndIf
	\EndIf
	\State \Return $(\alpha, \mub, \nu + \uvec j)$ \EndFunction
\end{algorithmic}
\end{algorithm}

It follows directly directly from the definition of the algorithm that \Call{Insert}{} does not violate \autoref{invar:recolouring}.

We rephrase the algorithm as a lemma, and directly show that it always computes a valid state, and that the resulting state has parameters whose difference to the parameters of the input state is bounded in $\poly(\pmax) \leq \poly(\epsinv,\pmax)$.
This will then significantly simplify the proofs of further properties.
\begin{lemma}[
	, name  = \textsc{Insert} Rephrased
	, label = stmt:algo_insert_rephrased
]
	Given a valid state $S = (\alpha, \mub, \nu)$ and $j \in \iv d$, the call \Call{Insert}{$j,S$}
	\begin{enumerate}[(i)]
\item if $(\alpha, \mub, \nu + \uvec j)$ is valid, returns $(\alpha, \mub, \nu + \uvec j)$ if $\OPT+(\nu) = \OPT+(\nu + \uvec j)$, and $((1 + \eps) \OPT-(\nu + \uvec j)  s^\intercal \mub), \mub, \nu + \uvec j)$ else, where $(1 + \eps) \OPT-(\nu + \uvec j)  s^\intercal \mub \leq \alpha + p_j$; otherwise
		\item returns a valid state $(\bar\alpha, \mub, \nu + \uvec j)$ with $\bar\alpha \in \linterval {\alpha} {\alpha + 2 p_j}$; or otherwise
		\item correctly concludes that $(\bar\alpha, \mub, \nu + \uvec j)$ is invalid for all $\bar\alpha$, and returns a valid state of the form $(\bar\alpha, \mub + \uvec {\crit(\OPT+(\nu))}, \nu + \uvec j)$ where $\bar \alpha < \alpha + \lb + \pmax + p_j$.
		\stopphier
	\end{enumerate}
\end{lemma}
\begin{proof}
	In the following, we can assume that there always exists a machine $i$ with $s_i = \smax$, and $\smax \OPT+(\nu + \uvec j) \geq \ell$, as otherwise, it holds for both the input and return state that $\alpha = 0$, $\mub = \zero$ by \autoref{invar:machine_colouring}, and thus, (i) is always fulfilled.
	\begin{enumerate}[]
		\item \underline{Case 1: $(\alpha, \mub, \nu + \uvec j)$ is valid.}
		\begin{enumerate}[]
			\item \underline{Case 1.1: $\OPT+(\nu + \uvec j) = \OPT+(\nu)$}.
			Then the if-branch of the first if-statement is entered and neither $\alpha$ nor $\mub$ are changed.
Since the algorithm then returns $(\alpha, \mub, \nu + \uvec j)$, this shows the claim, as (i) is satisfied.
			\item \underline{Case 1.2: $\OPT+(\nu + \uvec j) > \OPT+(\nu)$.}
			Then the else-branch of the first if statement is entered.
			Within this branch, the first if-branch of the next if-statement is entered due to the assumption of Case 1.
			Thus, the algorithm eventually returns
			$$\enpar*{(1 + \eps) \OPT-(\nu + \uvec j)  s^\intercal \mub, \mub, \nu + \uvec j} \enspace .$$
By \autoref{stmt:makespan_change} and using the fact that $\OPT+$ and $\OPT-$ are restricted to integer powers of $(1+\eps)$,
			\[
				(1 + \eps) \OPT-(\nu + \uvec j)  s^\intercal \mub = (1 + \eps) \OPT+(\nu)  s^\intercal \mub \geq \alpha \enspace ,
			\]
			where the latter follows from the fact that the initial state was valid and thus satisfied \autoref{invar:comp_colouring}.
Now, if $(1 + \eps) \OPT-(\nu + \uvec j)  s^\intercal \mub = (1 + \eps) \OPT+(\nu)  s^\intercal \mub > \alpha + p_j$, then
			$((1 + \eps) \OPT+(\nu)  s^\intercal \mub, \mub, \nu + \uvec j)$ would be a valid state for $\OPT+(\nu)$ by \autoref{stmt:kappa_free_add}, and thus $\OPT+(\nu) = \OPT+(\nu + \uvec j)$, a contradiction to the assumption of Case 1.2.
			Thus, $(1 + \eps) \OPT-(\nu + \uvec j)  s^\intercal \mub \leq \alpha + p_j$.
			Since $\mub$ remains unchanged, this shows the claim, as (i) is satisfied.
		\end{enumerate}
		
		\item \underline{Case 2: $(\alpha, \mub, \nu + \uvec j)$ is invalid and $\OPT+(\nu + \uvec j) \neq \OPT+(\nu)$.}
		Then only \lineref{alpha_second_pj} is executed. By definition of this line, $\alpha$ is increased by at most $p_j$.
		The returned state is valid: By the assumption, we know that $\OPT+(\nu + \uvec j) > \OPT+(\nu)$ and that there exists a machine $i$ with speed $\smax$ s.t. $\smax \OPT(\nu + \uvec j) \geq \lb$.
		This machine $i$'s capacity is then increased by at least $2 \pmax$, as here, $\OPT+(\nu + \uvec j) = (1+\veps) \OPT+(\nu)$ by definition, and therefore, $i$'s capacity is increased by at least $(1+\veps)^\inv \veps \lb \geq 2 \pmax$, since $i$ had capacity at least $(1+\veps)^\inv \lb$ before by \autoref{stmt:makespan_change}.
		
		Thus, if $i$ is coloured blue, \autoref{invar:comp_colouring} is maintained.
		If $i$ is coloured red instead, $j$ can be inserted without changing $\alpha$ at all, and $(\alpha, \mub, \nu + \uvec j)$ would be valid, a contradiction to the assumption of Case 2.
		It follows that (ii) is satisfied.

		\item \underline{Case 3: $(\alpha, \mub, \nu + \uvec j)$ is invalid and $\OPT+(\nu + \uvec j) = \OPT+(\nu)$.}
		Then the second if-branch is entered because of \fullref{stmt:kappa_free_add}.

		\begin{enumerate}[]
			\item \underline{Case 3.1: there is a valid state of the form $(\tilde\alpha, \mub, \nu + \uvec j)$.}
			Let $\alpha'$ be the value of $\alpha$ after \lineref{algo:insert:line:alpha_gets_min}.
			We prove that (ii) will always be satisfied.
			\begin{enumerate}[]
				\item \underline{Case 3.1.1: $\alpha' = \alpha + p_j$.}
				Then $(\alpha', \mub, \nu)$ is $p_j$-free by \fullref{defi:kappa_free} and hence $(\alpha', \mub, \nu)$ is valid by \fullref{stmt:kappa_free_add}.
				Moreover, $\bar\alpha \defeq \alpha' = \alpha + p_j$ satisfies the condition of (ii) on $\bar\alpha$.
				\item \underline{Case 3.1.2: $\alpha' = (1 + \eps) \OPT+(\nu + \uvec j)  s^\intercal \mub < \alpha + p_j$}.
				Then, with $\alpha'$, we use all the space available to us.
				By our case (Case 3.1), we know that there is a valid state of the form $(\tilde\alpha, \mub, \nu + \uvec j)$.
				Since $\alpha'$ is the largest possible value (otherwise \autoref{invar:comp_colouring} is violated), ${\alpha'} \geq \tilde\alpha$ .
				By \fullref{stmt:alpha_increase_valid}, $(\alpha', \mub, \nu + \uvec j)$ is valid.
Setting $\bar\alpha \defeq \alpha' < \alpha + p_j$ thus shows the claim.
			\end{enumerate}

			\item \underline{Case 3.2: $(\bar\alpha, \mub, \nu + \uvec j)$ is invalid for all $\bar\alpha$.}
			Then the second if-statement is entered with usable blue area $\alpha' = (1 + \eps) \OPT+(\nu + \uvec j)  s^\intercal \mub < \alpha + p_j$ (otherwise, $(\alpha', \mub, \nu)$ was $p_j$-free by \fullref{defi:kappa_free} \contradiction).
			We show that (iii) is satisfied.
\begin{enumerate}[]
				\item \underline{Case 3.2.1: $t \neq \bot \land \mub_t \neq \mu_t$.}
				Then $\bar\alpha \defeq \alpha' + \lb + \pmax$ and $(\bar \alpha, \mub + \uvec t, \nu + \uvec j)$ is returned by our algorithm, which is valid by \fullref{stmt:kappa_free_add} as $(\bar \alpha, \mub + \uvec {t}, \nu)$ is  $\pmax$-free.

				\item \underline{Case 3.2.2: $t = \bot \lor \mub_t = \mu_t$.}
				Then \Call{Insert}{} would fail, as it tries to recolour a machine of undefined type or a machine of a type where all machines are blue already.
				We show that this case cannot occur.
				First, we know that, since $\OPT+(\nu) = \OPT+(\nu + \uvec j)$, in this case there is at least one red machine by \fullref{stmt:exists_red}.
By \autoref{invar:recolouring} this implies that at some point before, a job $j'$ was added which lead to an increase of $\OPT+$ while there was still a red machine with capacity $s_{t'} \OPT+ = \lb$.
				This leads to a contradiction, as if at that step, this machine was recoloured \rtb, then $j'$ could have been added without increasing $\OPT+$ by \fullref{stmt:rtb_pmax_free}. \qedhere
			\end{enumerate}
		\end{enumerate}
	\end{enumerate}
\end{proof}
As an immediate consequence, since $(1+\eps) \lb \leq \poly(\epsinv,\pmax)$, we get that \Call{Insert}{} always computes a valid state $\bar{S} = (\bar \alpha, \mub[\bar], \allowbreak \bar \nu)$ such that $\abs{\Delta \alpha} + \norm{\Delta\mub}_1 + \norm{\Delta\nu}_1 \leq \poly(\epsinv,\pmax)$.
We can then achieve the desired migration bound and run time for the permanent setting using \fullref{stmt:sensitivity},~(\ref{stmt:sensitivity:param}) and \fullref{stmt:ipdyn_solve},~(\ref{stmt:ipdyn_solve:param}), which can be found in \appendixref{sec:ipsolve}.

\subsection{Dynamic Case: Handling Removal}
\label{sec:remove}

After successfully having constructed a robust algorithm for the static case, we now move on to the dynamic setting, for which we also need to be able to handle the removal of jobs.
We make use of the subroutine \Call{TryBTR}{} that simply tries to recolour up to $k$ machines \btr without violating \autoref{invar:comp_colouring}.
We then propose algorithm \Call{Remove}{} as a procedure to remove a job, using \Call{TryBTR}{} as a subroutine.
\begin{algorithm}[h]
\begin{algorithmic}[1]
	\Function{TryBTR}{$k, \alpha, \mub, \nu$}
		\If{$\mub \neq \zero$}
			\State $t \gets \crit(\OPT+(\nu))$
			\If{$t \neq \bot$}
				\State $k \gets \min \sing{ k, \mub_t }$
				\State $\mub \gets \mub - k \cdot \uvec t$
				\State $\alpha \gets \alpha - (1+\veps) \cdot k s_t \cdot \OPT+(\nu)$
			\EndIf
		\EndIf
		\Return $(\alpha, \mub, \nu)$	
	\EndFunction
	\end{algorithmic}
	\caption{Recolouring subroutine}
	\label{algo:tryBTR}
\end{algorithm}
\begin{algorithm}[h]
\begin{algorithmic}[1]
	\Function{Remove}{$j$, $S$}
	\State $S$ is of the form $(\alpha, \mub, \nu)$
	\State $t \gets \crit(\OPT+(\nu - \uvec j))$
\If{$\OPT+(\nu - \uvec j) < \OPT+(\nu)$}
		\State $S \gets \Call{TryBTR}{m, S}$ \label{algo:remove:line:btr_m}
		\Comment{Note that this modifies $\mub$}
		\State $\alpha \gets (1 + \eps) \cdot \OPT+(\nu - \uvec j)  s^\intercal \mub$ \label{algo:remove:line:alpha_after_btr} \EndIf
	\State $\nu \gets \nu - \uvec j$ \label{algo:remove:line:part}
	\If{$\text{$S$ is $(\veps \lb + \pmax)$-free} \land t \neq \bot \land \mub_t > 0$}
		\State $S \gets \Call{TryBTR}{1, S}$ \label{algo:remove:line:tryBTR1}
	\EndIf
	\If{$\text{$S$ is $\pmax$-free} \land \enpar*{t = \bot \lor \mub_t = 0}$}
		\State~$\alpha \gets \max \Sing{ \alpha - \pmax ,\, (1 + \eps) \cdot \OPT-(\nu) \cdot s^\intercal \mub) }$ \label{algo:remove:line:alpha_minus_pmax}
	\EndIf
	\State \Return $(\alpha, \mub, \nu)$
	\EndFunction
\end{algorithmic}
\caption{Removing a job $j$, given the current state $S$}
\label{algo:remove}
\end{algorithm}

We now want to show similar properties as in the insertion case, \eg prove that the state computed by the algorithm is valid for the new job vector, and the change of the parameters is bounded by $\poly(\epsinv,\pmax)$. The next lemma is almost the inverse of \fullref{stmt:rtb_pmax_free}, except that we require $S$ to be $(\veps \lb + \pmax)$-free instead of gaining $\pmax$-freeness.
\begin{lemma}[
	, name = Recolouring a Blue Machine
	, label = stmt:btr_2pmax_free
]
	If $S \defeq (\alpha, \mub, \nu)$ is $(\veps \lb + \pmax)$-free and $\mub_{\crit} > 0$, where $\crit = \crit(\OPT+(\nu))$, then $S' \defeq (\alpha - (1+\veps) \lb, \mub - \uvec {\crit} , \nu)$ is valid.
\end{lemma}
\begin{proof}
	Let $i$ be machine of type $\crit \defeq \crit(\OPT+(\nu))$ to be recoloured.
	Note that $\lb = s_{\crit} \OPT+(\nu)$ by definition of $\crit(\cdot)$.
	Now, let $(x, \nub)$ be the solution associated to $\bar S \defeq (\alpha - (\veps \lb + \pmax), \mub, \nu)$.
	By \fullref{defi:kappa_free} we know that $\bar S$ is valid.
	Moreover, as the job set is the same as in $S$, the values for $\OPT+$ do also coincide.
	Hence, the solution $(x, \nub)$ associated to $\bar S$ is a feasible solution to $S$ where the \kpconstr{} has slack at least $(\veps \lb + \pmax)$.

	Consider an assignment of blue jobs onto blue machines such that each blue machine of type $t \in \iv \tau$ is assigned jobs of total load at most $(1+\veps) s_t \OPT+(\nu) + \pmax$.
	Such an assignment exists by \fullref{stmt:greedy}, as $S$ is valid and thus \autoref{invar:comp_colouring} holds for $T \gets \OPT+(\nu)$.
	Let $\nu^\Delta$ be the jobs assigned $i$ of type $\crit$.
	Consider the following modifications to $(x, \nub)$ and $S$:
	\begin{enumerate}
		\item Set $\nu \gets \nu - \nu^\Delta$, $\nub \gets \nub - \nu^\Delta$ -- then $(x, \nub)$ has slack $(\veps \lb + \pmax) + p^\intercal \nu^\Delta \leq (1+\veps) \lb + \pmax$ under the current state.
		\item Set $\mub_{\crit} \gets \mub_{\crit} - 1$, $\mu_{\crit} \gets \mu_{\crit} - 1$, $\alpha \gets \alpha - (1+\veps) \lb$
		then the modified state is $T$-valid, as the removal of a machine together with the jobs assigned to it cannot increase the makespan of the solution.
		Moreover, our solution now has slack $\pmax + p^\intercal \nu^\Delta -\lb$.
		\item Let $\nu^+ \in \Interval 0 {\nu^\Delta}$ be an inclusion minimal selection of jobs such that $p^\intercal (\nu^\Delta - \nu^+) \leq \lb$.
		Then set $x_{\crit, (\nu^\Delta - \nu^+)} \gets x_{\crit, (\nu^\Delta - \nu^+)} + 1$, $\mu_{\crit} \gets \mu_{\crit} + 1$, $\nu \gets \nu + \nu^\Delta - \nu^+$.
		This solution then has makespan at most
		\[ \max \sing {T, s_{\crit}^\inv p^\intercal (\nu^\Delta - \nu^+) }
		\leq \max \sing {T, s_{\crit}^\inv \lb}
		= T \enspace . \]
		Moreover, it (still) has slack $\pmax + p^\intercal \nu^\Delta - \lb$.
		\item Note that
		\[
			\pmax + p^\intercal \nu^\Delta - \lb
			= \pmax + p^\intercal (\nu^\Delta - \nu^+) + p^\intercal \nu^+ - \lb
			\geq p^\intercal \nu^+ \enspace ,
		\]
		since $p^\intercal (\nu^\Delta - \nu^+) \geq \lb - \pmax$ by inclusion minimality of $\nu^+$. Thus, we can place $\nu^+$ as blue jobs, i.e. modify $\nub \gets \nub + \nu^+$ and $\nu \gets \nu + \nu^+$, without violating the \kpconstr{}.
	\end{enumerate}
	In the end, the value of $\nu$ is restored, i.e. is equal to the initial value, $\alpha$ is decreased by $(1+\veps) \lb$, and one more machine of critical type is now coloured red.
	\autoref{invar:machine_colouring} cannot be violated as we only recoloured a machine of the critical type $\crit$.
	\autoref{invar:comp_colouring} is not violated as recolouring a machine \btr decreases the total blue area by $(1+\veps) \lb$, but we also decreased the usable blue area $\alpha$ accordingly.
	Thus, the claim holds.
\end{proof}

With this result in hand, we can prove the first part: Our algorithm does indeed always compute a valid state.

\begin{lemma}[
	, name  = Correctness of \textsc{Remove}
	, label = stmt:algo_remove_correct
]Given a valid state, \Call{Remove}{} always computes a valid state.
\end{lemma}
\begin{proof}
	First, we show that the state computed after \lineref{algo:remove:line:part} is valid.
	Afterwards, we argue that the remaining lines will not change a valid state into an invalid one.
	Let $j$ be the job that is deleted, i.e. $\Delta \nu = \uvec j$.
	We make a case distinction on $\Delta\OPT$.
	If $\Delta \OPT+ = 0$, then simply removing the job already yields a valid state, which is exactly what the algorithm computes.
	If $\Delta \OPT+ < 0$, \lineref{algo:remove:line:btr_m} ensures that \autoref{invar:machine_colouring} is not violated.
	The line thereafter ensures, that \autoref{invar:comp_colouring} is not violated:
	$(1 + \eps) \OPT+(\nu - \uvec j)  s^\intercal \mub$ is exactly the blue area we get when all machines of $\crit(\OPT+(\nu - \uvec j))$ are coloured blue, which is true for the new $\crit(\OPT+(\nu - \uvec j))$, since machines of this type had capacity $> \lb$ for $\OPT+$ and where thus coloured blue by validity of the input state. For $\mub = \zero$, the equality above also holds as both sides are $0$.
	In consequence, setting $\alpha$ to the value above satisfies \autoref{invar:comp_colouring} for $T \gets \OPT+(\nu - \uvec j)$.
	Validity of the state then eventually follows from \fullref{stmt:opt_max}. Since \Call{TryBTR}{} only recolours machines of critical-type, \autoref{invar:recolouring} holds as well.

	Now, let us consider what happens after \lineref{algo:remove:line:part}.
	If a machine is recoloured due to an execution of \lineref{algo:remove:line:tryBTR1}, then $S$ was $(\veps \ell + \pmax)$-free before.
	Then the resulting state is also valid due to \fullref{stmt:btr_2pmax_free}.
	If \lineref{algo:remove:line:alpha_minus_pmax} is executed, then $\alpha$ is changed by at most $\pmax$, but the state was $\pmax$-free before, implying it is valid thereafter.
\end{proof}
Now, in order to prove bounded migration of our proposed procedure, we first need to introduce two lower bounds on the value of $\alpha$, and show that they hold for every step of our algorithm.
If both lower bounds hold for our starting state, this property is implied if they hold for any execution of \Call{Insert}{} and \Call{Remove}{} on a valid state as well.
These lower bounds will help us to first proof bounded $\kappa$-freeness of most states computed by our algorithm, and by that, help us show that the \enquote{critical lines} \lineref{algo:remove:line:btr_m} and \lineref{algo:remove:line:alpha_after_btr} of \Call{Remove}{} do not reduce $\mub$ and $\alpha$ too much, respectively.
This then will imply bounded migration. The first lower bound is \autoref{invar:lower_bound}.

\begin{invariant}[
	, label = invar:lower_bound
]
	\hfill\hfill$\displaystyle
		\crit \neq \bot \implies (1 + \eps) \OPT+(\nu)  s^\intercal (\mub - \uvec \crit) < \alpha
	\enspace.$\hfill\,\!
\end{invariant}

Note that $\mub - \uvec {\crit}$ can have a negative entry; in this case, for the sake of simplicity, we define $(1 + \eps) \OPT+(\nu)  s^\intercal (\mub - \uvec j)$ to be $-\infty$.
For the initial solution with no jobs, \autoref{invar:lower_bound} holds as $\OPT = \OPT+ = 0$, implying that there cannot be a blue machine, and hence $\crit = \bot$. As we only use procedures \Call{Insert}{} and \Call{Remove}{} to compute new states, if both procedure maintain the invariant, the whole algorithm does so.
\begin{proposition}[
	, label = stmt:algo_insert_maintains_lower_bound
]
	\Call{Insert}{} maintains \autoref{invar:lower_bound}
\end{proposition}
\begin{proof}
	Since the value for $\alpha$ is only increased by \Call{Insert}{}, there are only two scenarios where the claim does not become trivial:
	A change of $\OPT+$ and a recolouring of a red machine (both cannot occur at the same time due to \fullref{stmt:rtb_pmax_free}).
	Let $S \defeq (\alpha, \mub, \nu)$ be the state before insertion of the new job $j$ and $\bar S \defeq (\bar\alpha, \mub[\bar], \bar\nu)$ with $\bar \nu \defeq \nu + \uvec j$ the state afterwards.
	\begin{enumerate}[]
		\item \case [1] {$\mub[\bar] \neq \mub \land \OPT+(\nu) = \OPT+(\bar\nu)$.}
		This means that (iii) of \fullref{stmt:algo_insert_rephrased} \enquote{is executed}.
		Then $(\alpha', \mub, \bar\nu)$ is invalid for all $\alpha'$, so in particular for all $\alpha' \leq (1 + \eps) \OPT+(\bar \nu)  s^\intercal \mub $ which are all possible values that do not violate \autoref{invar:comp_colouring}.
		That means that the invalidity stems from a too large makespan of the associated solutions.
		Recolouring a machine \rtb without increasing the $\alpha$-value will only make the situation worse, i.e. cannot lead to a smaller makespan.
		In consequence, the state $(\bar\alpha, \mub + \uvec {\crit}, \bar\nu)$ returned from \Call{Insert}{} must satisfy $\bar\alpha > (1 + \eps) \OPT+(\bar \nu)  s^\intercal \mub = (1 + \eps) \OPT+(\bar \nu)  s^\intercal (\mub[\bar] - \uvec \crit)$.
	
		\item \case [2] {$\mub[\bar] = \mub \land \OPT+(\nu) \neq \OPT+(\bar\nu)$.}
		Let $\bar\crit \defeq \OPT+(\bar\nu)$ be the (new) critical type. 
		In particular, we have $\bar{\crit} \neq \crit$.
Due to \fullref{defi:critical_type} and \autoref{invar:machine_colouring}, we either have $\crit = \bot$ or $\mub_{\crit} = 0$.
		In the first case, the premise of the implications of \autoref{invar:lower_bound} evaluates to \enquote{false}, and \autoref{invar:lower_bound} holds.
		In the latter case, $(1 + \eps) \OPT+(\bar \nu)  s^\intercal (\mub - \uvec \crit)$ evaluates to  $-\infty$, and \autoref{invar:lower_bound} holds as well, for all possible $\alpha$.
	\qedhere\end{enumerate}
\end{proof}
\begin{proposition}[
	, label = stmt:tryBTR_maintains_lower_bound
]
	\Call{TryBTR}{} maintains \autoref{invar:lower_bound}.
\end{proposition}
\begin{proof}
	Consider a call \Call{TryBTR}{$k, \alpha, \mub, \nu$}.
	We may without loss of generality assume that $\crit = \crit(\OPT+) \neq \bot$, otherwise the function simply returns the input state.
	Moreover, we may without loss of generality assume that $k \leq \mub_{\crit}$ as internally $k$ is modified accordingly (for any $k' \geq \mub_{\crit}$ the function behaves identically).
	Now, assume that \autoref{invar:lower_bound} holds for the initial state, i.e.
	\[
		\alpha > (1 + \eps) \OPT+(\nu)  s^\intercal (\mub - \uvec \crit)
		\enspace ,
	\]
	and let $(\bar\alpha, \mub[\bar], \nu)$ be the state returned by the function.
	Then
	\[
		\bar\alpha = \alpha - k (1+\veps) s_{\crit} \OPT+{}
		= \alpha - k (1+\veps) \lb
		\enspace .
	\]
	Moreover,
	\begin{multline*}
		(1 + \eps) \OPT+(\nu)  s^\intercal \mub[\bar]
		= (1 + \eps) \OPT+(\nu)  s^\intercal (\mub - k \uvec \crit)
		\\ = (1 + \eps) \OPT+(\nu)  s^\intercal \mub - k (1+\veps) s_{\crit} \OPT+{}
		= (1 + \eps) \OPT+(\nu)  s^\intercal \mub - k (1+\veps) \lb
		\enspace .  
	\end{multline*}
	We can then show the claim, i.e. given that \autoref{invar:lower_bound} did hold initially prove that it holds afterwards.
	\begin{align*}
		\alpha >{} & (1 + \eps) \OPT+(\nu)  s^\intercal (\mub - \uvec \crit)
		\\
		& \iff
		\alpha - k (1+\veps) \lb > (1 + \eps) \OPT+(\nu)  s^\intercal (\mub - \uvec \crit) - k (1+\veps) \lb
		\\& \iff
		\bar \alpha > (1 + \eps) \OPT+(\nu)  s^\intercal (\mub - (k+1)\uvec \crit)
		\\& = (1 + \eps) \OPT+(\nu)  s^\intercal (\mub[\bar] - \uvec \crit) \qedhere
	\end{align*}
\end{proof}

\begin{proposition}[
	, label = stmt:remove_maintains_lower_bound
]
	\Call{Remove}{} maintains \autoref{invar:lower_bound}
\end{proposition}
\begin{proof}
	Since the value for $\alpha$ is only decreased by \Call{Remove}{}, there are only two scenarios where the claim does not become trivial:
	A change of $\OPT+$ or a recolouring of a blue machine.
Let $S \defeq (\alpha, \mub, \nu)$ be the state before removal of job $j$ and $\bar S \defeq (\bar\alpha, \mub[\bar], \bar\nu)$ with $\bar \nu \defeq \nu - \uvec j$ the state afterwards.
	\begin{enumerate}[]
		\item \case [1] {$\mub[\bar] \neq \mub \land \OPT+(\nu) = \OPT+(\bar\nu)$.}	
		The invariant can only be violated due to a decrease of $\alpha$.
		As in this case, \lineref{algo:remove:line:btr_m} and \lineref{algo:remove:line:alpha_after_btr} are not executed, this can only be due to \lineref{algo:remove:line:tryBTR1} or \lineref{algo:remove:line:alpha_minus_pmax}.
		The execution of \lineref{algo:remove:line:tryBTR1} maintains the invariant by \autoref{stmt:tryBTR_maintains_lower_bound}.
		Let $\crit \eqdef \crit(\OPT+(\nu))$.	
		\lineref{algo:remove:line:alpha_minus_pmax} is only executed if $\crit = \bot$ or $\mu_{\crit} = 0$; in both cases, the invariant trivially holds, as either its premise is false or the LHS of the inequality evaluates to $-\infty$.

		\item \case [2] {$(\mub[\bar] = \mub \lor \mub[\bar] \neq \mub) \land \OPT+(\nu) \neq \OPT+(\bar\nu)$.}
		Then, \lineref{algo:remove:line:btr_m} and \lineref{algo:remove:line:alpha_after_btr} are executed.
		Again, the execution of \lineref{algo:remove:line:btr_m} maintains the invariant by \autoref{stmt:tryBTR_maintains_lower_bound}.
		Let $\check \crit \eqdef \crit(\OPT+(\bar\nu))$.
		Then, after \lineref{algo:remove:line:alpha_after_btr}, if $\mub_{\check \crit} > 0$,
		\[
			\tilde\alpha \defeq (1 + \eps)  s^\intercal \OPT+(\bar\nu) \mub[\tilde] > (1+\eps) \OPT+(\bar\nu) s^\intercal (\mub[\tilde] - \uvec {\check \crit}) \enspace .
\]
		Otherwise, if $\mub_{\check \crit} = 0$ or $\check \crit = \bot$, the invariant trivially holds as argued above. Thus, \autoref{invar:lower_bound} always holds when \lineref{algo:remove:line:alpha_after_btr} is executed.

		In the current case of $\OPT+(\nu) \neq \OPT+(\bar\nu)$, $\bar S$ can not be $\pmax$-free (nor $(\veps \lb + \pmax)$-free), as otherwise, we could simply reinsert $j$ without a makespan increase, implying $\OPT+(\nu) = \OPT+(\bar\nu)$, a contradiction.
		Thus, neither \lineref{algo:remove:line:tryBTR1} nor \lineref{algo:remove:line:alpha_minus_pmax} are executed, no more changes to $\tilde \alpha$ are made, and \autoref{invar:lower_bound} always holds.		
\qedhere\end{enumerate}
\end{proof}

\begin{invariant}
\label{invar:alpha_geq_delta}
	\hfill\hfill$\displaystyle
		\delta \defeq (1+\eps) \OPT-(\nu) s^\intercal (\mub - \mub_{\crit} \uvec {\crit}) \leq \alpha
	\enspace.$\hfill\,\!
\end{invariant}

Our second lower bound is \autoref{invar:alpha_geq_delta}.
As $\alpha = \delta = 0$ initially, this lower bound already holds for our starting state.
Luckily, showing that \Call{Insert}{} or \Call{Remove}{} maintain \autoref{invar:alpha_geq_delta} is less involved.

\begin{proposition}[
	, label = stmt:algo_insert_maintains_agd
]
	\Call{Insert}{} maintains \autoref{invar:alpha_geq_delta}.
\end{proposition}
\begin{proof}
	If $\OPT+(\nu + \uvec j) = \OPT+(\nu)$, \Call{Insert}{} only increases $\alpha$, while $\delta$ does not change.
	Else if $\OPT+(\nu + \uvec j) \neq \OPT+(\nu)$, $\alpha$ is set to $(1+\eps) \OPT-(\nu + \uvec j) s^\intercal \mub \geq \delta$ if the input state is valid already, or $\alpha$ is increased by at least $p_j$ if it is not.
	In the latter case, $\alpha > (1+\eps) \OPT-(\nu + \uvec j) s^\intercal \mub \geq \delta$ after the increase, as otherwise, the resulting state would be valid for $\alpha = (1+\eps) \OPT-(\nu + \uvec j) s^\intercal \mub \geq \delta$ as well, a contradiction to the assumption that it was not.
\end{proof}

\begin{proposition}[
	, label = stmt:algo_remove_maintains_agd
]
	\Call{Remove}{} maintains \autoref{invar:alpha_geq_delta}.
\end{proposition}
\begin{proof}
	If $\OPT+(\nu - \uvec j) < \OPT+(\nu)$, $\alpha$ is set to at least $\delta$ in \lineref{algo:remove:line:alpha_after_btr}, as all machines of $\crit(\OPT+(\nu))$ are coloured \btr by \lineref{algo:remove:line:btr_m}, and all machines of the new  $\crit(\OPT+(\nu - \uvec j))$ are still coloured blue in the resulting state by virtue that \autoref{invar:machine_colouring} was maintained for the input state.
	While recolouring a blue machine in \lineref{algo:remove:line:tryBTR1}, only the capacity of the recoloured machine is deducted from $\alpha$, maintaining $\alpha \geq \delta$ by maintaining \autoref{invar:lower_bound}.
	 If \lineref{algo:remove:line:alpha_minus_pmax} is executed, $\alpha$ is set to at least $\delta$ as well.
\end{proof}

	Using these bounds given to us by these two invariants, we can prove upper bounds for the $\kappa$-freeness of states after the computation of a new state. We first consider \Call{Insert}{}.
	
\begin{proposition}[
	, label = stmt:algo_insert_kappa_ineq
]
	Let $S = (\alpha, \mub, \nu)$ be a valid state and $\kappa$ maximal such that $S$ is $\kappa$-free, and $\bar S = (\bar \alpha, \mub[\bar], \nu + \uvec j)$ be $\bar \kappa$-free state returned by \Call{Insert}{$j,S$}.
	Let $\crit \defeq \crit(\OPT+(\nu + \uvec j))$ and $\bar \delta \defeq (1+\eps) \OPT-(\nu + \uvec j) s^\intercal (\mub - \mub_{\crit} \uvec {\crit})$.
	Then
	\[
		\bar \alpha > \alpha \land \bar \alpha > \bar \delta
		\implies
		\bar\kappa \leq  \lb + 2 \pmax - 1
	\enspace . \stopphier \]
\end{proposition}
\begin{proof}
	Consider \fullref{stmt:algo_insert_rephrased}; read the $3$ cases as statements of an algorithm.
	If the algorithm terminates after (i), either $\bar \alpha = \alpha$ or $\bar \alpha = (1+\eps) \OPT-(\nu + \uvec j) s^\intercal \mub = \bar \delta$, where the last equality follows from the fact that no machines of type $\crit$ are coloured blue in this case, implying $\mub_{\crit} = 0$, and the statement is always true as its LHS terminates to false.
	If the algorithm terminates after (ii), then $\bar{\kappa} < p_j$, as we added up to $2 p_j$ usable blue area, but immediately occupy $p_j$ of it by the job $j$.
	Now assume that it terminates after (iii).
	Then, by the inequality $\bar \alpha < (1+\eps) \lb + p_j$, the statement follows directly, as this case only occurs if $S$ was less than $p_j$-free before, $p_j$ space is immediately occupied by the new job $j$, and thus it must be less than $(\lb + \pmax + p_j)$-free after.
\end{proof}

We show a similar statement for the slack of $\alpha$ after using \Call{Remove}{}.

\begin{proposition}[
	, label = stmt:algo_remove_kappa_ineq
]
	Let $S$ be a valid state and $\kappa$ maximal such that $S$ is $\kappa$-free, $\bar S = (\bar \alpha, \mub[\bar], \bar \nu)$ be $\bar\kappa$-free state returned by \Call{Remove}{$j,S$}.
	Let $\crit \defeq \crit(\OPT+(\nu))$ and $\delta \defeq (1+\eps) \OPT-(\nu) s^\intercal (\mub - \mub_{\crit} \uvec {\crit})$.
Then
	\begin{enumerate}[(i)]
	\item \label{remove_case1}	\[
		\bar \alpha > \delta
		\implies
		\bar{\kappa} \leq \max\Sing{\kappa \ssep (\veps \lb + \pmax)}
		\enspace.
	\]
	\item \label{remove_case2}	\[
		\OPT+(\nu - \uvec j) < \OPT+(\nu)
		\implies
		\bar{\kappa} < \pmax
		\enspace.\stopphier
	\]
	\end{enumerate}

\end{proposition}
\begin{proof}
(\ref{remove_case1})
		If $\bar \alpha > \delta$, then $\OPT+(\nu) = \OPT+(\nu - \uvec j)$, as otherwise, \lineref{algo:remove:line:btr_m} and \lineref{algo:remove:line:alpha_after_btr} would have been executed and set $\alpha = (1+\eps) \OPT+(\nu-\uvec j) s^\intercal (\mub - \mub_{\crit} \uvec \crit)= \delta$, where the last equality follows from \autoref{stmt:estimate_opt}.
		If $\bar\kappa < (\veps \lb + \pmax)$, the claim directly holds.
		Otherwise, $\bar \kappa \geq (\veps \lb + \pmax)$.
		We backtrack which lines have been executed within \Call{Remove}{}:
		If $\bar S$ is $(\veps \lb + \pmax)$-free even after \Call{Remove}{}, either \lineref{algo:remove:line:tryBTR1} or \lineref{algo:remove:line:alpha_minus_pmax} have been executed during this call, as both lines can only reduce the freeness, and thus, the state before these lines was at least $(\veps \lb + \pmax)$-free as well, and one of the conditions before the respective line is then always true.
		Since $\bar \alpha > \delta$ by assumption, and $\delta = (1+\eps) \OPT-(\nu - \uvec j) s^\intercal \mub$ if $\crit = \bot$ or $\mub_{\crit} = 0$, each of both lines reduce $\alpha$, and thus freeness, by at least $\pmax$ if they are executed, so the state $S'$ before these lines is at least $\bar\kappa + \pmax$-free.
		Since $S'$ is derived from $S$ by simply removing some job, we can conclude that $S$ is at least $\bar\kappa$-free, which shows the claim.
		
		(\ref{remove_case2}) $\OPT+(\nu-\uvec j) < \OPT+(\nu)$ holds by assumption. If $\bar S$ was at least $\pmax$-free, reinserting $j$ would yield a state $S'$ with $\OPT+(\nu-\uvec j) = \OPT+(\nu)$ by \autoref{stmt:kappa_free_add}, a contradiction.
\end{proof}

Consequently, we combine the two previous statements to upper bound the slack of all states computed by our algorithm using only \Call{Insert}{} and \Call{Remove}{} to compute new states.

\begin{proposition}[
	, label = stmt:max_freedom_of_states
]
	Let $S = (\alpha, \mub, \nu)$ be a valid state, $\crit \defeq \crit(\OPT+(\nu))$ and $\delta \defeq (1+\eps) \OPT-(\nu) s^\intercal (\mub - \mub_{\crit} \uvec {\crit})$.
	
	Let $\bar S = (\bar \alpha, \mub[\bar], \bar \nu)$ be a state computed either by $\bar S = \Call{Insert}{j,S}$ or $\bar S = \Call{Remove}{j,S}$.
	Let $\tilde \crit \defeq \crit(\OPT+(\bar \nu))$ and $\bar \delta \defeq (1+\eps) \OPT-(\bar \nu) s^\intercal (\mub[\bar] - \mub[\bar]_{\tilde \crit} \uvec {\tilde \crit})$.
	Then
		\[
		(\alpha > \delta
		\implies
		S \text{ is at most } (\lb + 2 \pmax -1)\text{-free})\\
		\implies\\
		(\bar \alpha > \bar \delta
		\implies
		\bar S \text{ is at most } (\lb + 2 \pmax -1)\text{-free})
		\enspace .
	\]
\end{proposition}
\begin{proof}
In the following, we assume that the LHS of the implication $(\alpha > \delta
\implies
S \text{ is at most } (\lb + 2 \pmax -1)\text{-free})$ as well as the the LHS of the implication on the RHS $(\bar \alpha > \bar \delta)$ both evaluate to true, as otherwise, the implication always evaluates to true.
We make a case distinction regarding $\OPT+(\bar \nu)$ in relation to $\OPT+(\nu)$.
	\begin{enumerate}[]
\item \case [1] {$\OPT+(\bar \nu) > \OPT+(\nu)$}. 
		This can only happen if $\bar S = \Call{Insert}{j,S}$.
		Then $\bar \alpha > \alpha$, as, if $\bar \alpha$ is not increased, $\bar \alpha = \bar \delta$ by \lineref{alpha_new_delta} of $\Call{Insert}{}$ and the fact that all machines of $\tilde \crit$ are coloured red in this case, a contradiction to the assumption that $\bar \alpha > \bar \delta$.
		Thus, $\bar S$ is at most $(\lb + 2 \pmax -1)$-free by \autoref{stmt:algo_insert_kappa_ineq}.
		\item \case [2] {$\OPT+(\bar \nu) < \OPT+(\nu)$}. 
		This can only happen if $\bar S = \Call{Remove}{j,S}$.
		Then $\bar S$ is at most $(\lb + 2 \pmax -1)$-free by \autoref{stmt:algo_remove_kappa_ineq},(\ref{remove_case2}).
		\item \case [3] {$\OPT+(\bar \nu) = \OPT+(\nu)$}.
		\begin{enumerate}[]
		\item \underline{Case 3.1: $\bar S = \Call{Insert}{j, S}$}
		It holds that $\bar \alpha \geq \alpha$ by definition of $\Call{Insert}$.
		If $\bar \alpha = \alpha$, then the statement is true:
		Either $\bar \alpha = \alpha = \delta$, which makes the statement always true, or $\bar \alpha = \alpha > \delta$, implying $(\lb + 2 \pmax -1)$-freeness for $S$, and thus $\bar S$ is at most $(\lb + 2 \pmax -1)$-free since insertion a job without increasing $\alpha$ or $\OPT+$ can only decrease the freeness.
		
		Else if $\bar \alpha > \alpha$, the statement is also true, as $\alpha \geq \delta$ by \autoref{invar:alpha_geq_delta} and $\delta = \bar \delta$ when $\OPT+(\bar \nu) = \OPT+(\nu)$, and thus $\bar S$ is at most $(\lb + 2 \pmax -1)$-free by \autoref{stmt:algo_insert_kappa_ineq}.
		\item \underline{Case 3.2: $\bar S = \Call{Remove}{j, S}$}
		It holds that $\bar \alpha \leq \alpha$ by definition of $\Call{Remove}$.
		Also, note that $\delta = \bar \delta$ when $\OPT+(\bar \nu) = \OPT+(\nu)$. Thus, $\bar \alpha \geq \bar \delta = \delta$, where the first inequality follows from \autoref{invar:alpha_geq_delta}.
		If $\bar \alpha = \bar \delta$, the statement is always true.
		Else if $\alpha \geq \bar \alpha > \bar \delta = \delta$, $S$ is at most $(\lb + 2 \pmax -1)$-free by our assumptions, and \autoref{stmt:algo_remove_kappa_ineq},(\ref{remove_case1}) then shows that $\bar S$ is at most $(\lb + 2 \pmax -1)$-free as well, proving the statement.
		
		\end{enumerate}
\end{enumerate}	
	This proves the claim.
\end{proof}

Since $(\alpha > \delta \implies S \text{ is at most } (\lb + 2 \pmax -1)\text{-free})$ is true for our starting state ($\nu = 0$ and the state is $0$-free), and we use only \Call{Insert}{} and \Call{Remove}{} to compute new states, \autoref{stmt:max_freedom_of_states} holds for every step our algorithm.
This implies the following corollary.

\begin{corollary}
\label{cor:max_freedom_of_states}
For each state $S = (\alpha,\mub,\nu)$ computed by our algorithm
$$\alpha > \delta \implies S \text{ is at most } (\lb + 2 \pmax -1)\text{-free}$$
is true, with $\delta = (1+\eps) \OPT-(\nu) s^\intercal (\mub - \mub_{\crit} \uvec {\crit})$ and $\crit \defeq \crit(\OPT+(\nu))$.
\end{corollary}

We can use this bound on the $\kappa$-freeness of states to show the following claim on the colouring and $\alpha$ change caused by \lineref{algo:remove:line:btr_m} and \lineref{algo:remove:line:alpha_after_btr}, respectively.

\begin{proposition}[
	, label = stmt:bounded_recolours_alpha
]
For any execution of \Call{Remove}{$j,S$} on a state $S$ produced by our algorithm and a job $j \in \nu$ the following holds:
\begin{enumerate}[(i)]
	\item On any execution of \lineref{algo:remove:line:btr_m}, at most $\frac{(\lb + 2 \pmax )}{\pmax}$ machines are recoloured \btr. \label{max_recolours_btr:m}
	\item On any execution of \lineref{algo:remove:line:alpha_after_btr}, $\alpha$ is reduced by at most $\lb + 3 \pmax$. \label{max_alpha_change}
\end{enumerate}
\end{proposition}
\begin{proof}
Let $S = (\alpha, \mub, \nu)$ and $\bar S = (\bar \alpha, \mub[\bar], \nu) = \Call{Remove}{j,S}$.
Also, let $\crit = \crit(\OPT+(\nu))$ and $\delta \defeq (1+\eps) \OPT-(\nu) s^\intercal (\mub - \mub_{\crit} \uvec {\crit})$.
\lineref{algo:remove:line:btr_m} and  \lineref{algo:remove:line:alpha_after_btr} are only executed during \Call{Remove}{j,S} if $\OPT+(\nu-\uvec j) < \OPT+(\nu)$.
Note that, in this case, $\delta =  (1+\eps) \OPT+(\nu - \uvec j) s^\intercal (\mub - \mub_{\crit} \uvec {\crit})$ by \autoref{stmt:makespan_change}.
Thus, $\bar{\alpha} = \delta$ after execution of \lineref{algo:remove:line:alpha_after_btr}.

\begin{enumerate}[]
\item \case {$\alpha > \delta$}
Then $S$ is at most $(\lb + 2 \pmax - 1)$-free by \autoref{cor:max_freedom_of_states}.
Reinserting $j$ with \Call{Insert}{$\bar S,j$} after its removal yields a valid state $S' = (\alpha', \mub{'}, \nu)$ with $\OPT+(\nu) > \OPT+(\nu - \uvec j)$.
For $S'$, $\mub{'}_{\crit} \leq 1$ and $\alpha' \leq \delta + p_j$, as, for $\OPT+(\nu) > \OPT+(\nu - \uvec j)$, \Call{Insert}{$\bar S,j$} sets $\alpha$ first to $\delta$, and then increases $\alpha$ by at most $p_j$ after.

(\ref{max_recolours_btr:m}).
Assume that \lineref{algo:remove:line:btr_m} is executed on state $S$ and job $j$ and more than $\frac{(\lb + 2 \pmax)}{\pmax}$ machines are recoloured \btr.
For this to occur, $\mub_{\crit} \geq \frac{(\lb + 2 \pmax + 1)}{\pmax}$ must hold at this point.

Then, since $S'$ is at least $0$-free due to its validity, $S$ is at least $(\lb + 2 \pmax)$-free, as each of the $\frac{(\lb + 2 \pmax + 1)}{\pmax}$ machines that are coloured blue in $S$ but not in $S'$ add at least $\pmax$ to $\alpha$, due to the maintaining of \autoref{invar:lower_bound}.
This is a contradiction to $S$ is at most $(\lb + 2 \pmax - 1)$-free if $\alpha > \delta$ by \autoref{cor:max_freedom_of_states}.
\end{enumerate}

(\ref{max_alpha_change}).
If $\alpha > \bar \alpha + \lb + 3 \pmax = \delta + \lb + 3 \pmax$, then $S$ is at least $(\lb + 2 \pmax)$-free for $\mub_{\crit} = 0$, as $\alpha' \leq \delta + \pmax$ for the state $S'$ with the same job set $\nu$.
This then holds for all $\mub_{\crit} \geq 0$, as any additional blue machine increases the $\kappa$-freeness by at least $\pmax$ while maintaining \autoref{invar:lower_bound}.
This again is a contradiction to $S$ is at most $(\lb + 2 \pmax - 1)$-free if $\alpha > \delta$, and thus, $\alpha \leq \delta + \lb + 3 \pmax$, and the statement follows as $\bar \alpha = \delta$.

\item \case{$\alpha = \delta$}
(\ref{max_recolours_btr:m}) and  (\ref{max_alpha_change}) are both true in this case, as $\alpha = \bar \alpha$, and recolouring one machine \btr with \Call{TryBTR}{} reduces $\alpha$ by at least $(1+\eps) \lb$, implying that no machine can be recoloured the case of $\alpha = \bar \alpha$.

\item \case{$\alpha < \delta$} can never occur by \autoref{invar:alpha_geq_delta}.
\end{proof}

We can now finally use this result to show that \Call{Remove}{} has indeed bounded migration.

\begin{lemma}[
	, name  = \textsc{Remove} Has Bounded Migration
	, label = stmt:algo_remove_migration_bounded
]
	Let $S = (\alpha, \mub, \nu)$ be a valid state and $j \in \supp(\nu)$.
	Then for $\bar S \defeq (\bar \alpha, \mub[\bar], \bar \nu) = \Call{Remove}{j,S}$, $1 \leq \abs{\Delta{\alpha}} + \norm{\Delta\nu}_1 + \norm{\Delta\mub}_1 \leq \poly(\epsinv,\pmax)$.
\end{lemma}
\begin{proof}
	$\norm{\Delta\nu}_1 = 1$ by definition of the problem.
	Due to \autoref{stmt:bounded_recolours_alpha}~(\ref{max_recolours_btr:m}), and the fact that in \lineref{algo:remove:line:tryBTR1}, the only other \btr recolouring, only $1$ machine is recoloured, $\norm{\Delta\mub}_1 \leq \frac{(\lb + 2 \pmax )}{\pmax} + 1 \leq \poly(\epsinv,\pmax)$.
	Thus, $\alpha$ is reduced at most by $\poly(\epsinv,\pmax) \cdot (\lb + \pmax) \leq \poly(\epsinv,\pmax)$ by recolouring.

	The only other times when $\alpha$ might be changed are \lineref{algo:remove:line:alpha_minus_pmax}, where $\alpha$ is reduced by at most $\pmax \leq \poly(\epsinv,\pmax)$, and \lineref{algo:remove:line:alpha_after_btr}. By \autoref{stmt:bounded_recolours_alpha}~(\ref{max_alpha_change}), the execution of \lineref{algo:remove:line:alpha_after_btr} reduces $\alpha$ by at most $\lb + 3 \pmax \leq \poly(\epsinv,\pmax)$.
	This shows the claim.
\end{proof}
As immediate consequence we get that \Call{Remove}{} always computes a valid state $\bar{S} = (\bar \alpha, \mub[\bar], \allowbreak \bar \nu)$ such that $\abs{\Delta \alpha} + \norm{\Delta\mub}_1 + \norm{\Delta\nu}_1 \leq \poly(\epsinv,\pmax)$.
As the same result has already been shown for \Call{Insert}{}, we can then achieve the desired migration bound of $\paramsens$ and run time of $\paramrt + |I|$ for the dynamic setting, using \Call{Insert}{} to add, and \Call{Remove}{} to remove jobs from step $q$ to $q+1$, via \fullref{stmt:sensitivity},~(\ref{stmt:sensitivity:param}) and \fullref{stmt:ipdyn_solve},~(\ref{stmt:ipdyn_solve:param}), which can be found in \appendixref{sec:ipsolve}.
We treat the initial job set $\mc J_0$ as adding $\abs{\mc J_0}$ jobs to the instance in arbitrary order, beginning with an empty set.
We thus start from a valid state $S = (0,\zero,\zero)$, proving \autoref{stmt:main_theorem},~(\ref{stmt:param_dynamic_algorithm}).

 	\section{An EPAS for dynamic $\QCmin$}
\label{sec:combine_online_cmin}

In this section, we give the full technical details necessary to proof \autoref{stmt:main_theorem},~(\ref{stmt:param_qcmin_algorithm}).

\subsection{Some Preliminaries}

Before we can formulate our procedure, we need to redefine some previously used definitions and statements to fit the new objective function.
We redefine $\realOPT$ to map to the maximum minimum completion time over all schedules.
The goal of dynamic $\QCmin$ is to find a schedule with maximum minimal completion time for every step $q$ over the duration, where either $\at {q+1} \nu = \at q \nu + \uvec j$, for some $j \in \iv d$, or $\at {q+1} \nu = \at q \nu - \uvec j$, for some $j \in \iv d$ and $\at q \nu_j > 0$.
We slightly increase our threshold $\lb$ to fit our new objective.
\[  \ell \defeq (1 + \eps)^{1 + \ceil {\log_{1+\eps} 3 \epsinv \pmax}} \enspace . \]
Note that
\[
	\eps (1 + \eps)^\inv \ell \geq \eps \ell \geq 3 \pmax \enspace , \qquad \tand \lb \leq \poly(\epsinv, \pmax) \enspace .
\]
We use the following ILP, a version of \ipdyn with modified dual knapsack constraint and modified objective $\lexobj'$.
The objective function $\lexobj'$ here encodes \enquote{Maximizing the minimum completion time} on the given set of configurations -- this can simply be done by modifying the order $\sqsubseteq$ on the elements as defined in \appendixref{sec:lexmin} to order primarily by ascending, instead of descending, completion time.

\vspace{-2em}
\begin{multicols}{2}
\begin{lpformulation}[{\mathrlap{\ipdyncmin}}]\lpeq*{}{}
	\lplabel{ip:dyn_cmin}
\lpobj*{min}{\lexobj'(x)}
	\lpeq[ip:dyn:constr:blue_area:cmin]{
		p^\intercal \super \tblue \nu
		\geq \alpha
	}{}
	\lpeq[ip:dyn:constr:red_count:cmin]{\sum_{c \in \mc C(\lb)}  x_{t,c} = \mu_t - \super \tblue \mu_t}{t \in \iv \tau}
	\lpeq[ip:dyn:constr:blue_red:cmin]{\sum_{\substack{t \in \iv \tau \\ c \in \mc C(\lb)}}  x_{t,c} \cdot c  + \super \tblue \nu = \nu}{}
\end{lpformulation}
\begin{lpformulation}
	\lpeq*{\text{\normalfont \underline{with variables}}}{}
	\lpeq*{\var x_{t,c} \in \N_0}{t \in \iv \tau, c \in \mc C(\lb)}
	\lpeq*{\var \super \tblue \nu \in \N_0^d}{}
	\lpeq*{\text{\normalfont \underline{with parameters}}}{}
	\lpeq*{\param \nu \in \N_0^d}{}
	\lpeq*{\param \super \tblue \mu \in \N_0^\tau}{}
	\lpeq*{\param \alpha \in \Q_{\geq 0}}{}
	\stopphier
\end{lpformulation}\end{multicols}
Since our ILP again is parameterized by some value $\alpha$, that now represents the part of the overall area on blue machines that \emph{must} be used by the ILP, our next step is to formalize this interaction, \ie ensure that we restrict ourselves to values of $\alpha$ that do not subceed the overall area of the blue machines. Here, we require the area of the blue machines to actually be a factor of $(1+\veps)$ higher than the current guess on the minimum completion time $T$.
\begin{invariant}[
	, name = Compatible Colouring Invariant
	, label = invar:comp_colouring_cmin
]\[ 
	\alpha \geq (1 + \eps) T s^\intercal \super \tblue \mu \enspace . \stopphier \]
\end{invariant}

For fixed $T$ and $\mub$, we will also call $(1 + \eps) T s^\intercal \super \tblue \mu$ the \enquote{required blue area (for $T$ and $\mub$)}.
Given a (approximate) minimum completion time $T$, we require the red schedule to not subceed $T$, and the blue schedule not to subceed $(1 + \eps) T$.
This, again, will later allow us to bound the number of machine recolourings by some kind of witness argument.
We use the definition of \enquote{states} as before, but redefine validity for a state
\begin{definition}[
	, name = {$T$-validity for $\QCmin$}
	, label = defi:state_valid_cmin
]
	The (unique) optimal solution $(x, \nub)$ of \ipdyncmin for these parameters is called \emph{the associated solution}.
	The \emph{(ILP) minimum completion time of a state} is the (ILP) minimum completion time  of the associated solution, \ie $\Cmin(x)$.
	If the ILP is infeasible, we define the makespan to be $\infty$.
	A state $S$ is \emph{$T$-valid} iff \autoref{invar:machine_colouring} and \autoref{invar:comp_colouring_cmin} both hold for $T$; and the associated solution of $S$ has (ILP) minimum completion time at least $T$.
\end{definition}

Note that if there is a $T$-valid state, there is also $T'$-valid state for any $T' \leq T$.
We again define the Optimal ILP Makespan as before, but redefine a state as valid if it is $\OPT-$-valid.
\begin{definition}[
	, name = Optimal ILP Minimum Completion Time
	, label = defi:opt_cmin
]
	Given a job vector $\nu \in \N_0^d$, $\OPT \in \Q_{\geq 0}$ is the largest number such that there is a $\OPT$-valid state $S$.
	Moreover, we define $\OPT-$ to be the largest power of $(1 + \eps)$ strictly smaller than $\OPT$; and $\OPT+ \defeq (1 + \eps) \OPT-$.
\end{definition}
\begin{definition}[
	, name = Valid States
	, label = defi:valid_cmin
]
	A state $S$ is valid if it is $\OPT-$-valid.
\end{definition}
Note that $(1+\eps)^{-1} \geq (1-\eps)$, for all $\eps > 0$. By the definition of \autoref{invar:comp_colouring_cmin}, $\OPT \geq \OPT*$ does not hold.
Instead, $\OPT \geq (1-\eps) \OPT*$: By definition of $\OPT*$ and \autoref{invar:comp_colouring_cmin}, there always exists a $(1+\eps)^{-1} \OPT*$-valid state. Thus, $\OPT \geq (1+\eps)^{-1} \OPT*$ by definition of $\OPT$.
As computing the assignment of blue jobs to blue machines introduced an additional multiplicative error of at most $(1-\eps) \OPT$, and $\OPT- \geq (1+\eps)^{-1} \OPT$, computing a solution to the ILP for any valid state always returns a at most $(1-\eps)^3 \in (1-\O(\eps))$-approximate solution.

We redefine the freeness of states: Here, it measures how much space can be additionally allotted to $\alpha$, without changing the minimum completion time of the associated solution.

\begin{definition}[
	, name  = $\kappa$-free States
	, label = defi:kappa_free_cmin
]
A valid state $(\alpha, \mub, \nu)$ is \emph{$\kappa$-free} if $(\alpha + \kappa, \mub, \nu)$ is valid.
\end{definition}
Machines of $\thecrit$ are defined as before, except that the shorthand $\crit$ now stands for $\crit = \thecrit(\OPT-)$.
Similarly, machines of this type are the ones that might be necessary to be recoloured during a sequence of steps where $\OPT-$ does not change.
This again follows from the fact that, as soon as a machine of capacity $\geq \lb$ exists, the change on the minimum completion time is bounded.
To see this, we need to show that $\OPT$ changes by a factor of at most $(1+\eps)$, which implies similar bounds for $\OPT+$ and $\OPT-$.
\begin{proposition}[
	, name = Bounding Minimum Completion Time Changes
	, label = stmt:makespan_change_cmin
]
Consider $\nu \in \N_0^d$ and $j \in \iv d$.
	Let $f(\cdot) \in \Sing{ \OPT(\cdot) \Ssep \OPT-(\cdot) \Ssep \OPT+(\cdot) }$.
	If $\smax\OPT-(\nu + \uvec j) \geq \lb$, then 
	\hfill$
		\frac{f(\nu + \uvec j)}{f(\nu)} \leq (1 + \eps)
	\enspace ,$ \hfill \, \\*
	and if $j \in \supp(\nu)$ and $\smax\OPT-(\nu) \geq \lb$, 
	\hfill$
		\frac{f(\nu - \uvec j)}{f(\nu)} \geq (1 + \eps)^\inv \geq (1 - \eps)
		\enspace .
		\stopphier
	$\hfill
\end{proposition}
\begin{proof}
As \autoref{stmt:estimate_opt} also holds for $\QCmin$, we can prove the statement by simply exchanging the assumptions $\smax\OPT+(\nu + \uvec j) \geq \lb$ and $\smax\OPT+(\nu) \geq \lb$ by $\smax\OPT-(\nu + \uvec j) \geq \lb$ and $\smax\OPT-(\nu) \geq \lb$, respectively, in the proof of \autoref{stmt:makespan_change}.
\end{proof}
This then implies that recolouring all machines of $\thecrit$ is enough to maintain \autoref{invar:machine_colouring} and feasibility of the solution.
We thus again adhere to \autoref{invar:recolouring} during the design of our algorithm.
We are now ready to describe how we want to modify a state each time we add or remove a job, such that the new state has distance bounded in $\poly(\epsinv,\pmax)$.

\subsection{Handling Removal}
\label{sec:removal_cmin}

For $\QCmin$, we begin with handling the removal of jobs. The change of our parameters $\mub$ and $\alpha$ depends on how much we have to increase the $\kappa$-freeness of the current state to offset the removal of a job. We again make a general observation about solutions with enough $\kappa$-freeness.

\begin{lemma}[
	, name = Removing a job from a $\kappa$-free state
	, label = {stmt:kappa_free_rem_cmin}
]
	Let $S = (\alpha, \mub, \nu)$ be valid and $\kappa$-free.
	Then for any $\nu^\Delta \in \N_0^d$ with $p^\intercal \nu^\Delta \leq \kappa$, $(\alpha, \mub, \nu - \nu^\Delta)$ is $\OPT-(\nu)$-valid.
\end{lemma}
\begin{proof}
	Let $S' \defeq (\alpha + p^\intercal \nu^\Delta, \mub, \nu)$.
	Then $S'$ is valid by \fullref{defi:kappa_free_cmin} and the fact that $p^\intercal \nu^\Delta \leq \kappa$.
	Let $i$ be the machine where $\nu^\Delta$ is to be removed.
	
	We first assume $i$ is a red machine.
	We now remove $\nu^\Delta$ from $i$, decreasing $i$'s completion time by $p^\intercal \nu^\Delta$.
	We now decrease the required blue area from $\alpha + p^\intercal \nu^\Delta$ back to $\alpha$.
	By this, we gain jobs of total processing time at least $\alpha + p^\intercal \nu^\Delta$ to be scheduled on the red machines.
	We schedule these jobs on $i$.
	Thus, $i$'s completion time is at least the same as before, \ie at least $\OPT-$.
	Therefore, since the completion times of all other machines remain at least $\OPT-$, there exists a solution with minimum completion time at least $\OPT-$.
	Since the associated solution of this new state maximizes the minimum completion time, it has minimum completion time at least $\OPT-$ as well.
	
	Now, assume $i$ is a blue machine.
	We can simply remove $\nu^\Delta$ from $S'$ and decrease the required blue area from $\alpha + p^\intercal \nu^\Delta$ back to $\alpha$ to get a valid state.
\end{proof}

So if a state is at least $p_j$-free, we can directly remove job $j$ from the ILP without changing any of the state parameters, and still stay valid.
If this is not possible, we may have increase the freeness somehow.
Luckily, when colouring a machine of $\crit(\OPT-(\nu))$ \btr, we gain at least $\pmax$ freeness due to the definition of $\lb$ and the additional error allowed on blue machines by \autoref{invar:comp_colouring_cmin}, leading to the following lemma.

\begin{lemma}[
	,name=Recolouring a Blue Machine
	,label=stmt:btr_pmax_free_cmin
]
	If $S \defeq (\alpha, \mub, \nu)$ is valid and there is a blue machine of type $t \in \iv \tau$ with $s_t \OPT- = \lb$, then $S' \defeq (\alpha - ((1+\eps) s_t \OPT- - \pmax), \mub - \uvec t, \nu)$ is valid and $\pmax$-free.
	Moreover, $\Cmin(S') \geq \Cmin(S)$.
\end{lemma}
\begin{proof}
	Let $i$ be machine to be recoloured and $t$ its type.
	Consider an assignment $\nu^\Delta$ of blue jobs $i$ such that each blue machine of type $t \in \iv \tau$ is assigned jobs of total load at least $(1+\veps) s_t \OPT-(\nu) - \pmax$.
	Such an assignment exists by \fullref{stmt:greedy}, as $S$ is valid and thus \autoref{invar:comp_colouring_cmin} holds for $T \gets \OPT-(\nu)$.
Then $(\alpha - ((1+\eps) s_t \OPT- - \pmax), \mub - \uvec t, \nu)$ has a minimum completion time no smaller than $(\alpha, \mub, \nu)$, as we can do the following modifications to the solution $(x, \nub)$ associated to $(\alpha, \mub, \nu)$:
	\begin{enumerate}
		\item Set $x_{t,\nu^\Delta} \gets x_{t,\nu^\Delta} + 1$, $\nu \gets \nu + \nu^\Delta$ and $\mu_t \gets \mu_t + 1$; then the modified solution is feasible and its minimum completion time did not decrease (as $\eps \lb \geq 3 \pmax$).
		\item Set $\alpha \gets \alpha - ((1+\eps) s_t \OPT-(\nu) - \pmax)$, $\nu \gets \nu - \nu^\Delta$, and $\nub \gets \nub - \nu^\Delta$; then the job set is the same as in the beginning, the minimum completion time did not decrease and the dual knapsack constraint is not violated as $p^\intercal \nub \geq \alpha$, so $p^\intercal (\nub - \nu^\Delta) \geq \alpha - p^\intercal \nu^\Delta \geq \alpha - ((1+\eps) s_t \OPT-(\nu) - \pmax)$.
		\item $\mub_t \gets \mub_t - 1$ and $\mu_t \gets \mu_t - 1$, which is a change of parameters, but does not change the right hand side of the ILP where only $\mu - \mub$ appears.
\end{enumerate}
	The resulting state is at least $\pmax$-free, since the newly red machine $i$ has completion time $(1+\eps) s_i \OPT- - \pmax$ (by \autoref{stmt:greedy}), even though its minimum required completion time for validity as a red machine is only $s_i \OPT-$.
	Thus, increasing $\alpha$ by $\pmax$ leads to a valid state, since $\eps s_i \OPT- - \pmax = \eps \lb - \pmax \geq 3 \pmax - \pmax = 2 \pmax$ and therefore, there is a combination of jobs currently placed on $i$ of total processing time at least $\pmax$ and at most $2 \pmax$, that can be placed on blue machines without violating validity, implying $\pmax$-freeness of the current sate.
	This implies $\Cmin(S') \geq \Cmin(S)$ as well.
\end{proof}

By \autoref{invar:recolouring} we may only recolour machines of critical type \btr to gain $\pmax$-freeness. What if all machines of this type have already been recoloured? We then can make the following observation about the validity and minimum completion time of such solutions.

\begin{lemma}[
	, label = stmt:opt_min_cmin
]
	Let $\nu$ be a job set and for all $t \in \iv \tau$, $\mub_t = \mu_t$ if $s_t \OPT- > \lb$ and $0$ otherwise.
	Then $\enpar[\Big]{\ceil*{(1 + \eps) \OPT-(\nu)  s^\intercal \mub}, \allowbreak\mub,\allowbreak \nu}$ is valid and its minimum completion time is \emph{at least} $\OPT$.
\end{lemma}
\begin{proof}
	At first glance it might seems like a contradiction that the minimum completion time could be more than $\OPT$.
	But this actually can happen, as \fullref{defi:opt_cmin} differs from \fullref{defi:valid_cmin} in terms of the parametrization of the invariants.
	In \autoref{defi:opt_cmin}, they have to hold for $\OPT$, while for a valid solution they are only required to hold for $\OPT-$.
	And since our definition of makespan of a state only considers $\Cmin(x)$ of the associated solution $(x, \nub)$, we can indeed have a minimum completion time above $\OPT$.
	Let $\bar S \defeq (\bar\alpha, \mub[\bar], \nu)$ be an $\OPT$-valid state, i.e. such that the invariants \ref{invar:machine_colouring} and \ref{invar:comp_colouring_cmin} both hold for $T \gets \OPT$, and the associated solution has minimum completion time at least $\OPT$.
	Such a state always exists by definition of $\OPT$.
	Then
	\begin{multline*}
		\bar\alpha
		\geq (1 + \eps) \OPT(\nu)  s^\intercal \mub[\bar] \geq (1 + \eps) \OPT-(\nu)  s^\intercal \mub[\bar]
		\\ = (1 + \eps) \OPT-(\nu)  s^\intercal \mub + (1 + \eps) \OPT-(\nu)  s^\intercal (\mub[\bar]-\mub)
		\enspace .
	\end{multline*}
	Note that $\mub \leq \mub[\bar]$, as otherwise, by assumption on the form of $\mub$, under $\mub[\bar]$ either less than $\mu_t$ machines are coloured blue for some $t \in \iv \tau$, or under $\mu_t$ a machine of some type $t \in \iv \tau$ with $s_t \OPT- < \lb$ is coloured blue, which would violate \autoref{invar:comp_colouring}.
	So in particular, by subtracting $(1 + \eps) \OPT-(\nu)  s^\intercal (\mub-\mub[\bar])$ on each side of the inequality and rounding up, we have 
	\[
		\alpha' \defeq \ceil*{\bar\alpha - (1 + \eps) \OPT-(\nu)  s^\intercal (\mub-\mub[\bar]) } \geq \ceil*{(1 + \eps) \OPT-(\nu)  s^\intercal \mub} \enspace .
	\]
	By inductively applying \fullref{stmt:btr_pmax_free_cmin} it follows that $\Cmin(\alpha', \mub, \nu) \allowbreak \geq \OPT$.
	Thus, as $\alpha' \geq \floor*{(1 + \eps) \OPT+(\nu)  s^\intercal \mub}$, also $\Cmin(\ceil*{(1 + \eps) \OPT-(\nu)  s^\intercal \mub} \skomma \mub \skomma \nu) \geq \OPT$.
\end{proof}
We can use this statement now to show the validity of states where a job is removed, and $\OPT-$ doesn't change.
\begin{lemma}[
	, name  = Existence of a Blue Machine
	, label = stmt:exists_blue_cmin
]
	Consider $j \in \iv d$.
	If $(\alpha, \mub, \nu)$ is valid and $\OPT-(\nu) = \OPT-(\nu - \uvec j)$, then either $(\alpha - \kappa, \mub, \nu - \uvec j)$ is valid for some $\kappa \in \iv {p_j}_0$, or there is a blue machine of capacity $\lb$. 
\end{lemma}
\begin{proof}
	Assume that $(\alpha - \kappa, \mub, \nu)$ is invalid for all $\kappa \in \iv {p_j}_0$.
	Then for some $\kappa \in \iv {p_j-1}_0$ we have $\alpha - \kappa = \ceil*{(1 + \eps) \OPT-(\nu)  s^\intercal \mub}$, as otherwise \fullref{defi:kappa_free_cmin} together with \fullref{stmt:kappa_free_rem_cmin} lead to a contradiction for $\alpha \gets \alpha - p_j$.
	If there was no blue machine of capacity $\lb$, we would have coloured all machines red that we are allowed to under \autoref{invar:machine_colouring}, and with $\alpha \gets \alpha - \kappa$ we would lowered the area as much as is allowed under \autoref{invar:comp_colouring_cmin}.
	This is a contradiction to \autoref{stmt:opt_min_cmin}. 
\end{proof}

To now prove correctness and bounded distance, we need the following proposition.
\begin{proposition}[
	, name =  Decreasing $\alpha$ for valid states
	, label = stmt:alpha_decrease_valid_cmin
]
	If $S = (\alpha, \mub, \nu)$ is valid, then $S' = (\alpha', \mub, \nu)$ is valid for any $\alpha' \in \interval[scaled] {(1 + \eps) \OPT-(\nu)  s^\intercal \mub} \alpha$.
\end{proposition}
\begin{proof}
	Let $(x, \nub)$ be the solution associated to $S$.
	Clearly, $(x, \nub)$ is a feasible solution to $S'$.
	Moreover, $\Cmin(x) \geq \OPT-$ as $S$ is valid, and since $\OPT-$ is the same for $S'$ (as it only depends on $\nu$), $S'$ is valid.
\end{proof}

We are now ready to describe our algorithm for removing a job for a state $S$, such that the new job set is valid to the returned state, and the distance from $S$ to the returned state is bounded.

\begin{algorithm}[h]
\caption{Removing a job $j$, given the current state $S$}
\label{algo:remove_cmin}
\begin{algorithmic}[1]
	\Function{Remove'}{$j$, $S$}
	\State $S$ is of the form $(\alpha, \mub, \nu)$
	\If{$\OPT-(\nu - \uvec j) = \OPT-(\nu)$}
		\If{$S$ is not $p_j$-free}
			\State$\alpha \gets \max \Sing{(1 + \eps) \OPT-(\nu - \uvec j)  s^\intercal \mub \ssep \alpha - p_j}$ \label{algo:remove:line:alpha_gets_max_cmin}
			\If{$S$ is (still) not $p_j$-free}
					\State $t \gets \crit(\OPT+(\nu))$ \Comment{Note that $t \neq \bot$}
					\State $\mub \gets \mub - \uvec t$ \Comment{Note that $\mub_t \neq \zero$}
					\State $\alpha \gets \alpha - (1+\eps) s_t \OPT-(\nu)$
			\EndIf
		\EndIf
	\Else
		\If{$(\alpha, \mub, \nu - \uvec j)$ is valid}
			\State$\alpha \gets (1 + \eps) \OPT+(\nu - \uvec j)  s^\intercal \mub$ \label{alpha_new_delta_cmin} \Comment{Only needed for \Call{Insert'}{}}
		\Else
			\State$\alpha \gets \alpha - p_j$ \label{alpha_second_pj_cmin}
		\EndIf
	\EndIf
	\State \Return $(\alpha, \mub, \nu - \uvec j)$ \EndFunction
\end{algorithmic}
\end{algorithm}

It follows directly directly from the definition of the algorithm that \Call{Insert}{} does not violate \autoref{invar:recolouring}.

Also here, we rephrase the algorithm as a lemma, and directly show that it always computes a valid state, and that the resulting state has parameters whose difference to the parameters of the input state is bounded in $\poly(\pmax) \leq \poly(\epsinv,\pmax)$.
This will then significantly simplify the proofs of further properties.
\begin{lemma}[
	, name  = \textsc{Remove'} Rephrased
	, label = stmt:algo_remove_rephrased_cmin
]
	Given a valid state $S = (\alpha, \mub, \nu)$ and $j \in \iv d$, the call \Call{Remove'}{$j,S$}
	\begin{enumerate}[(i)]
\item if $(\alpha, \mub, \nu - \uvec j)$ is valid, returns $(\alpha, \mub, \nu - \uvec j)$ if $\OPT-(\nu) = \OPT-(\nu - \uvec j)$, and $((1 + \eps) \OPT+(\nu - \uvec j)  s^\intercal \mub), \mub, \nu - \uvec j)$ else, where $(1 + \eps) \OPT+(\nu - \uvec j)  s^\intercal \mub \geq \alpha - p_j$; otherwise
		\item returns a valid state $(\bar\alpha, \mub, \nu - \uvec j)$ with $\bar\alpha \in \rinterval {\alpha - 2 p_j} {\alpha}$; or otherwise
		\item correctly concludes that $(\bar\alpha, \mub, \nu - \uvec j)$ is invalid for all $\bar\alpha$, and returns a valid state of the form $(\bar\alpha, \mub - \uvec {\crit(\OPT+(\nu))}, \nu - \uvec j)$ where $\bar \alpha > \alpha - (1+\eps) \lb - p_j$.
		\stopphier
	\end{enumerate}
\end{lemma}
\begin{proof}
	In the following, we can assume that there always exists a machine $i$ with $s_i = \smax$, and $\smax \OPT-(\nu) \geq \lb$, as otherwise, it holds for both the input and return state that $\alpha = 0$, $\mub = \zero$ by \autoref{invar:machine_colouring}, and thus, (i) is always fulfilled.
	\begin{enumerate}[]
		\item \underline{Case 1: $(\alpha, \mub, \nu - \uvec j)$ is valid.}
		\begin{enumerate}[]
			\item \underline{Case 1.1: $\OPT-(\nu - \uvec j) = \OPT-(\nu)$}.
			Then the if-branch of the first if-statement is entered and neither $\alpha$ nor $\mub$ are changed.
Since the algorithm then returns $(\alpha, \mub, \nu - \uvec j)$, this shows the claim, as (i) is satisfied.
			\item \underline{Case 1.2: $\OPT-(\nu - \uvec j) < \OPT-(\nu)$.}
			Then the else-branch of the first if statement is entered.
			Within this branch, the first if-branch of the next if-statement is entered due to the assumption of Case 1.
			Thus, the algorithm eventually returns
			$$\enpar*{(1 + \eps) \OPT+(\nu - \uvec j)  s^\intercal \mub, \mub, \nu - \uvec j} \enspace .$$
By \autoref{stmt:makespan_change_cmin} and using the fact that $\OPT+$ and $\OPT-$ are restricted to integer powers of $(1+\eps)$,
			\[
				(1 + \eps) \OPT+(\nu - \uvec j)  s^\intercal \mub = (1 + \eps) \OPT-(\nu)  s^\intercal \mub \leq \alpha \enspace ,
			\]
			where the latter follows from the fact that the initial state was valid and thus satisfied \autoref{invar:comp_colouring_cmin}.
Now, if $(1 + \eps) \OPT+(\nu - \uvec j)  s^\intercal \mub = (1 + \eps) \OPT-(\nu)  s^\intercal \mub < \alpha - p_j$, then 
			$((1 + \eps) \OPT-(\nu)  s^\intercal \mub, \mub, \nu - \uvec j)$ would be a valid state for $\OPT-(\nu)$ by \autoref{stmt:kappa_free_rem_cmin}, and thus $\OPT-(\nu) = \OPT-(\nu - \uvec j)$, a contradiction to the assumption of Case 1.2.
			Thus, $(1 + \eps) \OPT+(\nu - \uvec j)  s^\intercal \mub \geq \alpha - p_j$.
			Since $\mub$ remains unchanged, this shows the claim, as (i) is satisfied.
		\end{enumerate}
		
		\item \underline{Case 2: $(\alpha, \mub, \nu - \uvec j)$ is invalid and $\OPT+(\nu + \uvec j) \neq \OPT+(\nu)$.}
		Then only \lineref{alpha_second_pj_cmin} is executed. By definition of this line, $\alpha$ is decreased by at most $p_j$.
		The returned state is valid: By the assumption, we know that $\OPT-(\nu - \uvec j) < \OPT-(\nu)$ and that there exists a machine $i$ with speed $\smax$ s.t. $\smax \OPT(\nu) \geq \lb$.
		This machine $i$'s capacity is then decreased by at least $3 \pmax$, as here, $\OPT-(\nu - \uvec j) = (1+\veps) \OPT-(\nu)$ by definition, and therefore, $i$'s capacity is decreased by at least $(1+\veps)^\inv \veps \lb \geq 3 \pmax$, since $i$ had capacity at least $(1+\veps)^\inv \lb$ before by \autoref{stmt:makespan_change_cmin}.
		
		Thus, if $i$ is coloured blue, \autoref{invar:comp_colouring_cmin} is maintained.
		If $i$ is coloured red instead, $j$ can be removed without changing $\alpha$ at all, and $(\alpha, \mub, \nu - \uvec j)$ would be valid, a contradiction to the assumption of Case 2.
		It follows that (ii) is satisfied.

		\item \underline{Case 3: $(\alpha, \mub, \nu - \uvec j)$ is invalid and $\OPT-(\nu - \uvec j) = \OPT-(\nu)$.}
		Then the second if-branch is entered because of \fullref{stmt:kappa_free_rem_cmin}.

		\begin{enumerate}[]
			\item \underline{Case 3.1: there is a valid state of the form $(\tilde\alpha, \mub, \nu - \uvec j)$.}
			Let $\alpha'$ be the value of $\alpha$ after \lineref{algo:remove:line:alpha_gets_max_cmin}.
			We prove that (ii) will always be satisfied.
			\begin{enumerate}[]
				\item \underline{Case 3.1.1: $\alpha' = \alpha - p_j$.}
				Then $(\alpha', \mub, \nu)$ is $p_j$-free by \fullref{defi:kappa_free_cmin} and hence $(\alpha', \mub, \nu)$ is valid by \fullref{stmt:kappa_free_rem_cmin}.
				Moreover, $\bar\alpha \defeq \alpha' = \alpha - p_j$ satisfies the condition of (ii) on $\bar\alpha$.
				\item \underline{Case 3.1.2: $\alpha' = (1 + \eps) \OPT-(\nu - \uvec j)  s^\intercal \mub > \alpha + p_j$}.
				Then, with $\alpha'$, we reduce $\alpha$ as much as is allowed under \autoref{invar:comp_colouring_cmin}.
				By our case (Case 3.1), we know that there is a valid state of the form $(\tilde\alpha, \mub, \nu - \uvec j)$.
				Since $\alpha'$ is the smallest possible value (otherwise \autoref{invar:comp_colouring_cmin} is violated), ${\alpha'} \leq \tilde\alpha$ .
				By \fullref{stmt:alpha_decrease_valid_cmin}, $(\alpha', \mub, \nu - \uvec j)$ is valid.
Setting $\bar\alpha \defeq \alpha' < \alpha + p_j$ thus shows the claim.
			\end{enumerate}

			\item \underline{Case 3.2: $(\bar\alpha, \mub, \nu - \uvec j)$ is invalid for all $\bar\alpha$.}
			Then the second if-statement is entered with usable blue area $\alpha' = (1 + \eps) \OPT-(\nu - \uvec j)  s^\intercal \mub > \alpha - p_j$ (otherwise, $(\alpha', \mub, \nu)$ was $p_j$-free by \fullref{defi:kappa_free_cmin} \contradiction).
			We show that (iii) is satisfied.
\begin{enumerate}[]
				\item \underline{Case 3.2.1: $t \neq \bot \land \mub_t \neq \zero$.}
				Then $\bar\alpha \defeq \alpha' - (1+\eps) \lb$ and $(\bar \alpha, \mub - \uvec t, \nu - \uvec j)$ is returned by our algorithm, which is valid by \fullref{stmt:kappa_free_rem_cmin} as $(\bar \alpha, \mub - \uvec {t}, \nu)$ is $\pmax$-free.

				\item \underline{Case 3.2.2: $t = \bot \lor \mub_t = \zero$.}
				Then \Call{Remove'}{} would fail, as it tries to recolour a machine of undefined type or a machine of a type where all machines are red already.
				We show that this case cannot occur.
				First, we know that, since $\OPT-(\nu) = \OPT-(\nu - \uvec j)$, in this case there is at least one blue machine by \fullref{stmt:exists_blue_cmin}.
By \autoref{invar:recolouring} this implies that at some point before, a job $j'$ was removed which lead to a decrease of $\OPT-$ while there was still a blue machine with capacity $s_{t'} \OPT- = \lb$.
				This leads to a contradiction, as if at that time this machine was recoloured \btr, then $j'$ could have been removed without decreasing $\OPT-$ by \fullref{stmt:btr_pmax_free_cmin}. \qedhere
			\end{enumerate}
		\end{enumerate}
	\end{enumerate}
\end{proof}
As an immediate consequence, since $(1+\eps) \lb \leq \poly(\epsinv,\pmax)$, we get that \Call{Remove'}{} always computes a valid state $\bar{S} = (\bar \alpha, \mub[\bar], \allowbreak \bar \nu)$ such that $\abs{\Delta \alpha} + \norm{\Delta\mub}_1 + \norm{\Delta\nu}_1 \leq \poly(\epsinv,\pmax)$.
We can then achieve the desired migration bound and run time for the case of removing jobs using \fullref{stmt:sensitivity},~(\ref{stmt:sensitivity:param}) and \fullref{stmt:ipdyn_solve},~(\ref{stmt:ipdyn_solve:param}), which can be found in \appendixref{sec:ipsolve}. 
\subsection{Handling Insertion}
\label{sec:insertion_cmin}

We now incorporate the insertion of jobs into the algorithm.
We make use of the subroutine \Call{TryRTB}{} that simply tries to recolour up to $k$ machines \rtb without violating \autoref{invar:comp_colouring_cmin}.
We then propose algorithm \Call{Insert'}{} as a procedure to insert a job, using \Call{TryRTB}{} as a subroutine.
Let $\mur \defeq \mu - \mub$.
\begin{algorithm}[h]
\begin{algorithmic}[1]
	\Function{TryRTB}{$k, \alpha, \mub, \nu$}
		\If{$\mur \neq \zero$}
			\State $t \gets \crit(\OPT+(\nu))$
			\If{$t \neq \bot$}
				\State $k \gets \min \sing{ k, \mur_t }$
				\State $\mub \gets \mub + k \cdot \uvec t$
				\State $\alpha \gets \alpha + (1+\veps) \cdot k s_t \cdot \OPT-(\nu)$
			\EndIf
		\EndIf
		\Return $(\alpha, \mub, \nu)$	
	\EndFunction
	\end{algorithmic}
	\caption{Recolouring subroutine}
	\label{algo:tryBTR}
\end{algorithm}
\begin{algorithm}[h]
\begin{algorithmic}[1]
	\Function{Insert'}{$j$, $S$}
	\State $S$ is of the form $(\alpha, \mub, \nu)$
	\State $t \gets \crit(\OPT-(\nu + \uvec j))$
\If{$\OPT-(\nu + \uvec j) > \OPT-(\nu)$}
		\State $S \gets \Call{TryRTB}{m, S}$ \label{algo:insert:line:rtb_m_cmin}
		\Comment{Note that this modifies $\mub$}
		\State $\alpha \gets (1 + \eps) \cdot \OPT-(\nu + \uvec j)  s^\intercal \mub$ \label{algo:insert:line:alpha_after_rtb_cmin} \EndIf
	\State $\nu \gets \nu + \uvec j$ \label{algo:insert:line:part_cmin}
	\If{$\text{$S$ is $(\veps \lb)$-free} \land t \neq \bot \land \mur_t > 0$}
		\State $S \gets \Call{TryRTB}{1, S}$ \label{algo:insert:line:tryRTB1_cmin}
	\EndIf
	\If{$\text{$S$ is $\pmax$-free} \land \enpar*{t = \bot \lor \mur_t = 0}$}
		\State~$\alpha \gets \min \Sing{ \alpha + \pmax ,\, (1 + \eps) \cdot \OPT+(\nu) \cdot s^\intercal \mub) }$ \label{algo:insert:line:alpha_plus_pmax_cmin}
	\EndIf
	\State \Return $(\alpha, \mub, \nu)$
	\EndFunction
\end{algorithmic}
\caption{Inserting a job $j$, given the current state $S$}
\label{algo:insert_cmin}
\end{algorithm}

We now want to show similar properties as in the removal case, \eg prove that the state computed by the algorithm is valid for the new job vector, and the change of the parameters is bounded by $\poly(\epsinv,\pmax)$. The next lemma is almost the inverse of \fullref{stmt:btr_pmax_free_cmin}, except that we require $S$ to be $(\veps \lb)$-free instead of gaining $\pmax$-freeness.
\begin{lemma}[
	, name = Recolouring a Red Machine
	, label = stmt:rtb_lb_free_cmin
]
	If $S \defeq (\alpha, \mub, \nu)$ is $(\veps \lb)$-free and $\mur_{\crit} > 0$, where $\crit = \crit(\OPT-(\nu))$, then $S' \defeq (\alpha + (1+\veps) \lb, \mub + \uvec {\crit} , \nu)$ is valid.
\end{lemma}
\begin{proof}
	Let $i$ be machine of type $\crit \defeq \crit(\OPT-(\nu))$ to be recoloured.
	Note that $\lb = s_{\crit} \OPT-(\nu)$ by definition of $\crit(\cdot)$.
	Now, let $(x, \nub)$ be the solution associated to $S$, and $\sigma_i$ the assignment of jobs to $i$ according to $x$.
	Also note that $p^\intercal \sigma_i \geq \lb$.
	By \fullref{defi:kappa_free_cmin} we know that $\bar S$ is valid.
	Moreover, as the job set is the same as in $S$, the values for $\OPT-$ do also coincide.

Consider the following modifications to $(x, \nub)$ and $S$:
	\begin{enumerate}
		\item Set $\nu \gets \nu - \sigma_i$, $\nub \gets \nub + \sigma_i$
		\item Set $\mub_{\crit} \gets \mub_{\crit} + 1$, $\mur_{\crit} \gets \mur_{\crit} - 1$, $\alpha \gets \alpha + (1+\veps) \lb$
	\end{enumerate}
	We call the resulting state $\bar S$. This state $\bar S$ is valid: $\alpha$ can be increased by $(1+\veps) \lb$ compared to the valid state $S$, since we not only increase the load on the blue machines $\nub$ by $\sigma_i \geq \lb$, but $S$ was also $(\veps \lb)$-free before, \ie $\alpha$ can already be increased by $\veps \lb$ due to \fullref{defi:kappa_free_cmin} without violating the dual knapsack constraint.
	\autoref{invar:machine_colouring} cannot be violated as we only recoloured a machine of the critical type $\crit$.
	\autoref{invar:comp_colouring} is not violated as recolouring a machine \rtb increases the required total blue area by $(1+\veps) \lb$, but we also increased $\alpha$ accordingly.
	Thus, the claim holds.
\end{proof}

With this result in hand, we can prove the first part: Our algorithm does indeed always compute a valid state.

\begin{lemma}[
	, name  = Correctness of \textsc{Insert'}
	, label = stmt:algo_insert_correct_cmin
]Given a valid state, \Call{Insert'}{} always computes a valid state.
\end{lemma}
\begin{proof}
	First, we show that the state computed after \lineref{algo:insert:line:part_cmin} is valid.
	Afterwards, we argue that the remaining lines will not change a valid state into an invalid one.
	Let $j$ be the job that is inserted, i.e. $\Delta \nu = \uvec j$.
	We make a case distinction on $\Delta\OPT$.
	If $\Delta \OPT- = 0$, then simply inserting the job already yields a valid state, which is exactly what the algorithm computes.
	If $\Delta \OPT- > 0$, \lineref{algo:insert:line:rtb_m_cmin} ensures that \autoref{invar:machine_colouring} is not violated.
	The line thereafter ensures, that \autoref{invar:comp_colouring_cmin} is not violated:
	$(1 + \eps) \OPT-(\nu + \uvec j)  s^\intercal \mub$ is exactly the blue area we are required to fill when all machines of $\crit(\OPT-(\nu + \uvec j))$ are coloured red, which is true for the new $\crit(\OPT-(\nu + \uvec j))$, since machines of this type had capacity $< \lb$ for $\OPT-$ and where thus coloured red by validity of the input state. For $\mub = \zero$, the equality above also holds as both sides are $0$.
	In consequence, setting $\alpha$ to the value above satisfies \autoref{invar:comp_colouring_cmin} for $T \gets \OPT-(\nu + \uvec j)$.
	Validity of the state then eventually follows from \fullref{stmt:opt_min_cmin}. Since \Call{TryRTB}{} only recolours machines of critical-type, \autoref{invar:recolouring} holds as well.

	Now, let us consider what happens after \lineref{algo:insert:line:part_cmin}.
	If a machine is recoloured due to an execution of \lineref{algo:insert:line:tryRTB1_cmin}, then $S$ was $(\veps \lb)$-free before.
	Then the resulting state is also valid due to \fullref{stmt:rtb_lb_free_cmin}.
	If \lineref{algo:insert:line:alpha_plus_pmax_cmin} is executed, then $\alpha$ is changed by at most $\pmax$, but the state was $\pmax$-free before, implying it is valid thereafter.
\end{proof}
Now, in order to prove bounded migration of our proposed procedure, we first need to introduce two upper bounds on the value of $\alpha$, and show that they hold for every step of our algorithm.
If both upper bounds hold for our starting state, this property is implied if they hold for any execution of \Call{Remove'}{} and \Call{Insert'}{} on a valid state as well.
These upper bounds will help us to first proof bounded $\kappa$-freeness of most states computed by our algorithm, and by that, help us show that the \enquote{critical lines} \lineref{algo:insert:line:rtb_m_cmin} and \lineref{algo:insert:line:alpha_after_rtb_cmin} of \Call{Insert'}{} do not increase $\mub$ and $\alpha$ too much, respectively.
This then will imply bounded migration. The first upper bound is \autoref{invar:upper_bound_cmin}.

\begin{invariant}[
	, label = invar:upper_bound_cmin
]
	\hfill\hfill$\displaystyle
		\crit \neq \bot \implies (1 + \eps) \OPT-(\nu)  s^\intercal (\mub + \uvec \crit) > \alpha
	\enspace.$\hfill\,\!
\end{invariant}

Note that $\mub + \uvec {\crit}$ can have an entry in $\crit$ greater than the available number of machines $\mu_{\crit}$ of this type; in this case, for the sake of simplicity, we define $(1 + \eps) \OPT-(\nu)  s^\intercal (\mub + \uvec j)$ to be $\infty$.
For the initial solution with no jobs, \autoref{invar:upper_bound_cmin} holds as $\OPT = \OPT- = 0$, implying that there cannot be a blue machine, and hence $\crit = \bot$. As we only use procedures \Call{Remove'}{} and \Call{Insert'}{} to compute new states, if both procedure maintain the invariant, the whole algorithm does so.
\begin{proposition}[
	, label = stmt:algo_remove_maintains_upper_bound_cmin
]
	\Call{Remove'}{} maintains \autoref{invar:upper_bound_cmin}
\end{proposition}
\begin{proof}
	Since the value for $\alpha$ is only decreased by \Call{Remove'}{}, there are only two scenarios where the claim does not become trivial:
	A change of $\OPT-$ and a recolouring of a blue machine (both cannot occur at the same time due to \fullref{stmt:btr_pmax_free_cmin}).
	Let $S \defeq (\alpha, \mub, \nu)$ be the state before removal of job $j$ and $\bar S \defeq (\bar\alpha, \mub[\bar], \bar\nu)$ with $\bar \nu \defeq \nu - \uvec j$ the state afterwards.
	\begin{enumerate}[]
		\item \case [1] {$\mub[\bar] \neq \mub \land \OPT-(\nu) = \OPT-(\bar\nu)$.}
		This means that (iii) of \fullref{stmt:algo_remove_rephrased_cmin} \enquote{is executed}.
		Then $(\alpha', \mub, \bar\nu)$ is invalid for all $\alpha'$, so in particular for all $\alpha' \geq (1 + \eps) \OPT-(\bar \nu)  s^\intercal \mub $ which are all possible values that do not violate \autoref{invar:comp_colouring_cmin}.
		That means that the invalidity stems from a too small minimum completion time of the associated solutions.
		Recolouring a machine \btr without decreasing the $\alpha$-value will only make the situation worse, i.e. cannot lead to a larger minimum completion time.
		In consequence, the state $(\bar\alpha, \mub + \uvec {\crit}, \bar\nu)$ returned from \Call{Remove'}{} must satisfy $\bar\alpha < (1 + \eps) \OPT-(\bar \nu)  s^\intercal \mub = (1 + \eps) \OPT-(\bar \nu)  s^\intercal (\mub[\bar] + \uvec \crit)$.
	
		\item \case [2] {$\mub[\bar] = \mub \land \OPT+(\nu) \neq \OPT+(\bar\nu)$.}
		Let $\bar\crit \defeq \OPT+(\bar\nu)$ be the (new) critical type. 
		In particular, we have $\bar{\crit} \neq \crit$.
Due to \fullref{defi:critical_type} and \autoref{invar:machine_colouring}, we either have $\crit = \bot$ or $\mub_{\crit} = \mu_{\crit}$.
		In the first case, the premise of the implications of \autoref{invar:lower_bound} evaluates to \enquote{false}, and \autoref{invar:lower_bound} holds.
		In the latter case, $(1 + \eps) \OPT+(\bar \nu)  s^\intercal (\mub + \uvec \crit)$ evaluates to  $\infty$, and \autoref{invar:lower_bound} holds as well, for all possible $\alpha$.
	\qedhere\end{enumerate}
\end{proof}
\begin{proposition}[
	, label = stmt:tryRTB_maintains_upper_bound_cmin
]
	\Call{TryRTB}{} maintains \autoref{invar:upper_bound_cmin}.
\end{proposition}
\begin{proof}
	Consider a call \Call{TryRTB}{$k, \alpha, \mub, \nu$}.
	We may without loss of generality assume that $\crit = \crit(\OPT-) \neq \bot$, otherwise the function simply returns the input state.
	Moreover, we may without loss of generality assume that $k \leq \mur_{\crit}$ as internally $k$ is modified accordingly (for any $k' \geq \mur_{\crit}$ the function behaves identically).
	Now, assume that \autoref{invar:upper_bound_cmin} holds for the initial state, i.e.
	\[
		\alpha < (1 + \eps) \OPT-(\nu)  s^\intercal (\mub + \uvec \crit)
		\enspace ,
	\]
	and let $(\bar\alpha, \mub[\bar], \nu)$ be the state returned by the function.
	Then
	\[
		\bar\alpha = \alpha + k (1+\veps) s_{\crit} \OPT-{}
		= \alpha + k (1+\veps) \lb
		\enspace .
	\]
	Moreover,
	\begin{multline*}
		(1 + \eps) \OPT-(\nu)  s^\intercal \mub[\bar]
		= (1 + \eps) \OPT-(\nu)  s^\intercal (\mub + k \uvec \crit)
		\\ = (1 + \eps) \OPT-(\nu)  s^\intercal \mub + k (1+\veps) s_{\crit} \OPT-{}
		= (1 + \eps) \OPT-(\nu)  s^\intercal \mub + k (1+\veps) \lb
		\enspace .  
	\end{multline*}
	We can then show the claim, i.e. given that \autoref{invar:upper_bound_cmin} did hold initially prove that it holds afterwards.
	\begin{align*}
		\alpha <{} & (1 + \eps) \OPT-(\nu)  s^\intercal (\mub + \uvec \crit)
		\\
		& \iff
		\alpha + k (1+\veps) \lb < (1 + \eps) \OPT-(\nu)  s^\intercal (\mub + \uvec \crit) + k (1+\veps) \lb
		\\& \iff
		\bar \alpha < (1 + \eps) \OPT-(\nu)  s^\intercal (\mub + (k+1)\uvec \crit)
		\\& = (1 + \eps) \OPT-(\nu)  s^\intercal (\mub[\bar] + \uvec \crit) \qedhere
	\end{align*}
\end{proof}

\begin{proposition}[
	, label = stmt:insert_maintains_upper_bound_cmin
]
	\Call{Insert'}{} maintains \autoref{invar:upper_bound_cmin}
\end{proposition}
\begin{proof}
	Since the value for $\alpha$ is only increased by \Call{Insert'}{}, there are only two scenarios where the claim does not become trivial:
	A change of $\OPT-$ or a recolouring of a red machine.
Let $S \defeq (\alpha, \mub, \nu)$ be the state before insertion of job $j$ and $\bar S \defeq (\bar\alpha, \mub[\bar], \bar\nu)$ with $\bar \nu \defeq \nu + \uvec j$ the state afterwards.
	\begin{enumerate}[]
		\item \case [1] {$\mub[\bar] \neq \mub \land \OPT-(\nu) = \OPT-(\bar\nu)$.}	
		The invariant can only be violated due to an increase of $\alpha$.
		As in this case, \lineref{algo:insert:line:rtb_m_cmin} and \lineref{algo:insert:line:alpha_after_rtb_cmin} are not executed, this can only be due to \lineref{algo:insert:line:tryRTB1_cmin} or \lineref{algo:insert:line:alpha_plus_pmax_cmin}.
		The execution of  \lineref{algo:insert:line:tryRTB1_cmin} maintains the invariant by \autoref{stmt:tryRTB_maintains_upper_bound_cmin}.
		Let $\crit \eqdef \crit(\OPT-(\nu))$.	
		\lineref{algo:insert:line:alpha_plus_pmax_cmin} is only executed if $\crit = \bot$ or $\mur_{\crit} = 0$, the latter of which is equivalent to $\mub_{\crit} = \mu_{\crit}$; in both cases, the invariant trivially holds, as either its premise is false or the LHS of the inequality evaluates to $\infty$.

		\item \case [2] {$(\mub[\bar] = \mub \lor \mub[\bar] \neq \mub) \land \OPT-(\nu) \neq \OPT-(\bar\nu)$.}
		Then, \lineref{algo:insert:line:rtb_m_cmin} and \lineref{algo:insert:line:alpha_after_rtb_cmin} are executed.
		Again, the execution of \lineref{algo:insert:line:rtb_m_cmin} maintains the invariant by \autoref{stmt:tryRTB_maintains_upper_bound_cmin}.
		Let $\check \crit \eqdef \crit(\OPT-(\bar\nu))$.
		Then, after \lineref{algo:insert:line:alpha_after_rtb_cmin}, if $\mub_{\check \crit} < \mu_{\check \crit}$,
		\[
			\tilde\alpha \defeq (1 + \eps)  s^\intercal \OPT-(\bar\nu) \mub[\tilde] < (1+\eps) \OPT-(\bar\nu) s^\intercal (\mub[\tilde] + \uvec {\check \crit}) \enspace .
\]
		Otherwise, if $\mub_{\check \crit} = \mu_{\check \crit}$ or $\check \crit = \bot$, the invariant trivially holds as argued above. Thus, \autoref{invar:upper_bound_cmin} always holds when \lineref{algo:insert:line:alpha_after_rtb_cmin} is executed.

		In the current case of $\OPT-(\nu) \neq \OPT-(\bar\nu)$, $\bar S$ can not be $\pmax$-free (nor $(\veps \lb)$-free), as otherwise, we could simply remove $j$ again without a minimum completion time decrease, implying $\OPT-(\nu) = \OPT-(\bar\nu)$, a contradiction.
		Thus, neither \lineref{algo:insert:line:tryRTB1_cmin} nor \lineref{algo:insert:line:alpha_plus_pmax_cmin} are executed, no more changes to $\tilde \alpha$ are made, and \autoref{invar:upper_bound_cmin} always holds.		
\qedhere\end{enumerate}
\end{proof}
\begin{invariant}
\label{invar:alpha_leq_delta_cmin}
	\hfill\hfill$\displaystyle
		\delta \defeq (1+\eps) \OPT+(\nu) s^\intercal (\mub + \mub_{\crit} \uvec {\crit}) \geq \alpha
	\enspace.$\hfill\,\!
\end{invariant}

Our second upper bound is \autoref{invar:alpha_leq_delta_cmin}.
As $\alpha = \delta = 0$ initially, this upper bound already holds for our starting state.
Luckily, showing that \Call{Remove'}{} or \Call{Insert'}{} maintain \autoref{invar:alpha_leq_delta_cmin} is less involved.

\begin{proposition}[
	, label = stmt:algo_remove_maintains_ald_cmin
]
	\Call{Remove'}{} maintains \autoref{invar:alpha_leq_delta_cmin}.
\end{proposition}
\begin{proof}
	If $\OPT-(\nu - \uvec j) = \OPT-(\nu)$, \Call{Remove'}{} only decreases $\alpha$, while $\delta$ does not change.
	Else if $\OPT+(\nu + \uvec j) \neq \OPT+(\nu)$, $\alpha$ is set to $(1+\eps) \OPT+(\nu - \uvec j) s^\intercal \mub \leq \delta$ if the input state is valid already, or $\alpha$ is decreased by at least $p_j$ if it is not. In the latter case, $\alpha < (1+\eps) \OPT+(\nu - \uvec j) s^\intercal \mub \leq \delta$ after the decrease, as otherwise, the resulting state would be valid for $\alpha = (1+\eps) \OPT-(\nu) s^\intercal \mub \leq \delta$ as well, a contradiction to the assumption that it was not.
\end{proof}

\begin{proposition}[
	, label = stmt:algo_remove_maintains_agd
]
	\Call{Insert'}{} maintains \autoref{invar:alpha_leq_delta_cmin}.
\end{proposition}
\begin{proof}
	If $\OPT-(\nu + \uvec j) > \OPT-(\nu)$, $\alpha$ is set to at most $\delta$ in \lineref{algo:insert:line:alpha_after_rtb_cmin}, as all machines of $\crit(\OPT-(\nu))$ are coloured \rtb by \lineref{algo:insert:line:rtb_m_cmin}, and all machines of the new  $\crit(\OPT-(\nu + \uvec j))$ are still coloured red in the resulting state by virtue that \autoref{invar:machine_colouring} was maintained for the input state.
	While recolouring a blue machine in \lineref{algo:insert:line:tryRTB1_cmin}, only the capacity of the recoloured machine is added to $\alpha$, maintaining $\alpha \leq \delta$ by maintaining \autoref{invar:upper_bound_cmin}.
	 If \lineref{algo:insert:line:alpha_plus_pmax_cmin} is executed, $\alpha$ is set to at most $\delta$ as well.
\end{proof}

	Using these bounds given to us by these two invariants, we can prove upper bounds for the $\kappa$-freeness of states after the computation of a new state. We first consider \Call{Remove'}{}.

\begin{proposition}[
	, label = stmt:algo_remove_kappa_ineq_cmin
]
	Let $S = (\alpha, \mub, \nu)$ be a valid state and $\kappa$ maximal such that $S$ is $\kappa$-free, and $\bar S = (\bar \alpha, \mub[\bar], \nu + \uvec j)$ be $\bar \kappa$-free state returned by \Call{Remove'}{$j,S$}.
	Let $\crit \defeq \crit(\OPT-(\nu - \uvec j))$ and $\bar \delta \defeq (1+\eps) \OPT+(\nu - \uvec j) s^\intercal (\mub + \mub_{\crit} \uvec {\crit})$.
	Then
	\[
		\bar \alpha < \alpha \land \bar \alpha < \bar \delta
		\implies
		\bar\kappa \leq  (1+\eps) \lb + \pmax - 1
	\enspace . \stopphier \]
\end{proposition}
\begin{proof}
	Consider \fullref{stmt:algo_remove_rephrased_cmin}; read the $3$ cases as statements of an algorithm.
	If the algorithm terminates after (i), either $\bar \alpha = \alpha$ or $\bar \alpha = (1+\eps) \OPT-(\nu - \uvec j) s^\intercal \mub = \bar \delta$, where the last equality follows from the fact that no machines of type $\crit$ are coloured red in this case, implying $\mub_{\crit} = \mu_{\crit}$, and the statement is always true as its LHS terminates to false.
	If the algorithm terminates after (ii), then $\bar{\kappa} < p_j$, as we reduced $\alpha$ by at most $2 p_j$, but immediately remove $p_j$ total processing time from $p^\intercal \nub$.
	Now assume that it terminates after (iii).
	Then, by the inequality $\bar \alpha > (1+\eps) \lb - p_j$, the statement follows directly, as this case only occurs if $S$ was less than $p_j$-free before, $p^\intercal \nub$ is reduced by at least $p_j$ (reducing the freeness by this amount), and thus it must be less than $((1+\eps) \lb + p_j)$-free after.
\end{proof}

We show a similar statement for the slack of $\alpha$ after using \Call{Insert'}{}.

\begin{proposition}[
	, label = stmt:algo_insert_kappa_ineq_cmin
]
	Let $S$ be a valid state and $\kappa$ maximal such that $S$ is $\kappa$-free, $\bar S = (\bar \alpha, \mub[\bar], \bar \nu)$ be $\bar\kappa$-free state returned by \Call{Insert'}{$j,S$}.
	Let $\crit \defeq \crit(\OPT-(\nu))$ and $\delta \defeq (1+\eps) \OPT+(\nu) s^\intercal (\mub + \mub_{\crit} \uvec {\crit})$.
Then
	\begin{enumerate}[(i)]
	\item \label{insert_case1_cmin}	\[
		\bar \alpha < \delta
		\implies
		\bar{\kappa} \leq \max\Sing{\kappa \ssep \veps \lb}
		\enspace.
	\]
	\item \label{insert_case2_cmin}	\[
		\OPT-(\nu + \uvec j) > \OPT-(\nu)
		\implies
		\bar{\kappa} < \pmax
		\enspace.\stopphier
	\]
	\end{enumerate}

\end{proposition}
\begin{proof}
(\ref{remove_case1})
		If $\bar \alpha < \delta$, then $\OPT-(\nu) = \OPT-(\nu + \uvec j)$, as otherwise, \lineref{algo:insert:line:rtb_m_cmin} and \lineref{algo:insert:line:alpha_after_rtb_cmin} would have been executed and set $\alpha = (1+\eps) \OPT-(\nu+\uvec j) s^\intercal (\mub + \mub_{\crit} \uvec \crit)= \delta$, where the last equality follows from \autoref{stmt:estimate_opt}.
		If $\bar\kappa < \veps \lb$, the claim directly holds.
		Otherwise, $\bar \kappa \geq \veps \lb$.
		We backtrack which lines have been executed within \Call{Insert'}{}:
		If $\bar S$ is $(\veps \lb)$-free even after \Call{Insert'}{}, either \lineref{algo:insert:line:tryRTB1_cmin} or \lineref{algo:insert:line:alpha_plus_pmax_cmin} have been executed during this call, as both lines can only reduce the freeness, and thus, the state before these lines was at least $(\veps \lb)$-free as well, and one of the conditions before the respective line is then always true.
		Since $\bar \alpha < \delta$ by assumption, and $\delta = (1+\eps) \OPT+(\nu + \uvec j) s^\intercal \mub$ if $\crit = \bot$ or $\mub_{\crit} = \mu_ {\crit}$, each of both lines reduce $\alpha$, and thus freeness, by at least $\pmax$ if they are executed, so the state $S'$ before these lines is at least $\bar\kappa + \pmax$-free.
		Since $S'$ is derived from $S$ by simply inserting some job, we can conclude that $S$ is at least $\bar\kappa$-free, which shows the claim.
		
		(\ref{remove_case2}) $\OPT-(\nu+\uvec j) > \OPT-(\nu)$ holds by assumption. If $\bar S$ was at least $\pmax$-free, removing $j$ again would yield a state $S'$ with$\OPT-(\nu+\uvec j) = \OPT-(\nu)$ by \autoref{stmt:kappa_free_rem_cmin}, a contradiction.
\end{proof}

Consequently, we combine the two previous statements to upper bound the slack of all states computed by our algorithm using only \Call{Remove'}{} and \Call{Insert'}{} to compute new states.

\begin{proposition}[
	, label = stmt:max_freedom_of_states_cmin
]
	Let $S = (\alpha, \mub, \nu)$ be a valid state, $\crit \defeq \crit(\OPT-(\nu))$ and $\delta \defeq (1+\eps) \OPT+(\nu) s^\intercal (\mub + \mub_{\crit} \uvec {\crit})$.
	
	Let $\bar S = (\bar \alpha, \mub[\bar], \bar \nu)$ be a state computed either by $\bar S = \Call{Remove'}{j,S}$ or $\bar S = \Call{Insert'}{j,S}$.
	Let $\tilde \crit \defeq \crit(\OPT-(\bar \nu))$ and $\bar \delta \defeq (1+\eps) \OPT+(\bar \nu) s^\intercal (\mub[\bar] + \mub[\bar]_{\tilde \crit} \uvec {\tilde \crit})$.
	Then
		\[
		(\alpha < \delta
		\implies
		S \text{ is at most } ((1+\eps) \lb + \pmax - 1)\text{-free})\\
		\implies\\
		(\bar \alpha < \bar \delta
		\implies
		\bar S \text{ is at most } ((1+\eps) \lb + \pmax - 1)\text{-free})
		\enspace .
	\]
\end{proposition}
\begin{proof}
In the following, we assume that the LHS of the implication $(\alpha < \delta
\implies
S \text{ is at most } ((1+\eps) \lb + \pmax - 1)\text{-free})$ as well as the the LHS of the implication on the RHS $(\bar \alpha < \bar \delta)$ both evaluate to true, as otherwise, the implication always evaluates to true.
We make a case distinction regarding $\OPT-(\bar \nu)$ in relation to $\OPT-(\nu)$.
	\begin{enumerate}[]
\item \case [1] {$\OPT-(\bar \nu) < \OPT-(\nu)$}. 
		This can only happen if $\bar S = \Call{Remove'}{j,S}$.
		Then $\bar \alpha < \alpha$, as, if $\bar \alpha$ is not decreased, $\bar \alpha = \bar \delta$ by \lineref{alpha_new_delta_cmin} of $\Call{Remove'}{}$ and the fact that all machines of $\tilde \crit$ are coloured blue in this case, a contradiction to the assumption that $\bar \alpha < \bar \delta$.
		Thus, $\bar S$ is at most $((1+\eps) \lb + \pmax - 1)$-free by \autoref{stmt:algo_remove_kappa_ineq_cmin}.
		\item \case [2] {$\OPT-(\bar \nu) > \OPT-(\nu)$}. 
		This can only happen if $\bar S = \Call{Insert'}{j,S}$.
		Then $\bar S$ is at most $((1+\eps) \lb + \pmax - 1)$-free by \autoref{stmt:algo_insert_kappa_ineq_cmin},(\ref{insert_case2_cmin}).
		\item \case [3] {$\OPT-(\bar \nu) = \OPT-(\nu)$}.
		\begin{enumerate}[]
		\item \underline{Case 3.1: $\bar S = \Call{Remove'}{j, S}$}
		It holds that $\bar \alpha \leq \alpha$ by definition of $\Call{Remove'}$.
		If $\bar \alpha = \alpha$, then the statement is true:
		Either $\bar \alpha = \alpha = \delta$, which makes the statement always true, or $\bar \alpha = \alpha < \delta$, implying $((1+\eps) \lb + \pmax - 1)$-freeness for $S$, and thus $\bar S$ is at most $((1+\eps) \lb + \pmax - 1)$-free since removing a job without decreasing $\alpha$ or $\OPT-$ can only decrease the freeness.
		
		Else if $\bar \alpha < \alpha$, the statement is also true, as $\alpha \leq \delta$ by \autoref{invar:alpha_leq_delta_cmin} and $\delta = \bar \delta$ when $\OPT-(\bar \nu) = \OPT-(\nu)$, and thus $\bar S$ is at most $((1+\eps) \lb + \pmax - 1)$-free by \autoref{stmt:algo_remove_kappa_ineq_cmin}.
		\item \underline{Case 3.2: $\bar S = \Call{Insert'}{j, S}$}
		It holds that $\bar \alpha \geq \alpha$ by definition of $\Call{Insert'}$.
		Also, note that $\delta = \bar \delta$ when $\OPT-(\bar \nu) = \OPT-(\nu)$. Thus, $\bar \alpha \leq \bar \delta = \delta$, where the first inequality follows from \autoref{invar:alpha_leq_delta_cmin}.
		If $\bar \alpha = \bar \delta$, the statement is always true.
		Else if $\alpha \geq \bar \alpha > \bar \delta = \delta$, $S$ is at most $((1+\eps) \lb + \pmax - 1)$-free by our assumptions, and \autoref{stmt:algo_insert_kappa_ineq_cmin},(\ref{insert_case1_cmin}) then shows that $\bar S$ is at most $((1+\eps) \lb + \pmax - 1)$-free as well, proving the statement.
		
		\end{enumerate}
\end{enumerate}	
	This proves the claim.
\end{proof}

Since $(\alpha < \delta \implies S \text{ is at most } ((1+\eps) \lb + \pmax - 1)\text{-free})$ is true for our starting state ($\nu = 0$ and the state is $0$-free), and we use only \Call{Remove'}{} and \Call{Insert'}{} to compute new states, \autoref{stmt:max_freedom_of_states_cmin} holds for every step our algorithm.
This implies the following corollary.

\begin{corollary}
\label{cor:max_freedom_of_states_cmin}
For each state $S = (\alpha,\mub,\nu)$ computed by our algorithm
$$\alpha < \delta \implies S \text{ is at most } ((1+\eps) \lb + \pmax - 1)\text{-free}$$
is true, with $\delta = (1+\eps) \OPT+(\nu) s^\intercal (\mub + \mub_{\crit} \uvec {\crit})$ and $\crit \defeq \crit(\OPT-(\nu))$.
\end{corollary}

We can use this bound on the $\kappa$-freeness of states to show the following claim on the colouring and $\alpha$ change caused by \lineref{algo:insert:line:rtb_m_cmin} and \lineref{algo:insert:line:alpha_after_rtb_cmin}, respectively.

\begin{proposition}[
	, label = stmt:bounded_recolours_alpha_cmin
]
For any execution of \Call{Insert'}{$j,S$} on a state $S$ produced by our algorithm and a job $j \in \nu$ the following holds:
\begin{enumerate}[(i)]
	\item On any execution of \lineref{algo:insert:line:rtb_m_cmin}, at most $\frac{(1+\eps) \lb + \pmax}{\eps \lb}$ machines are recoloured \rtb. \label{max_recolours_rtb:m_cmin}
	\item On any execution of \lineref{algo:insert:line:alpha_after_rtb_cmin}, $\alpha$ is increased by at most $(1+\eps) \lb + 2 \pmax$. \label{max_alpha_change_cmin}
\end{enumerate}
\end{proposition}
\begin{proof}
Let $S = (\alpha, \mub, \nu)$ and $\bar S = (\bar \alpha, \mub[\bar], \nu) = \Call{Insert'}{j,S}$.
Also, let $\crit = \crit(\OPT-(\nu))$ and $\delta \defeq (1+\eps) \OPT+(\nu) s^\intercal (\mub + \mub_{\crit} \uvec {\crit})$.
\lineref{algo:insert:line:rtb_m_cmin} and  \lineref{algo:insert:line:alpha_after_rtb_cmin} are only executed during \Call{Insert'}{j,S} if $\OPT-(\nu+\uvec j) > \OPT-(\nu)$.
Note that, in this case, $\delta = (1+\eps) \OPT-(\nu + \uvec j) s^\intercal (\mub + \mub_{\crit} \uvec {\crit})$ by \autoref{stmt:makespan_change_cmin}.
Thus, $\bar{\alpha} = \delta$ after execution of \lineref{algo:insert:line:alpha_after_rtb_cmin}.

\begin{enumerate}[]
\item \case {$\alpha < \delta$}
Then $S$ is at most $((1+\eps) \lb + \pmax - 1)$-free by \autoref{cor:max_freedom_of_states_cmin}.
Removing $j$ again with \Call{Remove'}{$\bar S,j$} after its insertion yields a valid state $S' = (\alpha', \mub{'}, \nu)$ with $\OPT-(\nu) < \OPT-(\nu + \uvec j)$.
For $S'$, $\mub{'}_{\crit} \geq \mu_{\crit}-1$ and $\alpha' \geq \delta - p_j$, as, for $\OPT-(\nu) < \OPT-(\nu + \uvec j)$, \Call{Remove'}{$\bar S,j$} sets $\alpha$ first to $\delta$, and then decreases $\alpha$ by at most $p_j$ after.

(\ref{max_recolours_rtb:m_cmin}).
Assume that \lineref{algo:insert:line:rtb_m_cmin} is executed on state $S$ and job $j$ and more than $\frac{(1+\eps) \lb + \pmax}{\eps \lb}$ machines are recoloured \rtb.
For this to occur, $\mur_{\crit} \geq \frac{(1+\eps) \lb + \pmax + 1}{\eps \lb}$ must hold at this point.

Then, since $S'$ is at least $0$-free due to its validity, $S$ is at least $((1+\eps) \lb + \pmax)$-free, as each of the $\frac{(1+\eps) \lb + \pmax + 1}{\eps \lb}$ machines that are coloured red in $S$ but not in $S'$ subtract at least $\eps \lb$ from $\alpha$, due to the maintaining of \autoref{invar:upper_bound_cmin}.
This is a contradiction to $S$ is at most $((1+\eps) \lb + \pmax - 1)$-free if $\alpha < \delta$ by \autoref{cor:max_freedom_of_states_cmin}.
\end{enumerate}

(\ref{max_alpha_change}).
If $\alpha < \bar \alpha + (1+\eps) \lb + 2 \pmax = \delta + (1+\eps) \lb + 2 \pmax$, then $S$ is at least $((1+\eps) \lb + \pmax)$-free for $\mub_{\crit} = \mu_{\crit}$, as $\alpha' \geq \delta + \pmax$ for the state $S'$ with the same job set $\nu$.
This then holds for all $\mub_{\crit} \leq \mu_{\crit}$, as any additional red machine increases the $\kappa$-freeness by at least $\eps \lb$ while maintaining \autoref{invar:upper_bound_cmin}.
This again is a contradiction to $S$ is at most $((1+\eps) \lb + \pmax - 1)$-free if $\alpha < \delta$, and thus, $\alpha \geq \delta + (1+\eps) \lb + 2 \pmax$, and the statement follows as $\bar \alpha = \delta$.

\item \case{$\alpha = \delta$}
(\ref{max_recolours_btr:m}) and  (\ref{max_alpha_change}) are both true in this case, as $\alpha = \bar \alpha$, and recolouring one machine \rtb with \Call{TryRTB}{} increases $\alpha$ by at least $(1+\eps) \lb$, implying that no machine can be recoloured in the case of $\alpha = \bar \alpha$.

\item \case{$\alpha > \delta$} can never occur by \autoref{invar:alpha_leq_delta_cmin}.
\end{proof}

We can now finally use this result to show that \Call{Insert'}{} has indeed bounded migration.

\begin{lemma}[
	, name  = \textsc{Insert'} Has Bounded Migration
	, label = stmt:algo_remove_migration_bounded
]
	Let $S = (\alpha, \mub, \nu)$ be a valid state and $j \in \supp(\nu)$.
	Then for $\bar S \defeq (\bar \alpha, \mub[\bar], \bar \nu) = \Call{Insert'}{j,S}$, $1 \leq \abs{\Delta{\alpha}} + \norm{\Delta\nu}_1 + \norm{\Delta\mub}_1 \leq \poly(\epsinv,\pmax)$.
\end{lemma}
\begin{proof}
	$\norm{\Delta\nu}_1 = 1$ by definition of the problem.
	Due to \autoref{stmt:bounded_recolours_alpha_cmin}~(\ref{max_recolours_rtb:m_cmin}), and the fact that in \lineref{algo:insert:line:tryRTB1_cmin}, the only other \rtb recolouring, only $1$ machine is recoloured, $\norm{\Delta\mub}_1 \leq \frac{(1+\eps) \lb + \pmax}{\eps \lb} + 1 \leq \poly(\epsinv,\pmax)$.
	Thus, $\alpha$ is increased at most by $\poly(\epsinv,\pmax) \cdot (1+\eps) \lb \leq \poly(\epsinv,\pmax)$ by recolouring.

	The only other times when $\alpha$ might be changed are \lineref{algo:insert:line:alpha_plus_pmax_cmin}, where $\alpha$ is increased by at most $\pmax \leq \poly(\epsinv,\pmax)$, and \lineref{algo:insert:line:alpha_after_rtb_cmin}. By \autoref{stmt:bounded_recolours_alpha_cmin}~(\ref{max_alpha_change_cmin}), the execution of \lineref{algo:insert:line:alpha_after_rtb_cmin} increases $\alpha$ by at most $(1+\eps) \lb + 2 \pmax \leq \poly(\epsinv,\pmax)$.
	This shows the claim.
\end{proof}
As immediate consequence we get that \Call{Insert'}{} always computes a valid state $\bar{S} = (\bar \alpha, \mub[\bar], \allowbreak \bar \nu)$ such that $\abs{\Delta \alpha} + \norm{\Delta\mub}_1 + \norm{\Delta\nu}_1 \leq \poly(\epsinv,\pmax)$.
As the same result has already been shown for \Call{Remove'}{}, we can then achieve the desired migration bound of $\paramsens$ and run time of $\paramrt + |I|$ for the dynamic setting, using \Call{Remove'}{} to remove, and \Call{Insert'}{} to add jobs from step $q$ to $q+1$, via \fullref{stmt:sensitivity},~(\ref{stmt:sensitivity:param}) and \fullref{stmt:ipdyn_solve},~(\ref{stmt:ipdyn_solve:param}), which can be found in \appendixref{sec:ipsolve}.
We again treat the initial job set $\mc J_0$ as adding $\abs{\mc J_0}$ jobs to the instance in arbitrary order, beginning with an empty set.
We thus start from a valid state $S = (0,\zero,\zero)$, proving \autoref{stmt:main_theorem},~(\ref{stmt:param_qcmin_algorithm}).

 	\section{Robust Results without Parameterization}
\label{sec:noparam}
In this section, we give a formal description of our \ac{EPTAS} related results without the use of parameterization, as stated in \autoref{cor:noparam_nosmall} and \autoref{cor:noparam_general}.
A robust \ac{EPTAS} for minimization problems is an online algorithm, such that for any $\veps > 0$, the algorithm computes an $(1+\O(\veps))$-approximate solution with (amortized) migration factor dependent only on $\eps$, and running bounded by $f(\eps) \cdot \poly \abs I$, for the encoding size of the instance $\abs I$ and some computable function $f$.
For maximization problems, a robust \ac{EPTAS} is defined to be $(1-\O(\veps))$-competitive.
Again, assume \obda that for any occurring $\veps > 0$, $\epsinv$ is integral.
To this end, we bound the parameters $\pmax$ and $d$ in terms of $\epsinv$ by preprocessing the instance.
\label{sec:longroad}
\subsection{Preprocessing}

As here, we want to approximate the optimal value of a problem instance without use of parameterization, we do some preprocessing and rounding on the instance to simplify it, without increasing the error too much. We start by partitioning the job set into small jobs and large jobs: A job is large iff its processing time is at least $\eps\pmax$, and small otherwise. 
On identical machines, a standard approach to deal with small jobs is to replace them by jobs of size $\eps\pmax$ such that the area of the new jobs, \ie the sum of their processing times, just cover the area of the small jobs.
This causes an error of $< \eps \pmax$ per machine, which is $\leq \OPT*$ due to $\OPT* \geq \pmax$.
For uniform machines, this process is more involved, which leads to an additional load error of $\O(\eps \pmax)$ that cannot generally be estimated in terms of $\OPT*$.
We discuss the details of this process in \appendixref{sec:frames} when proving \fullref{stmt:dynamic_grouping}.
For now, it suffices to know that we can deal with small jobs if we can deal with large jobs -- the rough idea is that we group small jobs to chunks of size $\eps \pmax$ that we represent using place-holder jobs which we refer to as \emph{frames}.
We thus start by only considering instances without small jobs.

We adhere to the rounding scheme introduced by \textcite{BDJR22}, and first describe the approach used to approximate the makespan $\Cmax$.
First, the interval of all possible processing times $\iv{\eps \pmax, \pmax}$ is split into $(\log(\epsinv)$ growing intervals of size $2^i \eps \pmax$, for $i = \iv{0,\log(\epsinv)}$.
Each such interval $I_i$ is then split again into $\ceil*{\epsinv}$ subintervals $I_{i,k}=(b_{i,k},b_{i,k+1}]$ of equal size, with $b_{i,k} = 2^i \eps \pmax + k \eps^2 2^i \pmax$ and $k \in \iv{0, \ceil*{\epsinv-1}}$.
Each processing time is then rounded down to the next lower boundary $b_{i,k}$.
As this is the same rounding scheme as used by \cite{BDJR22}, with only the upper and lower bounds on the processing times $(1-2\eps)T$ and $\eps T$ being replaced by $\pmax$ and $\eps \pmax$, the following statement still holds with only slight modification:
\begin{lemma}[
	, name={\textcite[Lemma 6.2]{BDJR22}}
	, label=stmt:long_road_rounding
]
\begin{enumerate}
\item The number of different rounded processing times is at most $\ceil*{\epsinv} \ceil*{\log(\epsinv) + 1}$.
\item A schedule $\tilde \sigma$ of the rounded processing times implies a schedule $\sigma$ of the original processing time with $\Cmax(\sigma) \leq (1+\eps) \Cmax(\tilde \sigma)$.
\item We have $b_{i,k_1} + b_{i,k_2} = b_{i+1,(k_1+k_2)/2}$ for all $i$ and $k_1,k_2$ with $k_1 \mod 2 = k_2 \mod 2$. \stopphier
\end{enumerate}\leavevmode \end{lemma}

After the rounding, we scale each of the resulting processing times by $\epsinv \pmax^\inv$.
As each rounded processing time corresponds to some boundary $b_{i,k}$, and all boundaries are multiples of $\eps \pmax$, this scaling does not change any of the properties above.
After the scaling, each rounded processing times is in $\interval 1 \epsinv$, but not necessarily integral any more.
We again assume that machine speeds are all powers of $(1 + \eps)$ by rounding them up to the next power of $(1 + \eps)$, and let $\tau$ denote the number of different machine speeds after rounding, with $\smax$ being the largest speed after rounding and $\smin$ being the smallest.
In contrast do $d$, $\tau$ is not bounded in terms of $\epsinv$ after rounding.

After this procedure, it holds for the largest possible rounded processing time $\pmax = \epsinv$,
and for the smallest rounded processing time $\pmin = 1$.
Let here $p \in \interval \pmin \pmax^{d}$ be the vector of rounded job processing times and $\nu \in \N_0^{d}$ the corresponding multiplicity vector, \ie for $j \in \iv d$ there are $\nu_j$ jobs of rounded processing time $p_j$ in the instance. 
Analogously, as before, let $s \in \R^\tau$
be the vector of machine speeds and $\mu \in \N_0^\tau$ the corresponding multiplicity vector.
If not stated otherwise, we always assume that the instances we deal with in the upcoming section are instances after the application of this preprocessing procedure and are encoded by the vectors described above.
Note that this procedure (for large jobs) induces a multiplicative error of at most $(1 + \eps)^2 \in (1+\O(\veps))$, as each rounding step induces a multiplicative error of at most $(1 + \eps)$.
Exchanging $(1+\eps)$ by $(1-2\eps)$ in all previous steps, rounding up jobs instead of down and machine speeds down instead of up, and replacing $\leq$ by $\geq$ in \autoref{stmt:long_road_rounding}, 2., the procedure has same properties with an induced multiplicative error of at most $(1 - \eps)^2 \in (1-\O(\veps))$ for approximating the \enquote{maximize the minimum completion}-objective.

Now, with this rounding in hand, jobs with processing time $b_{i,k}$ and $b_{i,k'}$ with $k \mod k' = 2$ can be portrayed as a single job of processing time $b_{i+1,(k_1+k_2)/2} = b_{i,k} + b_{i,k'}$ according to \autoref{stmt:long_road_rounding}, 3.
This can now be used to replace configurations that contain more than $1$ job of the same parity in the same interval $i$ by configurations with a single job in the interval $i+1$ representing $2$ such jobs.
To correctly count the number of jobs distributed by such configurations for some given job set in an ILP formulation, a set $\hat{\mc C}$ of auxiliary configurations are introduced.
For $i',k'_1,k'_2$ with $k'_1 \mod 2 = k'_2 \mod 2$, let $\hat c = \hat c(i',k'_1,k'_2) \in \hat{\mc C}$ with $\hat c[i,k] = 2$ if $i = i'$ and $k = k'_1 = k'_2$, $\hat c[i,k] = 1$ if $i = i'$ and $k \in \{k'_1,k'_2\}$ and $k'_1 \neq k'_2$, and $\hat c[i,k] = -1$ if $i = i' + 1$ and $k = (k'_1 + k'_2)/2$. Else, let $\hat c[i,k] = 0$.
This implies that $\abs{\hat{\mc C}} \leq \epsinv[2] \log(\epsinv)$.
By definition, $\norm{\hat{c}}_1 \leq 3$, for all $\hat c \in \hat{\mc C}$.

Let $\mc C(\lb)$ be the set of configurations after rounding, for scaled $\lb \leq \poly(\epsinv)$.
Note that, for any $c \in \mc C(\lb)$ with $c[i,k_1] = 1$ and $c[i,k_2] = 1$, for some $i,k_1,k_2$ with $k_1 \mod 2 = k_2 \mod 2$, there exists some configurations $c' \in \mc C(\lb)$ and $\hat c \in \hat{\mc C}$, such that $c'$ has the same entries as $c$, except $c'[i,k_1] = c'[i,k_2] = 0$ and $c'[i+1,(k_1+k_2)/2] = c[i+1,(k_1+k_2)/2]+1$, such that $c = c' + \hat c$, \ie $c$ can be replaced by a combination of $c'$ and $\hat c$.
Now, let $\mc C_{red}(u) = \{c \in \mc C(u) : \norm{c}_1 \leq 2 \log(\epsinv)\}$.
As \textcite{BDJR22} have shown, one can transform a vector that encodes a schedule using only configurations $c \in \mc C(u)$ to a vector that encodes the same schedule using only configurations $c \in \mc C_{decr}(u) \cup \hat{\mc C}$ in linear time, and vice-versa (\cite[Lemma 6.3]{BDJR22}).
Note that $\norm c_1 \leq \Theta(\log(\epsinv))$, for all $c \in \mc C_{decr}(u) \cup \hat{\mc C}$, and thus, $\abs{C_{decr}(u) \cup \hat{\mc C}} \leq \log(\epsinv)^{\epsinv \log(\epsinv)} + \epsinv[2] \log(\epsinv)$.
We now use this configuration set in a modified \ipdyn to approximate dynamic $\QCmax$.
\vspace{-2em}
\begin{multicols}{2}
\begin{lpformulation}[{\mathrlap{\ipdynred}}]\lpeq*{}{}
	\lplabel{ip:dyn_red}
\lpobj*{min}{\lexobj(x_{decr})}
	\lpeq*{
		p^\intercal \super \tblue \nu
		\leq \alpha
	}{}
	\lpeq*{\sum_{c \in \mc C_{decr}(\ell)}  x_{t,c} = \mu_t - \super \tblue \mu_t}{t \in \iv \tau}
	\lpeq*{\sum_{\substack{t \in \iv \tau \\ c \in C_{decr}(\lb) \cup \hat{\mc C}}}  x_{t,c} \cdot c  + \super \tblue \nu = \nu}{}
\end{lpformulation}
\begin{lpformulation}
	\lpeq*{\text{\normalfont \underline{with variables}}}{}
	\lpeq*{\var x_{t,c} \in \N_0}{t \in \iv \tau, c \in C_{decr}(\lb) \cup \hat{\mc C}}
	\lpeq*{\var \super \tblue \nu \in \N_0^d}{}
	\lpeq*{\text{\normalfont \underline{with parameters}}}{}
	\lpeq*{\param \nu \in \N_0^d}{}
	\lpeq*{\param \super \tblue \mu \in \N_0^\tau}{}
	\lpeq*{\param \alpha \in \Q_{\geq 0}}{}
	\stopphier
\end{lpformulation}\end{multicols}
Let $x_{decr}$ be the vector of entries of $x$ of the form $x_{t,c}$, for $t \in \iv \tau$ and $c \in \mc C_{decr}(\lb)$.
As $\mc C_{decr} \subseteq \mc C$, applying $\lexobj$ to $x_{decr}$ still encodes the \enquote{minimizing the makespan} by the unique order of $\lexobj$ on $\mc C$.

As argued above, schedules encoded by configurations in $\mc C_{decr}(\lb) \cup \hat{\mc C}$ can be transformed to schedules encoded by configurations in $C(\lb)$ and vice-versa in linear time (\cite[Lemma 6.3]{BDJR22}).
This specifically implies that such transformed schedules have the same completion times on all machines, \ie we can transform a lexicographically minimal solution (with minimal maximum completion time) of \ipdynred to a schedule with configurations in $C(\lb)$ of minimal maximum completion time.
Now, in \ipdynred, the maximum absolute entry in every constraint besides the knapsack constraint
$$p^\intercal \super \tblue \nu \leq \alpha$$
is $\leq \Theta(\log(\epsinv))$, and every coefficient besides those in this constraint is integral.
We first make the coefficients of this constraint integral by using a different rounding scheme on the job processing in $\Interval {\eps \pmax} {\pmax}$ times here.
First, every jobs original processing time is rounded down to the value of the form $(1 + \eps)^i \cdot \eps \pmax$ for $i \in \N_0$.
Then, each processing time is rounded down again to the next integer multiple of $\eps \pmax$.
Eventually, we scale each processing time by $\epsinv \pmax^\inv$.
Thereafter, every such rounded and scaled processing time is in $\Interval {1} {\epsinv}$ and integral.
This procedure induces a multiplicative error of at most $(1 + \eps)^3 \in (1+\O(\veps))$, as each rounding step induces a multiplicative error of at most $(1 + \eps)$.
Moreover, there are again at most $d \in \O(\epsinv \cdot \log(\epsinv))$ possible processing times here.
To bring the total maximum absolute entry down to $\log(\epsinv)$, we use a technique introduced by \textcite{KPW20} called \enquote{Coefficient Reduction}.
\begin{lemma}[
	, name={\textcite[Lemma 8]{KPW20}}
	, label=stmt:coeff_red
]
Consider an instance $\{Ax = b, x\geq0\}$ of \textsc{ilp feasibility}, where $b \in \N^k$ and $A$ is a non-negative integer matrix with $k$ rows and $g$ columns. In polynomial time, this instance can be reduced to an equivalent instance $\{A'x = b', x\geq0\}$ of \textsc{ilp feasibility} where $A'$ is a $\{0,1\}$-matrix with $k' = \O(k \log(\norm{A}_{\infty}))$ rows and $g' = g + \O(k \log(\norm{A}_{\infty}))$ columns, and $b' \in \N^{k'}$ is a vector with $\norm{b'}_{\infty} = \O(\norm{b}_{\infty})$.
\end{lemma}

Here, \textsc{ilp feasibility} is defined as: given an ILP of the form $\{Ax = b, x\geq0\}$, where $A$ is an integer matrix with $k$ rows and $g$ columns, is there an $x \in \N^g$ that satisfies $Ax=b$?
The submatrix as given by the constraint $p^\intercal \super \tblue \nu \leq \alpha$ of \ipdynred is such an instance, with $A = p^\intercal$, $k = 1$, $g \in \O(\epsinv \log(\epsinv))$, $x = \nub$ and $b = \alpha$.
Here, $\norm{p^\intercal}_{\infty} \leq \epsinv$.
We directly apply \autoref{stmt:coeff_red} on this constraint to replace it by $\Theta(\log(\epsinv))$ constraints with coefficients in $\{0,1\}$ in \ipdynred, introducing an additional $\log(\epsinv)$ columns in the process.
For the new RHS $\alpha'$, $\norm{\alpha'}_{\infty} = \O(\norm{\alpha}_{\infty})$ holds.
Note that this $\alpha'$ may be not integral, but we deal with this while solving the ILP in \autoref{sec:ilpsolve}.
As the vector of variables $\nub$ itself is not changed by this replacement, it does change neither the feasibility nor optimality of solutions for \ipdynred.
The largest coefficient, and by extension the $\ell_1$-norm of any column, in the resulting ILP is now bounded by $\log(\epsinv)$.

Let $b$ be the RHS of the such further reduced \ipdynred at step $q$, $b'$ the RHS at step $q+1$.
As \Call{Insert}{} and \Call{Remove}{} are only dependent on the state parameters $\alpha$, $\mub$ and $\nu$, which are either not affected by the replacement of the configuration set and the replacement of constraints at all, or increased only by a constant factor in the case of $\alpha$, these two functions can be used as before to go from step $q$ to $q+1$ such that $\norm{b-b'}_1 \leq \poly(\epsinv,\pmax) \leq \poly(\epsinv)$, where the last inequality follows from the rounding and scaling.
An equivalent formulation with the same properties can be achieved for approximating dynamic $\QCmin$ by rounding down instead of up in each rounding step, changing \enquote{$\leq$} to \enquote{$\geq$} in the knapsack constraint, and using \Call{Insert'}{} and \Call{Remove'}{} to go from step $q$ to $q+1$, respectively.
The rest then follows equivalently, and we get solutions with an approximation guarantee of $(1-\O(\veps))$.
 \subsection{The Approximation Schemes}
If there are no small jobs ($p_j < \eps \pmax$) present in the instance, then we achieve an \ac{EPTAS} by using the preprocessing as described above to get \ipdynred, bounding the migration via \nfold sensitivity and solving the ILP with a state-of-the-art ILP algorithm.
Details on this can be found in \appendixref{sec:ilpsolve}.
With this, we prove \autoref{cor:noparam_nosmall}, which we restate here.
\noparamnosmall*
\begin{proof}
After preprocessing such an instance of dynamic $\QCmax$ (dynamic $\QCmin$) as described above, we use \autoref{stmt:ipdyn_solve},~(\ref{stmt:ipdyn_solve:rounded}) to solve the instance at any step $q$ in the claimed run time.
The migration between two solutions at steps $q$ and $q+1$ has the claimed bound by the bound on the sensitivity given in \autoref{stmt:sensitivity}~(\ref{stmt:sensitivity:rounded}).
\end{proof} For general instances, achieving an \ac{EPTAS} is not so easy. We still manage to achieve a result in the direction of an \ac{EPTAS}, by allowing an additional load error of $\O(\eps \pmax)$ on the solutions, and allowing the migration factor to be amortized.
We manage to bound the reassignment potential of the amortized migration factor in terms of $3 \eps \pmax$.
We do this by combining our \ac{EPAS} (\autoref{stmt:main_theorem}) with our general dynamic grouping framework (\autoref{stmt:dynamic_grouping}).
With this, we prove \autoref{cor:noparam_general}, which we restate here.
\noparamgeneral*
\begin{proof}
By \autoref{stmt:dynamic_grouping}, if there exists a robust algorithm for dynamic $\QCmax$ (dynamic $\QCmin$) with run time $g$ for jobs of size $\geq \eps \pmax$, where migration is bounded by a function $\beta(p_j)$ for each job $J$, then there is a robust algorithm for $\QCmax$ ($\QCmin$) for all jobs with amortized migration bounded by $2 \beta(3p_j)$, bounded reassignment potential $3 \eps \pmax$, run time at most $\O(g \cdot \poly(\epsinv))$, and with additional load error at most $(4 + \ceil*{\frac{3}{m}}) \eps \pmax$.
Such an algorithm exists by \autoref{cor:noparam_nosmall}, with $g = 2^{\O(\epsinv (\log \epsinv)^2)} + \O(\abs{I})$, and migration bounded by $\beta(p_j) \leq 2^{\O(\epsinv (\log \epsinv)^2)}$.
This proves the claim.
\end{proof}  	

\section{An Exact Approach}
\label{sec:simple}

It is well known that the Assignment-ILP, used to describe \eg offline $\QCmax$, is an \nfold ILP with block parameters $\nfoldr = d$, $\nfoldt = d$, $\nfolds = 1$, $\nfoldn = n$, and $\norm A_\infty = \pmax$ \cite{KK18}.
To use it for both dynamic $\QCmax$ and dynamic $\QCmin$, the idea is to add an objective (and auxiliary variables) to the Assignment-ILP in order to maintain solutions that are lexicographically minimal if converted to a solution of the Configuration-ILP.
However, we have to adjust the definition of lexicographically minimal solutions slightly: Configurations/Assignments to a machine of type $t$ with the same load are now assigned the same objective value.
This does not change the properties of lexicographically minimal solutions used in this paper, except for the uniqueness of solutions.
In this section, we refer to this modified variant of solutions that are lexicographically minimal after being converted to a solution to the Configuration-ILP simply by \emph{lexicographically minimal}.

In order to be able to incorporate this as a linear objective the idea is simple: For each machine and each possible load, we introduce would like to introduce an auxiliary $0$-$1$-variable that is $1$ iff the assignment on machine $i$ has the corresponding load.
This has two problems to overcome:
First, the number of variables necessary per machine could potentially be unbounded.
Second, the coefficients could become arbitrarily large.

To overcome the first problem, we proceed as follows:
For an instance at step $q$, we compute a solution where each machine $i$ has load at most $\at q \OPT s_i + \pmax$ and at least $\at q \OPT s_i - \pmax$ using a greedy algorithm.
Let $\tilde T$ denote its makespan.
Clearly, if we add or remove a job, the possible load values we need to consider are in $\interval {\at q \OPT s_i - 2\pmax} {\at q \OPT s_i + 2 \pmax}$.
Therefore, let us define 
\[
	\Gamma \colon \iv m \times \N_0 \to 2^{\N_0} \enspace , \qquad
	(i,T) \mapsto \Set { k \in \Interval{-2\pmax} {2\pmax} | s_i T + k \geq 0} \enspace ,
\]
i.e. $\Gamma(i,\tilde T)$ is the set of possible load values for machine $i$ if $\tilde T$ is the makespan of the greedy schedule.
Observe that 
\[ \abs*{\bigcup_{i \in \iv m} \Gamma(i, \tilde T)} \leq (4 \pmax + 1) \cdot \tau \enspace . \]
so there are only a few makespans we need to consider for our objective.
We can then introduce $4 \pmax + 1$ auxiliary variables for each machine, each representing a specific load/makespan.
Let $\lambda \colon \iv m \times \N_0 \to \N$ be the function that maps a machine and a load to the corresponding coefficient of (the modified variant of) $\lexobj$.

However, if we use $\floor{T s_i}$ as coefficient for a variable that represents makespan $T$ on machine $i$, the coefficients are unbounded, as stated before.
But, since there are only $4 \pmax + 1$ different possible makespans per machine, we can reduce the coefficients in each constraint where $\floor{T s_i}$ occurs by putting this value on the RHS of the ILP, and with our auxiliary variable measure how much the load assigned to any machine deviates from this value.
Since the deviation is at most $2 \pmax$, this reduces the largest coefficient to this value.
This approach is captured in the following ILP formulation.
\begin{lpformulation}{}
\lpobj*{min}{
		\sum_{i \in \iv m} \sum_{k \in \Gamma(i, \tilde T)} \lambda(i, \floor*{s_i \tilde T} + k) \cdot y_{i,k}
	}
\lpeq*{
		\sum_{i \in \iv m} x_{i,j} = \nu_j
	}{
		j \in \iv d
	}
\lpeq*{
		\sum_{ j \in \iv d} p_j x_{i,j} \leq s_i \tilde T + \sum_{k \in \Gamma(i, \tilde T)} y_{i,k} k
	}{
		i \in \iv m
	}
\lpeq*{
		\sum_{k \in \Gamma(i, \tilde T)} y_{i,k} = 1
	}{
		i \in \iv m
	}
	\lpeq*{\text{\normalfont \underline{with variables}}}{}
	\lpeq*{\var y_{i,k} \in \N_0}{i \in \iv m, k \in \Gamma(i, \tilde T)}
	\lpeq*{\var x_{i,j} \in \N_0}{(i,j) \in \iv m \times \iv d}
	\lpeq*{\text{\normalfont \underline{with parameters}}}{}
	\lpeq*{\param \tilde T \in \Q_{\geq 0}}{}
\end{lpformulation}
Note that for $\tilde T = \at q \OPT$, this ILP returns an optimal schedule for both $q$ and $q+1$, by setting $\nu = \at q \nu$ and $\nu = \at {q+1} \nu$ respectively.
Since the computed solutions are lexicographically minimal, and we allow loads up to a $\pmax$ deviation from $\at q \OPT$ in the ILP, the solution $x^*$ computed at step $q+1$ is also an optimal solution for $\tilde T = \at {q+1} \OPT$.
Thus, dynamic $\QCmax$ can be solved by, at each step $q$, guessing $\OPT$ and then solving the ILP for both steps $q$ and $q+1$ with $\tilde T = \at q \OPT$ and the respective job sets $\nu$.

Since we only added $\O(\pmax)$ auxiliary variables and one constraint per machine compared to the Assignment-ILP, the proposed ILP still is an \nfold ILP, with block parameters $\nfoldr = d$, $\nfoldt = d$, $\nfolds = 1$, $\nfoldN = n$, and $\norm A_\infty = \pmax$.
The ILP can thus be solved at any step $q$ in time $(\pmax)^{\O(d^2)} \O(n)$ using the algorithm by \textcite{DBLP:conf/soda/CslovjecsekEHRW21}.
Since, for computing solutions at steps $q$ and $q+1$, the only change in the RHS is the change in the job set, $\norm{b-b'}_1 = 1$ holds for RHS $b$ at $q$, $b'$ at $q+1$.
Therefore, the sensitivity, and by extension the migration factor, is bounded by $(\pmax)^{\O(d)}$ using \autoref{stmt:JLR}.
This result can easily applied to solve dynamic $\QCmin$ as well, by simply changing \enquote{$\leq$} to \enquote{$\geq$} in the second constraint. The rest then follows equivalently.
We again thank an anonymous reviewer for pointing out this approach and the necessary modifications to us.

\section{A (Dynamic) Grouping Technique}
\label{sec:frames}
	In this section, we generalize the grouping technique as know from offline scheduling on identical machines scheduling to online scheduling on uniform machines.
	There, for a makespan or minimum completion time guess $T$, the jobs of processing time $< \eps T$ are replaced by $\ceil{(\eps T)^\inv \cdot \sum_{j : p_j < \eps T} p_j}$ jobs of size $\eps T$.
We proceed similarly by replacing small jobs $\super * \tsmall J \defeq \set { j \in \mc J | p_j < \eps \pmax}$ by substitute jobs of size $\eps \pmax$.
	We call those substitute jobs \enquote{frames}.
	Let
\[
		F \defeq{\frac{\sum\limits_{j : p_j \leq \eps \pmax} p_j}{\eps \pmax}}
	\]
	We refer to the original instance by the term \emph{unframed instance}, whereas the one where small jobs are substituted by $f \geq F$ frames is called the \emph{$f$-framed instance}.
	We will later see how to choose $f$ appropriately.
	Naturally, $f \geq F$ should hold, but $f$ should not become too large in comparison to $F$.

	Before dealing with the more sophisticated problems rooted in the online setting, let us first verify that, if the set of small jobs is fixed, it does indeed suffice to work with frames and convert the resulting schedule in order to achieve the desired result. When dealing with the assignment of large jobs ($p_j \geq \eps \pmax$), we extend the vector of job types by one additional type for the frames, and use them as the $d+1$-th job type in our notation. 
	
Now, we show that if we have a schedule for the framed instance we can convert it into a schedule for the unframed instance with no larger makespan, while the overload increases by at most $\eps \pmax$.
\begin{lemma}[
	, label = stmt:convert_framed_to_unframed
]
		Given a schedule $\sigma$ with makespan $T$ (minimum completion time $T$) and for an $f$-framed instance, in polynomial time we can compute a schedule for the unframed instance with makespan $T$ (minimum completion time $T$) and additional load error $\eps \pmax$.
\end{lemma}
\begin{proof}
	This argument is analogous to the classical one for grouping on identical machines:
	A simple greedy algorithm (e.g. list-scheduling) yields the desired result.
\end{proof}

To be able to use the framed instance instead of the unframed one, we also need the other direction, i.e. prove that if there is a schedule with makespan $T$ (minimum completion time $T$) for the unframed instance, then there is also such a schedule with some additional load error for the framed instance.
Note that the additional load error does depend on $f - F$, which we assume $f-F$ to be upper bounded by a constant $k$, and lower bounded by $0$.
In order to be as precise as possible, in this statement we do not state the additional load error but instead allow a specific number of frames to be accounted with processing time $0$ on each machine.
The latter then does imply that the additional load error is at most that number times $\eps \pmax$ for approximating the makespan.
For approximating the minimum completion time, the lower bound on $f - F$ is enough to show a constant additional load error.
\begin{lemma}[
	, label = stmt:convert_unframed_to_framed
]
	Given a schedule $\sigma$ for the unframed instance with makespan $T$ (minimum completion time $T$), in polynomial time we can compute a schedule $\sigma'$ for $f$-framed instance for any $f \in \Interval F {F + k}$ for $k \in \N_0$, such that $\sigma'$ has makespan $T$ if $1 + \ceil*{\frac k m}$ frames per machine can be accounted with processing time $0$ ($\sigma'$ has minimum completion time $T - (1 + \ceil*{\frac k m}) \veps \pmax$).
\end{lemma}
\begin{proof}
	The argument here is also closely related to the classical one on identical machines.
	We show the statement by construction a schedule $\sigma'$ that uses \emph{at least} $f$ frames.
 	This then shows the claim on the makespan, as the removal of frames does not increase the makespan.
	In order to construct such a schedule $\sigma'$, for the non-small jobs, we use the very same assignment as in $\sigma$.
	Thus, all that is left to do is to find an assignment of the $f$ frames without exceeding the makespan $T$ (subceeding the minimum completion time $T$).
	For a machine $i$, let $\super * \tsmall J_i$ be the set of small jobs assigned to $i$ under $\sigma$.
	We then assign
	\[
		\phi_i \defeq 1 + \ceil*{\frac k m} + \floor*{\frac{\sum_{j \in \super * \tsmall J_i}p_j}{\veps \pmax}}
	\]
	frames onto machine $i$. 

	We now show that completion time of $i$ under $\sigma'$ is no larger than under $\sigma$ when accounting for $1 + \ceil*{\frac k m}$ frames per machine with processing time $0$, and no smaller than under $\sigma$ in any case.
	Therefor, consider the difference of the load between $\sigma'$ and $\sigma$:
	\[
		0 \geq \sum_{j \in \super * \tsmall J_i} p_j - \eps\pmax\enpar*{
			\enpar*{1 + \ceil*{\frac k m}} + 
			\floor*{\frac{\sum_{j \in \super * \tsmall J_i} p_j}{\eps\pmax}}
		} \geq -  \eps\pmax\enpar*{1 + \ceil*{\frac k m}} \enspace .
	\]
	Note that since we did not change the assignment of non-small jobs, we do not have to consider them here.
	Now, since we are allowed to account $1 + \ceil*{\frac k m}$ frames per machine with processing time $0$, we can conclude that $\sigma'$ then has makespan at most $T$.
	At the same time, any machine $i$ has at least the same load in $\sigma'$ as in $\sigma$, and removing at most $1 + \ceil*{\frac k m}$ frames per machine (until exactly $f$ frames are placed) reduces the load by at most $(1 + \ceil*{\frac k m}) \veps \pmax$ per machine, leading to a minimum completion time of at least $T - (1 + \ceil*{\frac k m}) \veps \pmax$.

	We now only have to show that by our procedure, at least $f$ frames are distributed:
	\begin{multline*}
		\sum_{i \in \iv m} \phi_i
		= \sum_{i \in \iv m} \enpar*{1 + \ceil*{\frac k m} + \floor*{\frac{\sum_{j \in \super * \tsmall J_i}p_j}{\eps \pmax}}}
		\\ \geq m \ceil*{\frac{k}{m}} + \sum_{i \in \iv m}\ceil*{\frac{\sum_{j \in \super * \tsmall J_i}p_j}{\eps \pmax}} 
		\geq k + F \geq f \qedhere
	\end{multline*}
\end{proof}

Now, let us focus on the online setting.
More precisely, how can we ensure that we have an appropriate number $F'$ of frames at any step, i.e. $F \leq f \leq F + \O(1)$, such that we have bounded amortized migration when small jobs are added/removed?
Therefor we have to ensure that \enquote{toggling} is not possible, \eg if some job is added, removed, added, removed, …, this may not cause a change of $f$ each time.
We do so by maintaining the following two invariants:
\begin{invariant}[
	, label = invar:frames_bounded
]
	\[ F \leq f \leq F + 3 \enspace . \stopphier \]
\end{invariant}
\begin{invariant}[
	, label = invar:frames_middle
]
	If $\Delta f \neq 0$, then $\at {q+1} f = \ceil*{\at {q+1} F}  + 1$.
\end{invariant}
This directly implies the algorithmic approach: we leave $f$ untouched until some step $q$, where the insertion/removal of a small job $j$ appears and afterwards \autoref{invar:frames_bounded} would be violated if $f$ was not changed.
We then set $\at {q} f \defeq \ceil*{\at q F} + 1$, which directly implies that both \autoref{invar:frames_middle} and \autoref{invar:frames_bounded} are always maintained.

Whenever $f$ changes, handling this change is equivalent to adding/removing a constant number of $\eps \pmax$ sized jobs, which is exactly our intent: in the following sections, we only have to deal with jobs whose size is at least $\eps \pmax$ and their insertion/removal.

\begin{lemma}[
	, label = stmt:frames_amortized_change
	, name  = Amortized Change Of The Number Of Frames
]
	There is a step-dependent potential $\at q \Phi \in \N_0$ such that $\at 0 \Phi = 0$ and for each step where a small job $j$ is either inserted removed we have
	\[
		\abs{\Delta\Phi} \leq 3 \eps \pmax
		\enspace , \qquad \tand \qquad
		\eps \pmax \abs {\Delta f} + \Delta\Phi \leq 3{p_j}
		\enspace . \stopphier
	\]
\end{lemma}
\begin{proof}
	For real numbers $a,b$, let $a \dotminus b \defeq \max\sing{a - b, 0}$.
	We set
	\[
		\at q \Phi \defeq 3 \eps \pmax \enpar*{
			\enpar*{\at q f \dotminus (\at q F + 2)}
			+
			\enpar*{\at q F + 1 \dotminus \at q f}
		}
		\enspace .
	\]		
	The first claim holds by construction of $\Phi$ and \autoref{invar:frames_bounded}.
	We now prove the second claim.
	Therefor, observe that at most one of the summands of $\Phi$ can be non-zero (or none, but not both).

	Consider some fixed step $q \to q+1$.

\begin{enumerate}[]
		\item \case[1]{$\Delta f = 0$,} i.e. we did not change the number of frames.
		Then $f \defeq \at q f = \at {q+1} f$ and $\at {q+1} F = \at q F \pm \frac{p_j}{\eps\pmax}$.
		We then have 
		\begin{align*}
			\frac {\abs{\Delta\Phi}} { 3 \eps \pmax } &= 
				\enpar*{f \dotminus (\at q F \pm \frac{p_j}{\eps\pmax} + 2)}
				+
				\enpar*{\at q F \pm \frac{p_j}{\eps\pmax} + 1 \dotminus f}
				\\ &\phantom{={}}\;
				-
				\enpar*{f \dotminus (\at q F + 2)}
				-
				\enpar*{\at q F + 1 \dotminus f}
			\intertext{and, since at most two of the summands in the definition of $\Phi$ can be non-zero (one positive and one negative)}
			&\leq \frac{p_j}{\eps\pmax} \enspace .
		\end{align*}
		Multiplying both sides with $3 \eps \pmax$ then shows the claim.
		
		\item \case[2]{$\Delta f \neq 0$.}
		Note that by \autoref{invar:frames_middle} we have $\at {q+1} f = \ceil*{\at {q+1} F} + 1$.
		This implies that
		\begin{multline*}
			\frac{\at {q+1} \Phi}{3 \eps \pmax} = 
				\underbrace{\enpar*{\ceil*{\at {q+1} F} + 1 \dotminus (\at {q+1} F + 2)}}_{= 0}
				+
				{\enpar*{\at {q+1} F + 1 \dotminus \at {q+1} f}}
				\\ = \at {q+1} F + 1 \dotminus \enpar*{\ceil*{\at {q+1} F} + 1}
				= 0
			\enspace .
		\end{multline*}
		In other words: whenever the number of frames is changed, the potential drops to $0$ thereafter.
\begin{enumerate}[]
			\item \case[2.1]{$3 \geq \Delta f > 0$,} i.e. we added some frames. Moreover, $\at {q+1} F = \at q F + \frac{p_j}{\eps\pmax}$, as if we needed to add a frame, this can only be due to some small job $j$ that was just inserted.

			Since we had to add frames, we know that otherwise \autoref{invar:frames_bounded} would have been violated.
			In particular, this implies
			\begin{equation*}
\at q f < \at {q+1} F = \at q F + \frac{p_j}{\eps \pmax} < \at q F + 1
			\enspace .
			\end{equation*}
			Thus,
			\[
				\frac{\at q \Phi}{3 \eps \pmax}
				= (\at q F + 1) \dotminus \at q f
				> (\at q F + 1) \dotminus \enpar*{ \at q F + \frac{p_j}{\eps\pmax} }
				\geq 1 - \frac{p_j}{\eps\pmax}
				\enspace , 
			\]
		which implies $\Delta \Phi < - 3 \eps \pmax + 3 p_j$.
		Moreover, due to \autoref{invar:frames_middle} we have
		\[ \at {q+1} f =  \ceil*{\at {q+1} F} + 1  \leq  \ceil*{\at {q} F} + 2 
		\leq \at q F + 3
		\enspace ,
		\]
		and thus $\Delta f \leq 3$ as $\at q f \geq \at q F$ by \autoref{invar:frames_bounded}.
This then show the claim due to
		\[
			\eps \pmax \abs {\Delta f} + \Delta \Phi \leq 3 \eps \pmax - 3 \eps \pmax + 3 p_j = 3 p_j
			\enspace .
		\]

			\item \case[2.2]{$\Delta f < 0$,} i.e we removed some frames. Moreover, $\at {q+1} F = \at q F - \frac{p_j}{\eps\pmax}$, as if we needed to remove a frame, this can only be due to some small job $j$ that was just removed.

			Since we had to remove frames, we know that otherwise \autoref{invar:frames_bounded} would have been violated.
			In particular, this implies
			\[
				\at q f  > \at {q+1} F + 3 = \at q F - \frac{p_j}{\eps \pmax} + 3 > \at q F + 2
			\enspace .
			\]
			Thus,
			\[
				\frac{\at q \Phi}{3 \eps \pmax}
				= \at q f \dotminus (\at q F + 2)
				> \enpar*{ \at q F - \frac{p_j}{\eps\pmax} + 3} \dotminus (\at q F + 2)
				\geq 1 - \frac{p_j}{\eps\pmax}
				\enspace ,
			\]
			
			which implies $\Delta \Phi < - 3 \eps \pmax + 3 p_j$.
			Moreover, due to \autoref{invar:frames_middle} we have
		\[ \at {q+1} f = \ceil*{\at {q+1} F} + 1 \geq \ceil*{\at {q} F} + 1 \geq \at q F + 1 \enspace ,
		\]
and thus $\Delta f \geq -2$ as $\at q f \leq \at q F + 2$ by \autoref{invar:frames_bounded}. 
\[
				\eps \pmax \abs {\Delta f} + \Delta \Phi \leq  2 \eps \pmax - 3 \eps \pmax + 3 p_j \leq 3 p_j
				\enspace ,
			\] as desired.
			\qedhere
		\end{enumerate}	
	\end{enumerate}

\end{proof}

Now, we need a more complex variant of \fullref{stmt:convert_framed_to_unframed} that ensures that a different assignment of frames and/or a different number of frames only leads to a bounded change of the assignment of the small jobs into the space occupied by frames.
\begin{lemma}[
	, label = , label = stmt:online_convert_framed_to_unframed
]
	Let $q$ be a step.
	Assume that \autoref{invar:frames_bounded} holds at $q$ and $q+1$.
	Let $\at q \sigma$ be a schedule for the $\at q f$-framed instance and $\at {q+1} \sigma$ for the $\at {q+1} f$-framed instance.
	Moreover, let $\at q {\super \tsmall \sigma}$ be an assignment of small jobs into the space occupied by frames in $\at q \sigma$, such that for all $i$ we have
	
	$$(1 + \at q \sigma_i(d+1))\eps \pmax \geq \sum_j{p_j \at q {\super \tsmall {\sigma_i}}(j)} \geq (\at q \sigma_i(d+1) - 3)\eps \pmax \enspace .$$
	
	Then, in polynomial time, we can find a schedule $\at {q+1} {\super \tsmall {\sigma}}$ for the small jobs at step $q+1$ such that such that for all $i$ we have 
	
	$$(1 + \at {q+1} \sigma_i(d+1))\eps \pmax \geq \sum_j{p_j \at {q+1} {\super \tsmall {\sigma_i}}(j)} \geq (\at {q+1} \sigma_i(d+1) - 3)\eps \pmax \enspace ,$$
	
	and migration for small jobs is bounded by $2 \sum_{i \in \mc M} |\Delta\sigma_i(d+1)|$.
\end{lemma}
\begin{proof}
	Let ${\super \tsmall \sigma} \defeq \at q {\super \tsmall \sigma}$. We modify ${\super \tsmall \sigma}$ in such a way that, after the modifications, setting $\at {q+1} {\super \tsmall \sigma} = {\super \tsmall \sigma}$ gives us $\at {q+1} {\super \tsmall \sigma}$ with the required properties.
	Let $\mc J^-$ be the small jobs to be removed, $\mc J^+$ the small jobs to be added.
	Consider the following procedure for modifying ${\super \tsmall \sigma}$.
	
	\begin{algorithm}[h]
\begin{algorithmic}[1]
	\Procedure{OnlineConvertFramedToUnframed}{$\at q {\super \tsmall \sigma}$, $\mc J^-$, $\mc J^+$}
		\State ${\super \tsmall \sigma} \gets \at q {\super \tsmall \sigma}$
		\State Remove all jobs in $\mc J^-$ from ${\super \tsmall \sigma}$
		\ForAll{$i \in \mc M \colon \sigma_i(d+1)\eps \pmax < \sum_j{p_j {\super \tsmall {\sigma_i}}(j)}$}
			\State Remove small jobs from $i$ until $\sigma_i(d+1)\eps \pmax \geq \sum_j{p_j {\super \tsmall {\sigma_i}}(j)}$
			\State Add these jobs to $\mc J^+$
		\EndFor
		\While{$\mc J^+ \neq \emptyset$}
			\State Remove a job $J$ from $\mc J^+$
			\State Place $J$ on a machine $i' = \max_{i \in \mc M}(\sigma_i(d+1) \eps \pmax - \sum_j{p_j {\super \tsmall {\sigma_i}}(j)})$
		\EndWhile
		\State $\at {q+1} {\super \tsmall \sigma} \gets {\super \tsmall \sigma}$
		
		\Return $\at {q+1} {\super \tsmall \sigma}$
	\EndProcedure
\end{algorithmic}
\caption{}
\label{algo:framed_to_unframed}
\end{algorithm}

	After this process, ${\super \tsmall \sigma}$ contains all small jobs of the instance.
	Moreover, by \autoref{invar:frames_bounded}, in every step of the iterative placement of jobs in $J^+$, there exists a machine $i$ with $\sigma_{i}(d+1)\eps \pmax \geq \sum_j{p_j {\super \tsmall {\sigma_i}}(j)}$.
	By definition, the algorithm always chooses such a machine $i$ to place the current $J \in \mc J^+$.
	Thus, after placing a small job of processing time $< \veps \pmax$ on $i$, $(1 + \at {q+1} \sigma_{i}(d+1))\eps \pmax \geq \sum_j{p_j \at q {\super \tsmall {\sigma_{i}}}(j)}$ holds.
	
	Also, if there was a machine $i'$ with $(\sigma_{i'}(d+1)-3)\eps \pmax >  \sum_j{p_j {\super \tsmall {\sigma_{i'}}}(j)}$ after the algorithm terminates, then $\sigma_{i}(d+1)\eps \pmax \geq \sum_j{p_j {\super \tsmall {\sigma_i}}(j)}$ for all $i \in \mc M$, as this implies that for all $i$, $\sigma_i(d+1) \eps \pmax - \sum_j{p_j {\super \tsmall {\sigma_i}}(j)}) \geq \sigma_{i''}(d+1) \eps \pmax - \sum_j{p_j {\super \tsmall {\sigma_{i'}}}(j)}) > \veps \pmax$ in every iteration of the while-loop, and one job of processing time $< \veps \pmax$ was added to any $i$ in each iteration.
	This would imply that $f > F+3$, a violation of \autoref{invar:frames_bounded}, and can therefore not occur.
	Thus, the process returns a schedule with
	
	$$(1 + \at {q+1} \sigma_i(d+1))\eps \pmax \geq \sum_j{p_j \at {q+1} {\super \tsmall {\sigma_i}}(j)} \geq (\at {q+1} \sigma_i(d+1) - 3)\eps \pmax \enspace .$$
	
	The only jobs to cause migration are the ones added to $\mc J^+$ after previously having been removed from some machine $i$.
	Such jobs are only removed if $(1 + \sigma_i(d+1))\eps \pmax < \sum_j{p_j {\super \tsmall {\sigma_i}}(j)}$ for machine $i$.
	This only occurs if $\Delta\sigma_i(d+1) < 0$ for $i$, and removing jobs of total processing time at most $|\Delta\sigma_i(d+1)| + \veps \pmax$ is enough to get $(1 + \sigma_i(d+1))\eps \pmax \geq \sum_j{p_j {\super \tsmall {\sigma_i}}(j)}$.
	Therefore, since $p_{\delta+1} = \veps \pmax$, at most $|\Delta\sigma_i(d+1)| + \veps \pmax \leq 2 |\Delta\sigma_i(d+1)|$ are migrated per machine $i$, which sums up to a total migration of at most $2 \sum_{i \in \mc M} |\Delta\sigma_i(d+1)|$.
	This shows the claim.
\end{proof}

We are now ready to prove the main theorem on dynamic grouping, which is basically just a chaining of the results above.

\stmtDynamicGrouping*
\begin{proof}
In the following, we describe our algorithmic approach.
At any step $q$, we set $f$ such that \autoref{invar:frames_bounded} and \autoref{invar:frames_middle} are maintained.
		Consequently, at step $q = 0$, let $f = 1$.
		
We first show the statement for $\QCmax$.
		Let $\sigma$ be a schedule for $\at q \nu$ for the unframed instance with makespan $T$. Due to \autoref{stmt:convert_unframed_to_framed}, there exists a schedule $\sigma'$ for the framed instance with makespan $T$, if $1+\ceil*{\frac{k}{m}}$ frames per machine can be accounted with processing time $0$. As $k \leq 3$ according to the invariants, this implies an additional load error of at most $1+\ceil*{\frac{3}{m}}$ per machine. We now want to go from $q$ to $q+1$, adding or removing some job $j$. If $j$ is large, we simply use the robust algorithm to add or remove it. If $j$ is small, we adjust $f$ accordingly. Due to \autoref{stmt:frames_amortized_change}, there is a step-dependent potential $\at q \Phi \in \N_0$ such that $\at 0 \Phi = 0$ and for each step where a small job $j$ is either inserted removed we have $\abs{\Delta\Phi} \leq 3 \eps \pmax$ and $\eps \pmax \abs {\Delta f} + \Delta\Phi \leq 3{p_j}$. Thus, incorporating the frames added or removed by the change $\Delta f$ using the robust algorithm has amortized migration at most $\beta(3 p_j)$ and bounded reassignment potential $3 \veps \pmax$, resulting in a schedule $\bar{\sigma}'$ for the framed instance at step $q+1$ with an additional load error at most $1+\ceil*{\frac{k}{m}}$. Since, by assigning the frames at step $q$ accordingly, all prerequisites of \autoref{stmt:online_convert_framed_to_unframed} are met, we can use the Lemma to construct a schedule $\bar{\sigma}$ for the unframed instance at step $q+1$ in polynomial time, migration bounded in $2 \sum_{i \in \mc M} |\Delta\sigma_i(d+1)| \leq 2 \beta(p_j)$. Since  $(1 + \at {q+1} \sigma_i(d+1))\eps \pmax \geq \sum_j{p_j \at q {\super \tsmall {\sigma_i}}(j)}$, this assignment here introduces an additional load error of at most $\veps \pmax$. Thus, we have found a schedule $\bar{\sigma}$ with amortized migration bounded in $2 \beta(3 p_j)$, bounded reassignment potential $3 \veps \pmax$, additional load error at most $\veps \pmax (2 + \ceil*{\frac{3}{m}})$, and in time $\O(g \cdot \poly(\pmax)) \in \O(g \cdot \poly(\epsinv))$.
		
We now show the statement for $\QCmin$.
	Let $\sigma$ be a schedule for $\at q \nu$ for the unframed instance with minimum completion time $T$. Due to \autoref{stmt:convert_unframed_to_framed}, there exists a schedule $\sigma'$ for the framed instance with minimum completion time $T -(1+\ceil*{\frac{k}{m}}) \geq T -(1+\ceil*{\frac{3}{m}})$, as $k \leq 3$ according to the invariants. 
	We now want to go from $q$ to $q+1$, adding or removing some job $j$.
	If $j$ is large, we simply use the robust algorithm to add or remove it.
	If $j$ is small, we adjust $f$ accordingly.
	Due to \autoref{stmt:frames_amortized_change}, here is a step-dependent potential $\at q \Phi \in \N_0$ such that $\at 0 \Phi = 0$ and for each step where a small job $j$ is either inserted removed we have $\abs{\Delta\Phi} \leq 3 \eps \pmax$ and $\eps \pmax \abs {\Delta f} + \Delta\Phi \leq 3{p_j}$.
	Thus, incorporating the frames added or removed by the change $\Delta f$ using the robust algorithm has amortized migration at most $\beta(3 p_j)$ and bounded reassignment potential $3 \veps \pmax$, resulting in a schedule $\bar{\sigma}'$ for the framed instance at step $q+1$.
	Since, by assigning the frames at step $q$ accordingly, all prerequisites of \autoref{stmt:online_convert_framed_to_unframed} are met, we can use the Lemma to construct a schedule $\bar{\sigma}$ for the unframed instance at step $q+1$ in polynomial time, migration bounded in $2 \sum_{i \in \mc M} |\Delta\sigma_i(d+1)| \leq 2 \beta(p_j)$.
	Since  $\sum_j{p_j \at q {\super \tsmall {\sigma_i}}(j)} \geq (\at {q+1} \sigma_i(d+1) - 3)\eps \pmax$, this assignment here introduces an additional load error of at most $3 \veps \pmax$. Thus, we have found a schedule $\bar{\sigma}$ with amortized migration bounded in $2 \beta(3 p_j)$, bounded reassignment potential $3 \veps \pmax$, additional load error at most $\veps \pmax (4 + \ceil*{\frac{3}{m}})$, and in time $\O(g \cdot \poly(\pmax)) \in \O(g \cdot \poly(\epsinv))$.

\end{proof}

\section{Solving \ipdyn Efficiently}
\label{sec:ipsolve}\label{sec:ilpsolve}
We present a way to solve both \ipdyn and \ipdynred efficiently.
To be more precise: We propose algorithms that, given a state, efficiently either compute an optimal solution or correctly conclude that the given state is invalid.
We use the underlying core observation to prove a sensitivity bound for \ipdyn and \ipdynred, which is important for online scheduling with bounded migration, and argue how we can use the aforementioned algorithms, or more precisely, any algorithms solving the aforementioned problems, to find a new solution for a changed right hand side with distance bounded as described in the sensitivity theorem.
We then refine our approach to speed up the computation.

One core observation is the following: If we assign any job to a machine $i \in \iv m$ with $s_i \OPT < \pmin$, then this machine would have a completion time $> \OPT$.
Hence, we can a priori exclude all machines that are too slow by preassigning the $\zero$ configuration to them:
\begin{observation}[
	, name = Support of $T$-valid solutions
	, label = obsi:Tvalid_supp
]
	If $(\alpha, \mub, \nu)$ is valid, then the associated solution $(x, \nub)$ has
	\[
		x_{t,c} = \begin{dcases}
			\mu_t & \tif c = \zero \\
			0 & \totherwise
		\end{dcases}
	\]
	for all $t \in \iv \tau$ with $s_t T < \pmin$.
\end{observation}

As a direct consequence, by temporarily removing all red machines of types where the corresponding variables are a priori known to be zero and all machine types that cannot be coloured red without violating \autoref{invar:machine_colouring} with $T \gets \OPT+$ (for $\QCmax$) $T \gets \OPT-$ (for $\QCmin$) before setting up the ILP formulation, the number of (red) machine types $\super \tred \tau$ is now bounded by $\O(\log_{1 + \eps}(\frac \lb \pmin)) \subseteq \O(\epsinv \log(\epsinv \pmax))$ in \ipdyn and by $\O(\epsinv \log(\epsinv))$ in \ipdynred.
\begin{lemma}[
	, name  = Sensitivity Bounds
	, label = stmt:sensitivity
]
	Let $S = (\alpha, \mub, \nu)$ and $\bar S = (\bar\alpha, \mub[\bar], \bar\nu)$ be valid states with $\norm{\Delta\nu}_1 + \abs{\Delta\alpha} + \norm{\Delta\mub}_1 \in \Q_{\geq 1} \ucup \sing 0$.
Then
	\begin{enumerate}[(i)]
	\item \label{stmt:sensitivity:param} The associated solutions of \ipdyn have a bounded distance from each other, i.e.
	\[
		\norm{\Delta x}_1 + \norm{\Delta\nub} 
		\leq \paramsens \cdot \enpar*{\norm{\Delta\nu}_1 + \abs{\Delta\alpha} + \norm{\Delta\mub}_1}
		\enspace .
	\]
	\item \label{stmt:sensitivity:rounded} The associated solutions of \ipdynred have a bounded distance from each other, i.e.
	\[
		\norm{\Delta x}_1 + \norm{\Delta\nub} 
		\leq \roundedsens \cdot \enpar*{\norm{\Delta\nu}_1 + \abs{\Delta\alpha} + \norm{\Delta\mub}_1}
		\enspace . \stopphier
	\]
	\end{enumerate}

\end{lemma}
\begin{proof}
\ipdyn is an \nfold ILP, with block parameters $\nfoldr = d + 1$, $\nfoldt = \max(d,\abs{\mc C(\lb)})$, $\nfolds = 1$, $\nfoldN = d + \tau$, and $\norm A_\infty = \epsinv \pmax$.
	\ipdynred is an \nfold ILP, with block parameters $\nfoldr \in \O(\epsinv \log(\epsinv))$, $\nfoldt \in \roundedsens$, $\nfolds = 1$, $\nfoldN \in \O(\epsinv \log(\epsinv))$, and $\norm A_\infty \leq \log(\epsinv)$.
	The theorem by \textcite{JLR20} (\autoref{stmt:JLR}) states that given an optimal integer solution $x$ to an \nfold ILP with constraint matrix $A$ and RHS $b$, \ie $Ax = b$, if changing the RHS to $b'$ is feasible with a finite optimum, then there is an optimal solution $x'$ to the latter, \ie $Ax' = b'$, such that
	\[
		\norm { x - x' }_1 \leq \norm { b - b'}_1  \O(\nfoldr \nfolds\norm A_\infty)^{\nfoldr \nfolds} \cdot \norm{b - b'}_1
		\enspace ,
	\]
	where $\nfoldr,\nfolds$ are block parameters and $\norm A_\infty$ is the largest absolute value of any entry in $A$.
	In our case, for \ipdyn we have $\nfoldr = d+1$, $\nfolds = 1$, and $\norm A_\infty \leq \epsinv \pmax$, which yields
	\[
		\norm { x - x' }_1 \leq \norm { b - b'}_1  \paramsens \enspace .
	\]
	and, for \ipdynred, we have $\nfoldr \in \O(\epsinv \log(\epsinv))$, $\nfolds = 1$, and $\norm A_\infty \leq \log(\epsinv)$, which yields
	\[
		\norm { x - x' }_1 \leq \norm { b - b'}_1  \roundedsens \enspace .
	\]
	We then substitute $\alpha$ by $\floor \alpha$ and $\bar \alpha$ by $\floor {\bar \alpha}$, which -- due to the integrality of the matrix -- does neither change the (unique) optimal solution nor validity; and then apply the bound we just derived.
	To see that this shows the claim, note that, unless $b = b'$, which is trivial, we have 
	\begin{multline*}
		\norm { b - b'}_1 \leq \norm{\Delta\nu}_1 + \abs{\Delta\floor{\alpha}} + \norm{\Delta\mub}_1
		\leq 1 + \norm{\Delta\nu}_1 + \abs{\Delta\alpha} + \norm{\Delta\mub}_1
	\\	\leq 2 \cdot \enpar*{\norm{\Delta\nu}_1 + \abs{\Delta\floor{\alpha}} + \norm{\Delta\mub}_1}
	\end{multline*}
	This proofs both claims. \qedhere
\end{proof}
Also recall that for valid states, the associated solution is unique.
Thus, due to this uniqueness of associated solutions, the application of the algorithm by \textcite{JR22} to \ipdyn and \ipdynred directly yields an efficient algorithm to \enquote{solve} both ILPs, i.e., to either compute the associated solution or decide that the given state is invalid:
\begin{lemma}[
	, name = Solving The ILPs Efficiently
	, label = stmt:ipdyn_solve 
]
	For a given state $S$ and makespan $T$, we can decide if $S$ is valid, and if so, compute the solution associated to $S$, that is, the optimal solution
	\begin{enumerate}[(i)]
	\item \label{stmt:ipdyn_solve:param} in time $\paramrt \allowbreak+ \O(\abs{I})$ for \ipdyn.
	\item \label{stmt:ipdyn_solve:rounded} in time $\roundedrt \allowbreak+ \O(\abs{I})$ for \ipdynred. \stopphier
	\end{enumerate}\leavevmode \end{lemma}
\begin{proof}
	As argued before, we can (temporarily) reduce the number of machine types we need to consider for red machines, which amounts to $\super \tred \tau \in \O(\epsinv \cdot \log(\epsinv \pmax))$ red machines in \ipdyn, and to $\super \tred \tau \in \O(\epsinv \cdot \log(\epsinv))$ in \ipdynred.
	We then apply the algorithm by \textcite{JR22}, respectively.
	Its run time of $\O(H)^{2M} \log^2(\norm b_{\infty})+ \O(MN)$ for an ILP of the form $Ax = b$, $x \geq \zero$ with a linear objective, $N$ variables, $M$ constraints, and $H$ being a bound on the hereditary discrepancy of the constraint matrix matrix $A$.
	In the case of \ipdyn, we have
	\[
		\hspace*{0.75cm}M \leq 1 + d + \super \tred \tau \in \O(d + \epsinv \cdot \log(\epsinv \pmax))
		\enspace , \qquad
		H \in \poly(\epsinv \cdot \pmax)
		\enspace ,
	\]
	\[
		\hspace*{+2.5cm}N \leq \super \tred \tau \lb^d \leq 2^{\O(d \log(\epsinv \pmax) + \log(\epsinv) \log \log (\epsinv \pmax))}
		\enspace ,
	\]
	where the second statement follows from the $\ell_1$-norm of any column in the matrix of \ipdyn being bounded in $\O(\lb) \subseteq (\pmax/\eps)^{\O(1)}$, and a bound of $H$ in the $\ell_1$-norm on columns by \textcite{BF81}.
	In the case of \ipdynred, we have
	\[
		\hspace*{1.5cm}M \leq 1 + d + \super \tred \tau \in \O(\epsinv \cdot \log(\epsinv))
		\enspace , \qquad
		H \leq \log(\epsinv)
		\enspace ,
	\]
	\[
		\hspace*{+1.5cm}N \leq \super \tred \tau \abs{C_{red}(u) \cup \hat{\mc C}} + \log(\epsinv) \leq \roundedrt
		\enspace ,
	\]
	where the second statement follows from the $\ell_1$-norm of any column in the matrix of \ipdynred being bounded in $\log(\epsinv)$ after rounding, scaling and reducing, and the same bound on $H$ in the $\ell_1$-norm of columns \cite{BF81}.
	As $a \cdot b \leq a^2 + b^2$ for all numbers $a$ and $b$, and $\log^2(\norm b_{\infty}) \in \O(\abs{I})$, we get the claimed run times.
\end{proof}

\section{Lexicographically Minimal Vectors}
\label{sec:lexmin}
\newcommand\lexleq{\mathbin{\super {\mathrm {lex}} \preceq}}
We make use of a lexicographic order on (partial) solution vectors, which was also used in prior work of \textcite{SV16}.
We write $a \lexleq b$ if $a$ is lexicographically smaller than $b$.
It is well-known that $\lexleq$ is a total order, and thus for any finite, non-empty set of vectors $M$ there is a unique minimal element in $M$.

Now, let us consider $n \in \N$, $\Delta \in \N$, and some set 
\[
	M \subseteq \Set { v \in \N_0^n | \norm v_1 \leq \Delta} \enspace .
\]	
Optimizing the following linear objective does result in the lexicographically minimal solution:
\[
	\sum_{i \in \iv n} (\Delta + 1)^{n - i} x_i \enspace . 
\]
To prove this more formally, let $x^\star$ be the lexicographically minimal vector in $M$ and $x$ an vector in $M$ that optimizes our objective.
We prove the claim by contradiction, assume that $x \neq x^\star$.
Then there is an index $j$ such that $x_i = x^\star_i$ for all $i \in \iv {j-1}$, but $x_j \neq x^\star_j$.
If $x^\star_j > x_j$, then $x \prec x^\star$ which contradicts the fact that $x^\star$ is lexicographically minimal.
If $x^\star_j < x_j$, then
\begin{align}
	\label{eq:Delta}
	&\sum_{i \in \iv n} (\Delta + 1)^{n - i} x^\star_i - \sum_{i \in \iv n} (\Delta + 1)^{n - i} x_i\\
	&= 	\sum_{i \in \Interval j n} (\Delta + 1)^{n - i} x^\star_i - \sum_{i \in \Interval j n} (\Delta + 1)^{n - i} x_i
	\\ &= 	(\Delta + 1)^{n-j} (x^\star_j - x_j) + \sum_{i \in \oInterval j n} (\Delta + 1)^{n - i} x^\star_i - \sum_{i \in i \in \oInterval j n} (\Delta + 1)^{n - i} x_i
	\\& \leq - (\Delta + 1)^{n-j} + \sum_{i \in \oInterval j n} \Delta (\Delta + 1)^{n-i}
	\\& < 0 \enspace .
\end{align}
Thus, we get that $x^\star$ has a better objective value that $x$, which is a contradiction to $x$ being an optimal solution.

It is easy to see that this result can be extended to any set $\mathcal D$ of indices with a given total order on them.
For our scheduling problem, our indices will be of the form $t,c$, where $t \in \iv \tau$ is a machine type and $c$ is a configuration.
The order $\sqsubseteq$ we use for these indices is as follows:
\begin{align*}
	(s,t) \sqsubseteq (s',t')
	\iff & \phantom{{}\lor{}}
	s_t^\inv p^\intercal c > s_{t'}^\inv p^\intercal {c'}
		\\ & \lor
		(s_t^\inv p^\intercal c = s_{t'}^\inv p^\intercal {c'} \land t < t')
		\\ & \lor 
		(s_t^\inv p^\intercal c = s_{t'}^\inv p^\intercal {c'} \land t = t' \land c \lexleq c')
		\enspace .
\end{align*}
This means that we primarily order the indices by the corresponding completion time descendingly, that is by $(t,c) \mapsto s_t ^\inv p^\intercal c$.
We then break ties by preferring the machine type of smaller index, and for same types the configuration that is lexicographically smaller.

Now, consider \ipdyn (but without objective, as we are constructing it here).
Let
\[
	X \defeq \Set { x | \exists \nub\qsep \text{$(x, \nub)$ is feasible}}
	\enspace .
\]
Since for any $x \in X$ we have $\norm x_1 \leq m$ (we cannot choose more configurations than we have machines), we can construct a linear objective as before that models lexicographic minimization with respect to our ordering over $X$.
We call this function $\lexobj$ and use it as objective for our \ipdyn.

Since for any $x \in X$ there is a unique value $\nub$ such that $(x, \nub)$ is feasible to \ipdyn, we get that \ipdyn has a unique optimal solution.
And this solution has a very nice property: when only considering chosen configurations, it has minimal makespan.
We prove this by contradiction.
Let $x^\star$ be the lexicographically minimal solution of \ipdyn.
Assume that there was a feasible solution $x$ to \ipdyn with makespan strictly smaller than $x^\star$, i.e. $\Cmax(x) < \Cmax(x^\star)$.
Let $(t,c) \in \supp(x^\star)$ be the smallest element of the support of $x^\star$ with respect to our given ordering of indices.
Then for all predecessors $(t', c')$ of $(t,c)$ we have $x^\star_{t',c'} = 0$.
As we primarily ordered the indices descendingly by completion time, and $\Cmax(x) < \Cmax(x^\star)$ by assumption, we can derive that $x_{t,c} = 0$ and $x_{t',c'} = 0$ for all predecessors $(t',c')$ of $(t,c)$.
This implies that $x$ is lexicographically strictly smaller that $x^\star$, which contradicts the minimality of $x^\star$.

\section{Robust Scheduling With Additive Error $\pmax$}
\label{sec:additive} \label{sec:greedy}
We first want to proof the following statement:
Given a schedule $\sigma$ for some job set where the completion time of every machine differs by at most $\pmax$, we can transform this schedule into a schedule $\bar \sigma$ for a different job set, where the completion time of every machine differs by at most $\pmax$, such that the migration is bounded only in the total processing time of difference in the two job sets plus $\pmax$ times the different number of jobs.
For this, we use a simple procedure quite similar to List Scheduling \cite{G69}, which suffices to return us a schedule with the claimed properties.
Let $\realOPT{(\Cmax)}$ the optimal objective value of an instance with $\Cmax$ objective function, and $\realOPT{(\Cmin)}$ that of an instance with $\Cmin$ objective.
We then proceed to show that such schedules have makespan at most $\realOPT{(\Cmax)} + \pmax$ and minimum completion time at least $\realOPT{(\Cmin)} - \pmax$, bounding the error sufficiently on machines of large capacity.
Let $C(\sigma,i) \defeq  p^\intercal \sigma_i / s_i$ is the completion time of machine $i$ in $\sigma$.
\begin{lemma}[Robust $\pmax$-additive Approximation]
\label{stmt:greedy}

Let $\mc M$ be a set of machines, and let $\nu$, $\bar {\nu}$ be two job sets.
	Let $\Delta \nu \defeq \nu - \bar \nu$ be the difference in jobs between multiplicity vectors $\nu$ and $\bar {\nu}$. 
	Let $\sigma \colon \iv m \times \iv d \to \N$ be a schedule with
	
	$$\max_{i,i' \in \mc M} (|C(\sigma,i)-C(\sigma,i')|) \leq \pmax \enspace .$$ 
	
	Then, we can find a schedule $\bar{\sigma} \colon \iv m \times \iv d \to \N$ with
	
	$$\max_{i,i' \in \mc M} (|C(\bar \sigma,i)-C(\bar \sigma,i')|) \leq \pmax \enspace ,$$
	
	such that migration between $\sigma$ and $\bar{\sigma}$ is bounded by $\sum_{j \in \iv d} \abs{\Delta \nu_j} (p_j + \pmax)$.
\end{lemma}

\begin{proof}\leavevmode
Consider the following procedure.
\begin{algorithm}[H]
\begin{algorithmic}[1]
	\Procedure{RobustGreedy}{$\nu, \sigma, \bar {\nu}$} \label{algo:greedy:line:begalgo}
		\State $\nu^- = \min(\nu - \bar \nu, 0)$
		\State $\nu^+ = \min(\bar \nu - \nu, 0)$
		\State For all $j \in \iv d$, remove $\nu^-_j$ of type $j$ from $\sigma$ \label{algo:greedy:line:remj-}
		\While{$\nu^+_j > 0$ for some $j \in \iv d$}
			\State $\nu^+_j = \nu^+_j - 1$
			\State $\sigma(i',j) = \sigma(i',j) + 1$, for some machine $i' = \{i \colon \min_{i \in \mc M}(C(\sigma,i))\}$ \label{algo:greedy:line:addj+}
		\EndWhile
		\While{$i' = \max_{i \in \mc M}(C(\sigma,i)),i'' = \min_{i \in \mc M}(C(\sigma,i)) \colon C(\sigma,i') - C(\sigma,i'') > \pmax$} \label{algo:greedy:line:loop_2}
			\State For some $j \in \iv d$
			\State $\sigma(i',j) = \sigma(i',j) - 1$
			\State $\sigma(i'',j) = \sigma(i'',j) + 1$
		\EndWhile
	\EndProcedure
\caption{Greedy procedure for scheduling jobs into machines with changed capacities}
\end{algorithmic}
\end{algorithm}
	Let $\bar \sigma$ be the schedule returned by the procedure.
	$\nu^+$ is the set of jobs to be added, and $\nu^-$ the set of jobs to be removed, respectively. Observe that $\Delta \nu = \nu^+ + \nu^-$.
	Jobs are migrated in $\sigma$ in \lineref{algo:greedy:line:remj-}, \lineref{algo:greedy:line:addj+}, and during each loop of \lineref{algo:greedy:line:loop_2}.
	\lineref{algo:greedy:line:remj-} and \lineref{algo:greedy:line:addj+} cause no migration by the definition of migration.
	For measuring the migration caused by \lineref{algo:greedy:line:loop_2}, note the following: If $C(\sigma',i') - C(\sigma',i'') > \pmax$ holds for $i',i''$ in the modified schedule $\sigma'$ before execution of the lines after \lineref{algo:greedy:line:loop_2}, this can only be due to the fact that jobs have been removed from $i'$ in \lineref{algo:greedy:line:remj-} or added to $i''$ in \lineref{algo:greedy:line:addj+}, as $|C(\sigma,i') - C(\sigma,i'')| \leq \pmax$ held before execution of \lineref{algo:greedy:line:loop_2} by assumption.
	As jobs are only added to a machine $i''$ with minimum completion time $C(\sigma,i'')$ over all machines, and the maximum processing time of each job is at most $\pmax$, \lineref{algo:greedy:line:addj+} at most only added jobs to $i''$ without violating  $|C(\sigma,i') - C(\sigma,i'')| \leq \pmax$ if it held before.
	
	Thus, $C(\sigma',i') - C(\sigma',i'') > \pmax$ can only occur due to jobs being removed from $i''$ by \lineref{algo:greedy:line:remj-}.
	Let $\nu^{i''}$ be the jobs removed from $i''$. It holds that $\zero < \nu^{i''} \leq \nu^-$, and the total processing time of jobs removed from $i''$ is $\sum_{j \in \iv d} \nu^{i''}_j p_j$.
	Again, as $|C(\sigma,i') - C(\sigma,i'')| \leq \pmax$ held before, it follows that  $C(\sigma',i') - C(\sigma',i'') \leq \pmax + \sum_{j \in \iv d} \nu^{i''}_j p_j$.
	Thus, moving a total of at most $\pmax + \sum_{j \in \iv d} \nu^{i''}_j p_j$ from machine $i' = \max_{i \in \mc M}(C(\sigma',i))$ to $i''$ in every execution of the loop at \lineref{algo:greedy:line:loop_2} implies $C(\sigma',i) - C(\sigma',i'') \leq \pmax$ for all $i \in \mc M$.
	As this is done for every $i'' = \min_{i \in \mc M}(C(\sigma',i))$, and there are at most $\sum_{j \in \iv d} |\nu^-_j| \leq \sum_{j \in \iv d} |\Delta \nu_j|$ such machines with $\nu^{i''} > \zero$, \lineref{algo:greedy:line:loop_2} terminates after moving jobs of total processing time at most $\sum_{j \in \iv d} \abs{\Delta \nu_j} (p_j + \pmax)$, returning a schedule $\bar \sigma$ with $\max_{i,i' \in \mc M} (|C(\bar \sigma,i)-C(\bar \sigma,i')|) \leq \pmax$.
	This shows the claim.
\end{proof}

Now, let $\fracOPT$ be the optimal fractional schedule to an instance of $\QCmax$ and $\QCmin$ respectively, \ie each machine of type $t$ machines has load exactly $s_t \frac{p^\intercal \nu}{s^\intercal \mu}$ in $\fracOPT$.
Note that $\realOPT{(\Cmin)} \leq \fracOPT \leq \realOPT{(\Cmax)}$.

\begin{lemma}
	\label{stmt:greedy_approx}
	Let $\sigma$ be a schedule for a multiplicity vector of jobs $\nu$ and a set of machines $\mc M$, where $\max_{i,i' \in \mc M} \allowbreak (|C(\sigma,i)-C(\sigma,i')|) \leq \pmax$.
	Then $\Cmax(\sigma) \leq \realOPT{(\Cmax)} + \pmax$ and $\Cmin(\sigma) \geq \realOPT{(\Cmin)} - \pmax$.
\end{lemma}
\begin{proof}
	Let $i \defeq \max_{i \in \mc M}(C(\sigma,i))$ and $i' \defeq \min_{i \in \mc M}(C(\sigma,i))$.
	By definition, $\Cmax(\sigma) = C(\sigma, i)$ and $\Cmin(\sigma) = C(\sigma, i')$.
	Also, by definition of $\fracOPT$, we know that $C(\sigma,i') \leq \fracOPT \leq C(\sigma,i)$.
	
	Since $C(\sigma,i) \leq C(\sigma,i') + \pmax$ by assumption, $\Cmax(\sigma) = C(\sigma,i) \leq C(\sigma,i') + \pmax \leq \fracOPT + \pmax \leq \realOPT{(\Cmax)} + \pmax$.
	Equivalently, $C(\sigma,i') \geq C(\sigma,i) - \pmax$ by assumption, and $\Cmin(\sigma) = C(\sigma,i') \geq C(\sigma,i) - \pmax \geq \fracOPT - \pmax \geq \realOPT{(\Cmin)} - \pmax$.
	This shows the claim.
\end{proof}

Thus, using \autoref{stmt:greedy} on machine sets where each machine has capacity at least $\ell \geq \epsinv \pmax$, the returned schedule is a $(1 + \eps)$--approximate solution for $\Cmax$ and a $(1 - \eps)$--approximate solution for $\Cmin$, at every time point $q$ with migration bounded by $p_j + \pmax$ for $j$ being the job to be added/removed, due to \autoref{stmt:greedy_approx} and,
\[
(1 + \eps) s_t \realOPT{(\Cmax)}
\geq s_t \realOPT{(\Cmax)} + \eps \ell
\geq s_t \realOPT{(\Cmax)} + \pmax \enspace ,
\]
\[
(1 - \eps) s_t \realOPT{(\Cmin)}
\leq s_t \realOPT{(\Cmin)} - \eps \ell
\leq s_t \realOPT{(\Cmin)} - \pmax \enspace ,
\]
for every $t \in \iv \tau$.
 	\section{Construction of legacy schedules}
\label{sec:conversion}
In this section, we want to show that a robust algorithm for the high-multiplicity encoding of an instance implies a robust algorithm for the original, non high-multiplicity instance, with the same migration factor and approximation guarantee.

We call a schedule $\sigma \colon \mc J \rightarrow \iv m$ for the original instance \emph{legacy schedule}, and a schedule $\sigma' \colon \iv m \times [d] \rightarrow \N_0$ for the high-multiplicity instance simply \emph{schedule}.
Let $type(J)$ be a function that returns the type $j \in \iv d$ of a job $J \in \mc J$.
We now say a legacy schedule $\sigma$ follows a schedule $\sigma'$ iff, for each $i \in [m]$ and $j \in [d]$, $\sigma'(i,j) = |\{J \colon \sigma^{-1}(i) = J \wedge type(J) = j\}|$, \ie each machine $i$ has the same number of jobs of each type $j$ in both schedules $\sigma,\sigma'$.
Thus, if $\sigma$ follows $\sigma'$, both schedules have the same objective value regarding $\Cmax$ and $\Cmin$.
We are now ready to prove a bound on the migration for legacy schedules in the the online setting as well.

\begin{lemma}
\label{lem:followingsched}
Let $\mc J$ be the job set for the non high-multiplicity instance at time $q$, $\bar{\mc J}$ the job set for the non high-multiplicity instance at time $\bar q$.
For a legacy schedule $\sigma \colon \mc J \to \iv m$ that follows some schedule $\sigma'$ for $q$, a schedule $\bar{\sigma}'$ for $\bar q$ with (amortized) migration bounded by $\beta$ compared to $\sigma'$, we can construct a legacy schedule $\bar{\sigma} \colon \bar {\mc J} \to \iv m$ that follows $\bar{\sigma}'$ in polynomial time, with (amortized) migration (compared to $\sigma$) bounded by $\beta$.
\end{lemma}
\begin{proof}
Consider \autoref{algo:convert}.
\begin{algorithm}[h]
\begin{algorithmic}[1]
	\Procedure{LegacyScheduleConvert}{$\sigma$,$\sigma'$,$\bar{\sigma}'$} \label{algo:leg:line:begalgo}
		\State $\mc J^- \gets \mc J \setminus  \bar {\mc J}$
		\State $\mc J^+ \gets \bar {\mc J}\setminus  \mc J$ \label{algo:leg:line:j+def}
		\State $\bar{\sigma} \leftarrow \sigma$
		\ForAll{$i \in \iv m, J \in \mc J^-$}
			\If{$\sigma(J) = i$}
				\State Remove $J$ from $i$ in $\bar{\sigma}$
			\EndIf
		\EndFor
		\State $\Delta(i,j):=\bar{\sigma}'(i,j)-\sigma'(i,j)$
		\ForAll{$i \in \iv m, j \in \iv d$}
			\If{$\Delta(i,j) < 0$}
				\State Let $X$ be a set of $\abs{\Delta(i,j)}$ jobs $J$ of type $j$ scheduled on $i$ in $\sigma$ \label{algo:leg:line:constrj}
				\State Remove all $J \in X$ from $i$ in $\bar{\sigma}$
				\State $\mc J^+ = \mc J^+ \cup X$ \label{algo:leg:line:j+add}
			\EndIf
		\EndFor
		\ForAll{$i \in \iv m, j \in \iv d$}
			\If{$\Delta(i,j) \geq 0$}
				\State Schedule $\abs{\Delta(i,j)}$ jobs of type $j$ on $i$ in $\bar{\sigma}$, and remove these jobs from $\mc J^+$ \label{algo:leg:line:sched}
			\EndIf
		\EndFor
		\Return $\bar{\sigma}$
	\EndProcedure
\end{algorithmic}
\caption{Procedure for constructing a legacy schedule $\bar{\sigma}$ that follows a schedule $\bar{\sigma}'$, given a legacy schedule $\sigma$ that follows a schedule $\sigma'$ which has bounded migration $\beta$ compared to $\bar{\sigma}'$}
\label{algo:convert}
\end{algorithm}

First note that \lineref{algo:leg:line:constrj} and \lineref{algo:leg:line:sched} can always be executed without an error:
As $\sigma$ is following $\sigma'$, there are exactly $\sigma'(i,j)$ jobs $J$ of type $j$ scheduled on $i$ in $\sigma$.
As $\bar{\sigma}'(i,j) \geq 0$ for all $i,j$, if $\Delta(i,j)< 0$, $\abs{\Delta(i,j)} \leq \sigma'(i,j)$ follows.
Similarly, for all $\Delta(i,j) \geq 0$, $\abs{\Delta(i,j)}$ many jobs of type $j$ are scheduled additionally on $i$ in $\bar{\sigma}'$ compared to $\sigma'$. These jobs are either jobs migrated from some other machines in $\sigma'$, or jobs that are in $\bar{\mc J}$ but not in $\mc J$. By \lineref{algo:leg:line:j+def} and \lineref{algo:leg:line:j+add}, all such jobs are included in $\mc J^+$ when \lineref{algo:leg:line:sched} is executed, and, since $\sigma$ follows $\sigma'$, there are always at least $\abs{\Delta(i,j)}$ jobs $J$ of type $j$ in $\mc J^+$ at the time of execution for each pair $i$ and $j$.

This resulting legacy schedule $\bar{\sigma}$ then follows $\bar{\sigma}'$ by construction of the procedure.
In total, the total processing time of jobs migrated from $\sigma$ to $\bar{\sigma}$ is at most $\beta$, since, for every job of type $j$ migrated from $\sigma'$ to $\bar{\sigma}'$, we migrated at most one job of type $j$ in our construction of $\bar{\sigma}$, proving the claim.
The run time of this procedure is linear in $|\mc J|$ and $m$, and thus linear in the size $\abs I$ of the non high-multiplicity instance $I$.
\end{proof}

Thus, using \autoref{lem:followingsched} to compute the legacy schedule of a schedule at each time $q$, we can at each time $q$ return a legacy schedule with the same objective value regarding $\Cmax$ and $\Cmin$ and bound on the migration as the schedule at time $q$. 	\end{appendix}

\end{document}